\documentclass[amsmath,amssymb,showpacs,showkeywords,twocolumn]{revtex4}
\usepackage{graphicx}
\usepackage{times}
\usepackage{mathrsfs}
\usepackage{amsmath}
\usepackage{graphicx}
\usepackage{epsfig}
\usepackage{dcolumn}
\usepackage{bm}
\usepackage{multirow}
\usepackage{pdfpages}
\usepackage[utf8x]{inputenc}
\DeclareUnicodeCharacter{2212}{-}
\begin{document}
\title{Unconventional spin valve effect in altermagnets induced by Rashba spin–orbit coupling and triplet superconductivity}

\author{Saumen Acharjee\footnote{saumenacharjee@dibru.ac.in}, Aklanta Dihingia\footnote{aklanta450@gmail.com}, Nayanav Sonowal\footnote{sonowalnayanav@gmail.com} and  Abyoy Anan Kashyap\footnote{kashyapabyoyanan@gmail.com}  }
\affiliation{Department of Physics, Dibrugarh University, Dibrugarh 786004, 
Assam, India}

\begin{abstract}
We theoretically investigate spin-dependent transport in altermagnet/triplet-superconductor/altermagnet (AM/TSC/AM) junctions in the presence of interfacial Rashba spin–orbit coupling (RSOC). Within a microscopic Bogoliubov–de Gennes scattering formalism, we compute angle and energy-resolved conductance, spin polarization, zero-bias response, and tunneling magnetoresistance (TMR) for nodal $p_x$ and chiral $p_x+ip_y$ superconductors. Although altermagnets carry no net magnetization, the momentum-dependent spin splitting, combined with RSOC, enables a pronounced spin valve effect without ferromagnetic electrodes. We show that conductance, spin polarization, and TMR exhibit distinct and robust fingerprints of the triplet pairing symmetry. For nodal $p_x$ superconductor, sign change induced surface Andreev bound states dominate subgap transport, producing strongly anisotropic conductance, giant zero-bias spin polarization, and a monotonic enhancement of TMR with increasing RSOC. In contrast, the chiral $p_x+ip_y$ state exhibits smoother conductance and polarization profiles governed by topological edge modes, resulting in broader, lobe-like TMR patterns with weaker sensitivity to interface transparency. Moreover, RSOC can acts as an electrically tunable spin-mixing knob, while the barrier strength controls coherence and energy selectivity, together enabling large, symmetry controlled spin filtering and magnetoresistance. Our results establish AM/TSC/AM junctions as a symmetry sensitive transport platform for realizing electrically tunable spin valve functionality and probing triplet pairing without ferromagnetic components.
\end{abstract}

\maketitle

\section{Introduction}
The spin valve effect arises from control of electronic transport via the relative orientation of magnetic order parameters.
It has received significant attention in  spintronics due to its wide range of applications in modern technologies, from magnetic sensing to nonvolatile memory devices \cite{Baibich1988,Binasch1989,Zutic2004,Linder2015,Wolf2001,Dieny1991, Li2026,Lagarrigue2026,Liu2026}. Conventionally, spin valves are realized in ferromagnetic multilayers, where magnetoresistance (MR) arises from spin-dependent scattering and the relative alignment of the uniform exchange fields \cite{Parkin2004,Fert2001,Butler2001,Bass2007}. In ferromagnet (FM)/superconductor (SC) hybrid junctions, the spin valve effect originates from spin dependent quasiparticle spectra in the FM leads together with the pair-breaking action of exchange fields on singlet Cooper pairs. Andreev reflection and proximity effects mediate the resulting spin-dependent transport across the interfaces~\cite{Zheng2000,Devizorova2019,StoddartStones2022,Devizorova2017,Leksin2015,Buzdin2005}.  However, FM-based spin valves intrinsically rely on net magnetization, giving rise to stray fields, slow magnetic dynamics, and limited symmetry tunability. So, it has limitation in their scalability and coherent control in nanoscale architectures \cite{Baltz2018,Chen2024,Liu2025}. A central challenge in this context is therefore to realize spin valve functionality without invoking macroscopic magnetization. 

Recent progress in magnetic quantum materials has uncovered altermagnets (AM), a distinct class of magnets characterized by anisotropic spin-split electronic bands despite vanishing net magnetization~\cite{feng1,occhialini,betancourt,smejkal1,smejkal2,smejkal3, smejkal4, Boruah2025}. Unlike antiferromagnets (AFM), AMs exhibit momentum-dependent spin splitting of electronic bands
protected by crystalline symmetries, producing FM-like transport responses in absence  macroscopic magnetization~\cite{smejkal1,smejkal2,smejkal3, smejkal4, Boruah2025}. Microscopically, the effective exchange field is odd under momentum inversion, $\mathbf{h}(\mathbf{k}) = -\mathbf{h}(-\mathbf{k})$, reflecting the interplay of noncollinear magnetic order and lattice symmetries~\cite{smejkal4}. As a result, AMs exhibit large momentum-space spin splitting while preserving zero macroscopic magnetization and suppressing stray fields, making them a promising platform for next-generation superconducting spintronic heterostructures~\cite{cheng11,sun1,ouassou11,papaj,beenakker1}. In this context, superconductors that efficiently couple to spin-polarized quasiparticles while remaining robust against exchange fields are therefore of particular interest~\cite{Tanaka1995,Hu1994}. Spin-triplet superconductors (TSCs) provide such a platform, supporting equal-spin Cooper pairs and spin-polarized supercurrents that are naturally compatible with spin-split electronic states~\cite{meng,trifunovic,annunziata,Tanaka2000}.
In contrast to singlet superconductors, triplet systems allow equal-spin Andreev reflection and exhibit pronounced sensitivity to spin–orbit coupling (SOC), thereby intertwining spin and momentum degrees of freedom and enabling efficient conversion of momentum-dependent spin textures into measurable spin-polarized transport signals~\cite{Eschrig2011,Linder2015}.

The absence of uniform magnetization in AMs raises the fundamental question of whether a spin valve effect can be realized without FM electrodes~\cite{Xiao2025}. Although the lack of a global magnetization axis might seem to preclude conventional magnetoresistive behavior, the momentum dependent spin texture of AMs provides an alternative route to spin selectivity~\cite{Zhang2025,Dou2025,Mukasa2025}. When quasiparticles traverse interfaces that mix momentum and spin, most notably those with interfacial Rashba spin–orbit coupling (RSOC), the intrinsic spin splitting of AMs can be converted into spin-polarized transport and MR responses~\cite{Manchon2015,Cheng2013,Cheng2014}. RSOC, arising from structural inversion asymmetry, generates momentum-dependent spin rotation and enables electrically tunable control of spin-dependent scattering~\cite{Bychkov1984,Buzdin2008,Manchon2015,Acharjee2023}. In superconducting hybrids, RSOC converts singlet correlations into triplet components, induces anomalous Josephson effects~\cite{acharjee2022,Acharjee2023,assouline1,yuan1}, and enables electrically controllable spin filtering~\cite{Eschrig2011,Buzdin2008}. Furthermore, RSOC combined with AM can provide a unique symmetry protected momentum-dependent spin polarized and spin-resolved transport system, enabling a magnetization-free and electrically tunable spin valve. Moreover, the introduction of a TSC further enriches this framework: unconventional $p$-wave pairing supports equal-spin Cooper pairs and spin-polarized supercurrents~\cite{Sigrist1991,Mackenzie2003,Ikegaya2018,Kallin2012}. Nodal $p_x$ states host zero-energy surface Andreev bound states that dominate subgap transport and enhance interfacial spin mixing~\cite{Ikegaya2018}, whereas chiral $p_x+ip_y$ superconductors support topological edge modes that govern low-energy transport~\cite{Kallin2012}. 

Motivated by these considerations, we propose and theoretically analyze a magnetization-free superconducting spin valve based on AM/TSC/AM junctions with interfacial RSOC. This platform enables an unconventional spin-valve effect in which electrically tunable RSOC converts the momentum-dependent spin texture of AMs into spin-polarized superconducting transport carried by triplet Cooper pairs. In contrast to conventional spin valves governed by the relative orientation of uniform magnetizations, the spin valve response here is controlled by the relative orientation of AM spin textures and RSOC. As a consequence, AM/TSC/AM junctions provide a symmetry-sensitive and electrically tunable platform for magnetization-free superconducting spintronics, with transport signatures that sharply discriminate between nodal and chiral triplet pairing.

The paper is organized as follows: In Sec. II, we discuss the theoretical framework for the AM/TSC/AM system with interfacial RSOC. In Sec. III, we analyze the tunneling conductance, spin polarization and tunneling magnetoresistance (TMR) and investigate the resulting spin valve effect in our proposed setup. Finally, Sec. IV presents our conclusions.

\section{Model and Formalism} 
\begin{figure}[t] 
\centerline 
\centerline{ 
\includegraphics[scale=0.26]{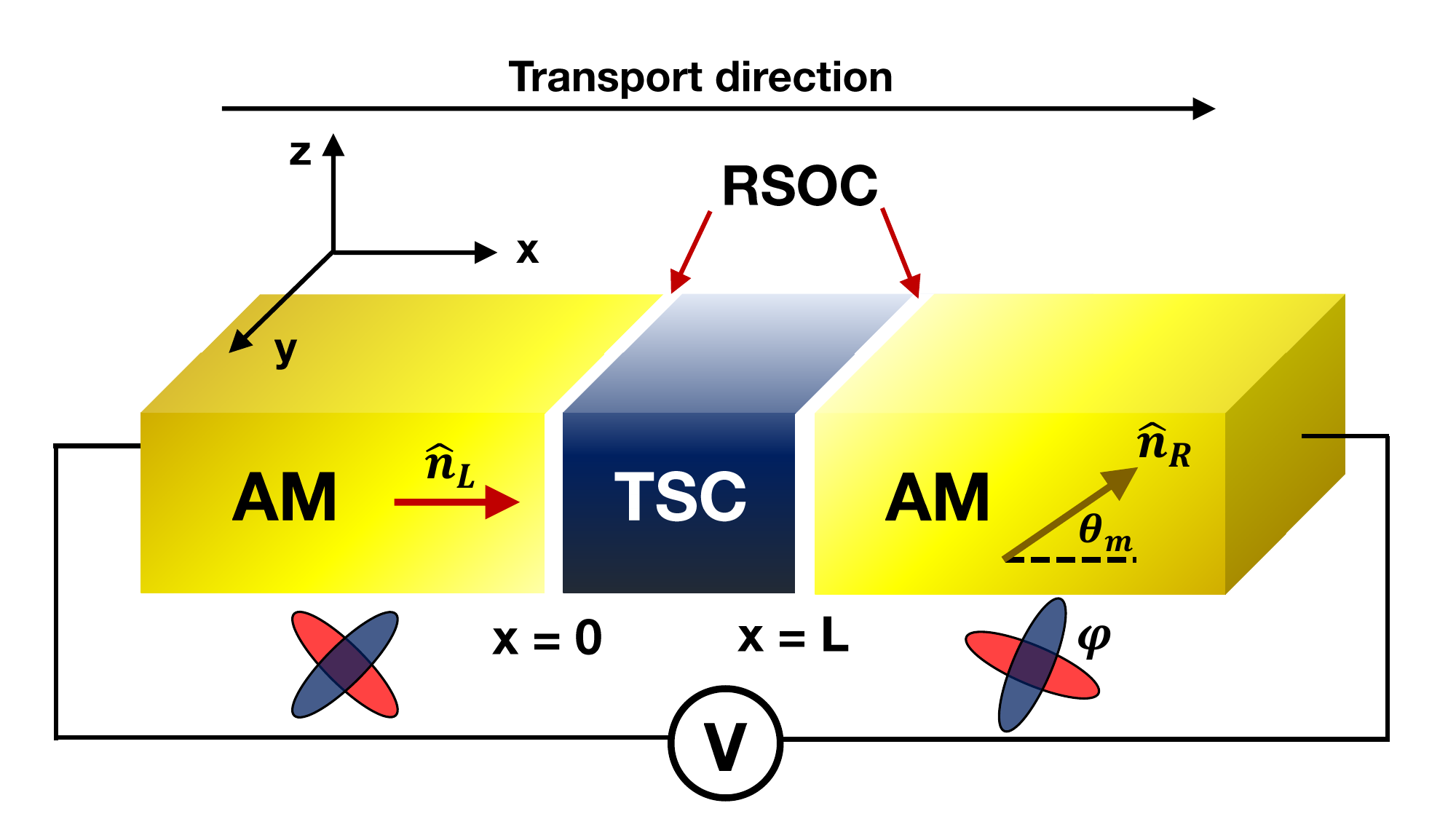} } 
\caption{ Schematic illustration of the altermagnet-triplet superconductor-altermagnet junction where a finite-length spin-triplet $p$-wave superconductor (TSC) of length $L$ is sandwiched between two altermagnetic leads (AM). The transport direction is taken along the $x$ axis, while the interface planes lie in the $yz$ plane. Rashba spin--orbit coupling (RSOC) is present at both AM/TSC interfaces and is modeled by delta-function barriers. The relative N\'eel order parameter in the left and right AM regions are misaligned by an angle $\theta_m$. A bias voltage $V$ is applied across the junction to drive quasiparticle transport. } 
\label{fig1} 
\end{figure}

We consider a two-terminal spin valve junction composed of an altermagnet-triplet $p$-wave superconductor-altermagnet (AM/TSC/AM) heterostructure, as schematically illustrated in Fig.~\ref{fig1}. The transport direction is chosen along the $x$ axis, with the two interfaces located at $x= 0$ and  $L$. Due to translational invariance along the $y$ direction, the transverse momentum $k_y$ is conserved. The central region ($0 < x < L$) consists of a spin-triplet $p$-wave superconductor characterized by equal-spin Cooper pairing. The AMs possess zero net magnetization but exhibit spin-split Fermi surfaces protected by crystalline symmetries, resulting in spin-dependent charge transport. The left ($x < 0$) and right ($x > L$) AMs are characterized by distinct orientations of their N\'eel vectors, parameterized by an angle $\theta_m$ in spin space. We define the N\'eel vectors in the left and right AMs as \begin{align} 
\mathbf{\hat{n}}_{j} &= (\cos\theta_{j},\sin\theta_{j},0), \quad j\in\{\text{L,\,\,R}\} 
\label{eq1} 
\end{align} 
with the relative misalignment angle $\theta_m \equiv \theta_R - \theta_L$. Despite the absence of net magnetization, the relative angle $\theta_m$ controls the spin valve configuration through the momentum-dependent spin texture of the AMs. In this work we consider the N\'eel vectors in the $xy$ plane, which is necessary to capture the full spin valve physics of AMs and ensures compatibility with equal-spin triplet pairing in the central TSC region. 

At each AM/TSC interface, inversion symmetry is locally broken, giving rise to RSOC induced by strong interfacial atomic spin-orbit coupling. Thus, the proposed system combines momentum-dependent spin polarization intrinsic to AMs, equal-spin triplet superconductivity in the central region, spin–momentum locking induced by interfacial RSOC, and N\'eel-vector-misalignment dependent spin filtering. Importantly, in contrast to conventional FMs, AMs break time-reversal symmetry without developing a net magnetization. Instead, their spin polarization alternates in momentum space, leading to a qualitatively distinct form of superconducting spin valve physics that has no direct analogue in FM-based heterostructures. 
To describe superconducting proximity effects and Andreev reflection at the AM/TSC interfaces, we introduce the Nambu field-operator basis $ \Psi = \left(\psi_{\uparrow} , \psi_{\downarrow}, \psi^\dagger_{\uparrow}, \psi^\dagger_{\downarrow}\right)^\text{T}$ where $\psi_s$ ($\psi_s^\dagger$) annihilates (creates) an electron with spin $s$. In this basis, the quasiparticle excitation spectrum of the heterostructure is governed by the Bogoliubov--de Gennes (BdG) Hamiltonian $\check{\mathcal{H}}_{\mathrm{BdG}} = \check{\mathcal{H}}_{\mathrm{AM}} + \check{\mathcal{H}}_{\mathrm{TSC}}$ where, $\check{\mathcal{H}}_{\mathrm{AM}}$ and $\check{\mathcal{H}}_{\mathrm{TSC}}$ describe the AM and the TSC regions, respectively. The BdG Hamiltonian for an altermagnetic region with an arbitrarily oriented N\'eel vector takes the block-diagonal form~\cite{Boruah2025,papaj,cheng11} 
\begin{equation} 
\check{\mathcal{H}}_{\mathrm{AM}}(\mathbf{k}) = \begin{pmatrix} 
\hat{\mathcal{H}}_0(\mathbf{k}) & 0 \\ 0 & -\sigma_y \hat{\mathcal{H}}_0^{*}(-\mathbf{k}) \sigma_y 
\end{pmatrix}, 
\label{eq2} 
\end{equation} 
which explicitly reflects particle--hole symmetry in the absence of intrinsic superconducting pairing within the AM region. 

The normal-state single-particle Hamiltonian in spin space is given by~\cite{cheng11}
\begin{equation} 
\hat{\mathcal{H}}_0(\mathbf{k}) = \left( \frac{\hbar^2 k^2}{2m} - \mu_{\mathrm{AM}} \right)\hat{I} + \mathbf{h}^j_{\mathrm{AM}}(\mathbf{k}) \cdot \boldsymbol{\sigma}, \label{eq3} 
\end{equation} 
where $\hat{I}$ is an identity matrix, $\boldsymbol{\sigma}=(\sigma_x,\sigma_y,\sigma_z)$ are the Pauli matrices, $m$ is the effective mass, and $\mu_{\mathrm{AM}} = \mu_{\mathrm{AM}}^L \Theta(-x) + \mu_{\mathrm{AM}}^R \Theta(x-L)$ represents the chemical potential of the AM region. The vector $\mathbf{h}^j_{\mathrm{AM}}(\mathbf{k})$ represents the momentum-dependent single particle AM exchange field with $j\in\{\text{L,\,\,R}\}$ corresponds to left and right AM regions. A defining property of AMs is the absence of a net magnetization despite explicit breaking of time-reversal symmetry. Microscopically, this follows from the antisymmetry condition~\cite{smejkal1} \begin{equation} 
\mathbf{h}^j_{\mathrm{AM}}(-\mathbf{k}) = - \mathbf{h}^j_{\mathrm{AM}}(\mathbf{k}), \label{eq4} 
\end{equation} 
which ensures that the Brillouin-zone-averaged magnetization vanishes identically. Nevertheless, at each individual momentum $\mathbf{k}$ the spin degeneracy is lifted, giving rise to spin-split Fermi surfaces with opposite spin polarization at $\pm\mathbf{k}$. The single-particle AM exchange field in region can be written as 
\begin{equation} 
\mathbf{h}_{\mathrm{AM}}^{j}(\mathbf{k}) = h_0\, f(\mathbf{k})\,\hat{\mathbf n}_j. \label{eq5} 
\end{equation} 
\noindent where $h_0$ sets the overall strength of the AM splitting and the dimensionless function $f(\mathbf{k})$ gives the crystal symmetry allowed momentum dependence of the AM exchange field and satisfies the antisymmetry condition $f(-\mathbf{k})=-f(\mathbf{k})$, ensuring the absence of a net magnetization upon Brillouin-zone averaging. In terms of this form factor, the normal-state single-particle Hamiltonian of the AM region can be written compactly as 
\begin{equation} 
\hat{\mathcal{H}}_0(\mathbf{k}) = \left( \frac{\hbar^2 k^2}{2m} - \mu_{\mathrm{AM}} \right)\hat{I} + h_0\, f(\mathbf{k})\, \hat{\mathbf{n}}_j \cdot \boldsymbol{\sigma}, \label{eq6} 
\end{equation} 
The Eq.~\eqref{eq6} highlights that spin splitting in an AM is intrinsically momentum selective and locked to the crystal symmetry through $f(\mathbf{k})$, rather than being spatially uniform as in conventional FMs. 
The parameter $h_0$ denotes the amplitude of the AM exchange field, which produces a momentum-dependent spin splitting $h_0 f(\mathbf{k})$ that yields zero net magnetization but finite spin polarization on the Fermi surface. Unlike the momentum-independent exchange field in a FM, $h_0$ governs symmetry-driven spin-selective transport in AM. Moreover, it controls the strength of unconventional spin filtering and the tunnelling conductance in the AM/TSC/AM junction.

For tetragonal or orthorhombic AMs the exchange field is a $d$-wave–like, for which the form factor can be expressed by the symmetry allowed form as~\cite{smejkal1}
\begin{equation} 
f(\mathbf{k}) =
\tilde{\alpha}_1\,\frac{k_x k_y}{k_F^2}
+
\frac{\tilde{\alpha}_2}{2}\,
\frac{k_x^2 - k_y^2}{k_F^2},
\label{eq7}
\end{equation} 
where, $k_F = \sqrt{2m\mu_{\mathrm{AM}}}
$. Here the two contributions transform as the $B_{2g}$ and $B_{1g}$ irreducible representations of the underlying point group, respectively~\cite{smejkal2}. 
The dimensionless coefficients $\tilde{\alpha}_1$ and $\tilde{\alpha}_2$
control the relative weight of the symmetry-allowed channels.
To ensure a stable metallic state in the AM region, they must satisfy
$\tilde{\alpha}_1^2 + \tilde{\alpha}_2^2 < \tilde{\alpha}_c$,
where $\tilde{\alpha}_c$ is a constant of order unity, which guarantees an elliptic Fermi surface and avoids unphysical hyperbolic band dispersions. 

The BdG Hamiltonian for the central TSC region is defined as~\cite{trifunovic, annunziata,Eschrig2011,Cheng2013,Tanaka2000}
\begin{equation} 
\check{\mathcal{H}}_{\mathrm{TSC}}(\mathbf{k}) = \begin{pmatrix} \hat{\mathcal{H}'}_0(\mathbf{k}) & \hat{\Delta}(\mathbf{k}) \\[4pt] \hat{\Delta}^{\dagger}(\mathbf{k}) & -\hat{\mathcal{H}}_0^{'*}(-\mathbf{k}) 
\end{pmatrix} 
\label{eq8} 
\end{equation} 
where $\hat{\mathcal{H'}}_0(\mathbf{k})$ is the single-particle Hamiltonian in the TSC region which is assumed to be spin degenerate and is given by 
\begin{equation} 
\hat{\mathcal{H'}}_0(\mathbf{k}) = \left( \frac{\hbar^2 k^2}{2m} - \mu_{\mathrm{S}} \right)\hat{I} \equiv \xi_{\mathbf{k}}\,\hat{I} , 
\label{eq9} 
\end{equation} 
where $m$ is the effective mass, $\mu_{\mathrm{S}}$ is the chemical potential of the TSC region, and $\xi_{\mathbf{k}}$ denotes the normal state dispersion. The gap matrix $\hat{\Delta}(\mathbf{k})$ in spin space for the TSC can be defined as~\cite{Cheng2013}
\begin{equation} 
\hat{\Delta}(\mathbf{k}) = \begin{pmatrix} 
\Delta_{\uparrow\uparrow}(\mathbf{k}) & \Delta_{\uparrow\downarrow}(\mathbf{k}) \\ \Delta_{\downarrow\uparrow}(\mathbf{k}) & \Delta_{\downarrow\downarrow}(\mathbf{k}) \end{pmatrix} = \big[ \mathbf{d}(\mathbf{k}) \cdot \boldsymbol{\sigma} \big]\, i\sigma_y , 
\label{eq10} 
\end{equation} 
where $\mathbf{d}(\mathbf{k})$ is the vector order parameter characterizing the spin structure of the triplet Cooper pairs. The Fermionic antisymmetry of the pairing potential, $\hat{\Delta}(\mathbf{k}) = -\hat{\Delta}^{\mathrm{T}}(-\mathbf{k})$, requires the $\mathbf{d}$ vector to be odd in momentum, $\mathbf{d}(\mathbf{k}) = -\mathbf{d}(-\mathbf{k})$, for unitary triplet states. We focus on equal-spin pairing, for which the $\mathbf{d}$ vector is chosen to be aligned along the $\hat{\mathbf{z}}$ axis~\cite{Cheng2014}, 
\begin{equation} 
\mathbf{d}(\mathbf{k}) = \Delta_0\, \hat{\mathbf{z}}\, g(\mathbf{k}), \label{eq11} 
\end{equation} 
where $\Delta_0$ is the superconducting gap amplitude and $g(\mathbf{k})$ is an odd-parity orbital form factor. For $p_x$-wave pairing, $g(\mathbf{k})\propto k_x$, while for chiral $p_x \pm i p_y$ pairing, $g(\mathbf{k})\propto k_x \pm i k_y$. Since the AM eigenstates are spin polarized in the $xy$ plane, the choice of equal-spin triplet pairing with $\mathbf{d} \parallel \hat{\mathbf{z}}$ permits nonvanishing Andreev reflection processes without requiring spin-flip scattering at the AM/TSC interface. This alignment maximizes the coupling between spin-polarized quasiparticles in the AM leads and the triplet Cooper pairs in the TSC region. The quasiparticle spectrum can be obtained by diagonalizing Eq.~\eqref{eq8}, which can be read as 
\begin{equation} 
E_{\mathbf{k}} = \pm \sqrt{ \xi_{\mathbf{k}}^{\,2} + |\mathbf{d}(\mathbf{k})|^2 } = \pm \sqrt{ \xi_{\mathbf{k}}^{\,2} + \Delta_0^2 |g(\mathbf{k})|^2 }, 
\label{eq12} 
\end{equation} 
so that the momentum-dependent superconducting gap magnitude is $\Delta_{\mathbf{k}} = \Delta_0 |g(\mathbf{k})|$. 

\subsection{Scattering formalism} For an electron incident from the left AM region with spin $\sigma=\pm1$ and excitation energy $E$, the quasiparticle wave function in the left AM region can be written as~\cite{Boruah2025} 
\begin{align} 
\Psi^{\mathrm{L}}_{\mathrm{AM},\sigma}(x<0) = &\; \Phi^{+}_{e \sigma}\, e^{i k_{e,\sigma} x} + \sum_{\sigma'} b_{\sigma \sigma'}\, \Phi^{-}_{e \sigma'}\, e^{-i k_{e,\sigma'} x} \nonumber\\ &+ \sum_{\sigma'} a_{\sigma \sigma'}\, \Phi^{-}_{h \sigma'}\, e^{i k_{h,\sigma'} x}, 
\label{eq13} 
\end{align} 
where $b_{\sigma \sigma'}$ and $a_{\sigma \sigma'}$ denote the normal and Andreev reflection amplitudes respectively. 

Here the spin index $\sigma=\pm1$ labels eigenstates of the AM exchange field rather than projections along a fixed laboratory axis. Since the AM single-particle Hamiltonian $\hat{\mathcal{H}}_0(\mathbf{k})$ does not commute with $\sigma_z$ for a generic N\'eel-vector orientation $\theta_m$, the natural spin basis in the AM regions is given by the eigenstates of $\hat{\mathbf{n}}_j\cdot\boldsymbol{\sigma}$, The corresponding spinors can be written as \begin{equation} 
\chi^{j}_{\sigma} = \frac{1}{\sqrt2}\begin{pmatrix}1\\\sigma e^{i\theta_j}\end{pmatrix}, 
\label{eq14} 
\end{equation} 
which represent spin quantization along the N\'eel vector direction $\hat{\mathbf{n}}$. Thus, the electron and hole-like spinors in the AM regions are given by 
\begin{equation} 
\Phi^{\pm}_{e \sigma} = \begin{pmatrix} \chi^j_\sigma\\ 0 \end{pmatrix}, \qquad \Phi^{\pm}_{h \sigma} = \begin{pmatrix} 0\\ \chi_\sigma^{j*} 
\end{pmatrix}. 
\label{eq15} 
\end{equation} 
The longitudinal wave vectors of electron (hole) in the AM region can be written as~\cite{Boruah2025} 
\begin{equation} 
k^{(\pm)}_{e(h),\sigma} = \frac{ - \sigma \alpha_1 k_y \pm \sqrt{ 2\mathcal{Q}_1\left(\mu_{\mathrm{AM}} +(-) E\right) + \mathcal{Q}_2 k_y^2 } }{ \mathcal{Q}_1 }, \label{eq16} 
\end{equation} 
where, $\mathcal{Q}_1 = \frac{\hbar^2}{m} + \sigma \alpha_2$ and $ \mathcal{Q}_2 = \alpha_1^2 + \alpha_2^2 $ Similarly, the wave function in the right AM region can be written as 
\begin{align} 
\Psi^{\mathrm{R}}_{\mathrm{AM},\sigma}(x>L) = &\sum_{\sigma'} t^{e}_{\sigma \sigma'}\, \Phi^{+}_{e \sigma'}\, e^{i k_{e,\sigma'}x} \nonumber\\ &+ \sum_{\sigma'} t^{h}_{\sigma \sigma'}\, \Phi^{+}_{h \sigma'}\, e^{-i k_{h,\sigma'}x}, 
\label{eq17} 
\end{align} 
where $t^{e}_{\sigma \sigma'}$ and $t^{h}_{\sigma \sigma'}$ denote the transmission amplitudes for electrons and holes in the right AM region. Throughout this work, we employ the Andreev approximation $\mu_{\mathrm{AM}} \gg E,\Delta_0$, under which the longitudinal momenta may be approximated as $k_{e,\sigma} \simeq k_{h,\sigma} \simeq k_F \cos\theta$, for the slowly-varying envelope factors, with $\theta$ is the angle of incidence measured from the interface normal~\cite{Acharjee2023}. Nonetheless, the exact expressions in Eq.~(\ref{eq17}) are retained when evaluating the velocity operator and interface matrix for quantitative calculations. The quasiparticle wave function in the TSC region ($0<x<L$), corresponding to an incident quasiparticle from the AM lead with spin index $\sigma$, can be written as~\cite{Cheng2014}
\begin{align} 
\Psi_{\mathrm{TSC},\sigma}(x) = \sum_{\eta=\pm} \sum_{s=\uparrow,\downarrow} \Big[ & c^{\eta}_{\sigma s}\, \Lambda^{\eta}_{e s}\, e^{ i \eta q_e x} + d^{\eta}_{\sigma s}\, \Lambda^{\eta}_{h s}\, e^{- i \eta q_h x} \Big], \label{eq18} 
\end{align} 
Here, the index $\sigma$ labels the incoming spin channel of the AM. The coefficients $c_{\sigma s}^{\eta}$ and $d_{\sigma s}^{\eta}$ are the scattering amplitudes for electron and hole-like quasiparticles propagating in the TSC region. The longitudinal wave vectors of the quasiparticles in the TSC region are given by 
\begin{equation} 
q_{e(h)} = \sqrt{ \frac{2m}{\hbar^2} \left( \mu_{\mathrm{S}} +(-) \sqrt{E^2-|\Delta|^2} \right) - k_y^2 }, 
\label{eq19} 
\end{equation} 
\noindent The BdG eigenvectors $\Lambda^{\eta}_{e(h)s}$ in the TSC region are spin-degenerate and do not carry an intrinsic spin quantum number, can be defined as 
\begin{align} \nonumber \Lambda_{e\uparrow}^{+} &= ( u, \,\, 0, \,\, 0, \,\, v )^\text{T}, \quad \Lambda_{e\downarrow}^{+} = ( 0, \,\, u , \,\, -v , \,\,0)^\text{T},\\ 
\label{eq20} \quad \Lambda_{h\uparrow}^{+} &= ( v , \,\, 0, \,\, 0 , \,\, u )^\text{T}, \quad \Lambda_{h\downarrow}^{+} = ( 0, -v , \,\, u , \,\, 0 )^\text{T}. 
\end{align} 
\begin{figure*}[t]
\centering
\hspace{-6mm}
\vspace{-3mm}
\centerline{
\includegraphics[scale=0.21]{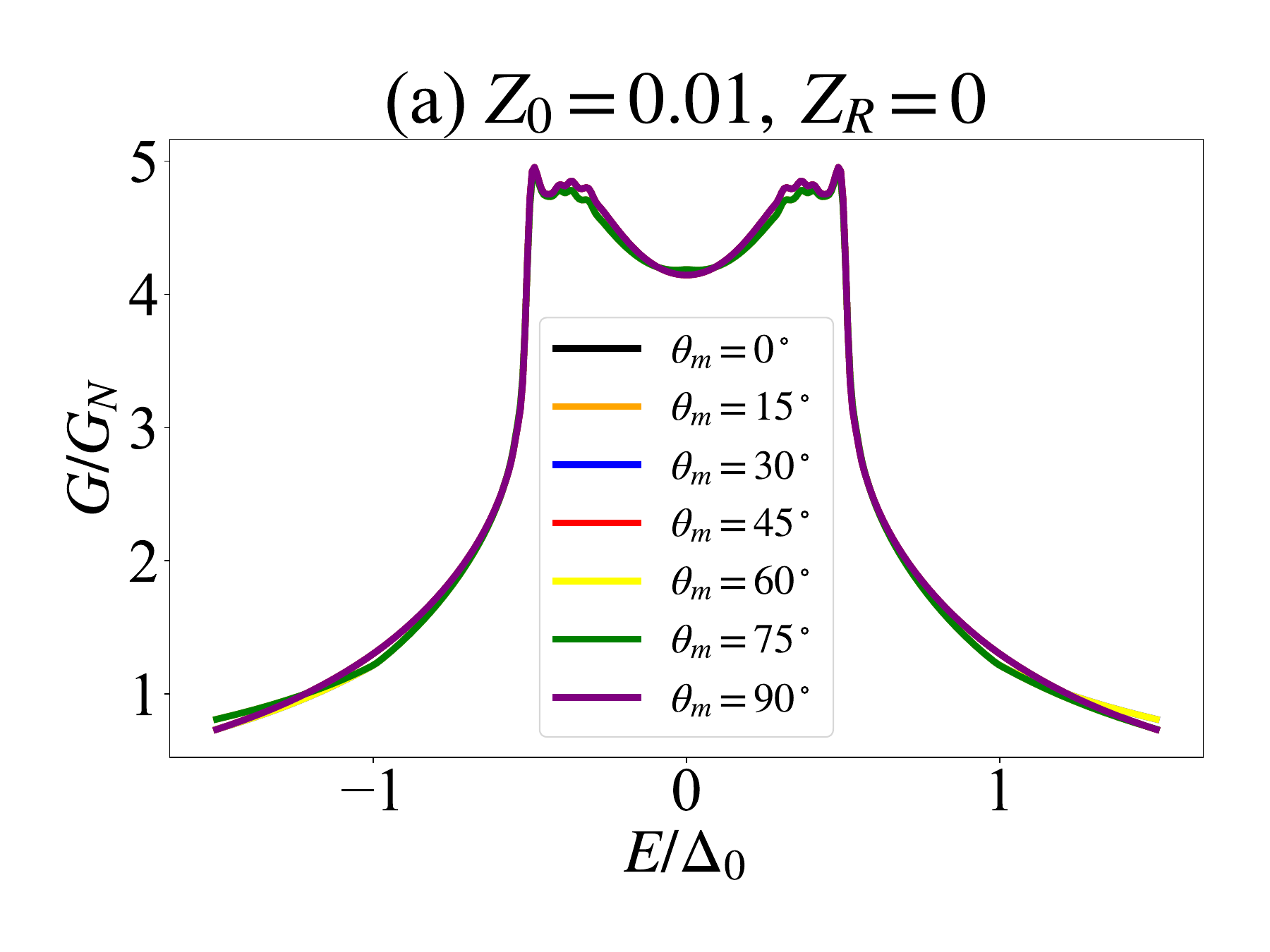}
\hspace{-6mm}
\vspace{-3mm}
\includegraphics[scale=0.21]{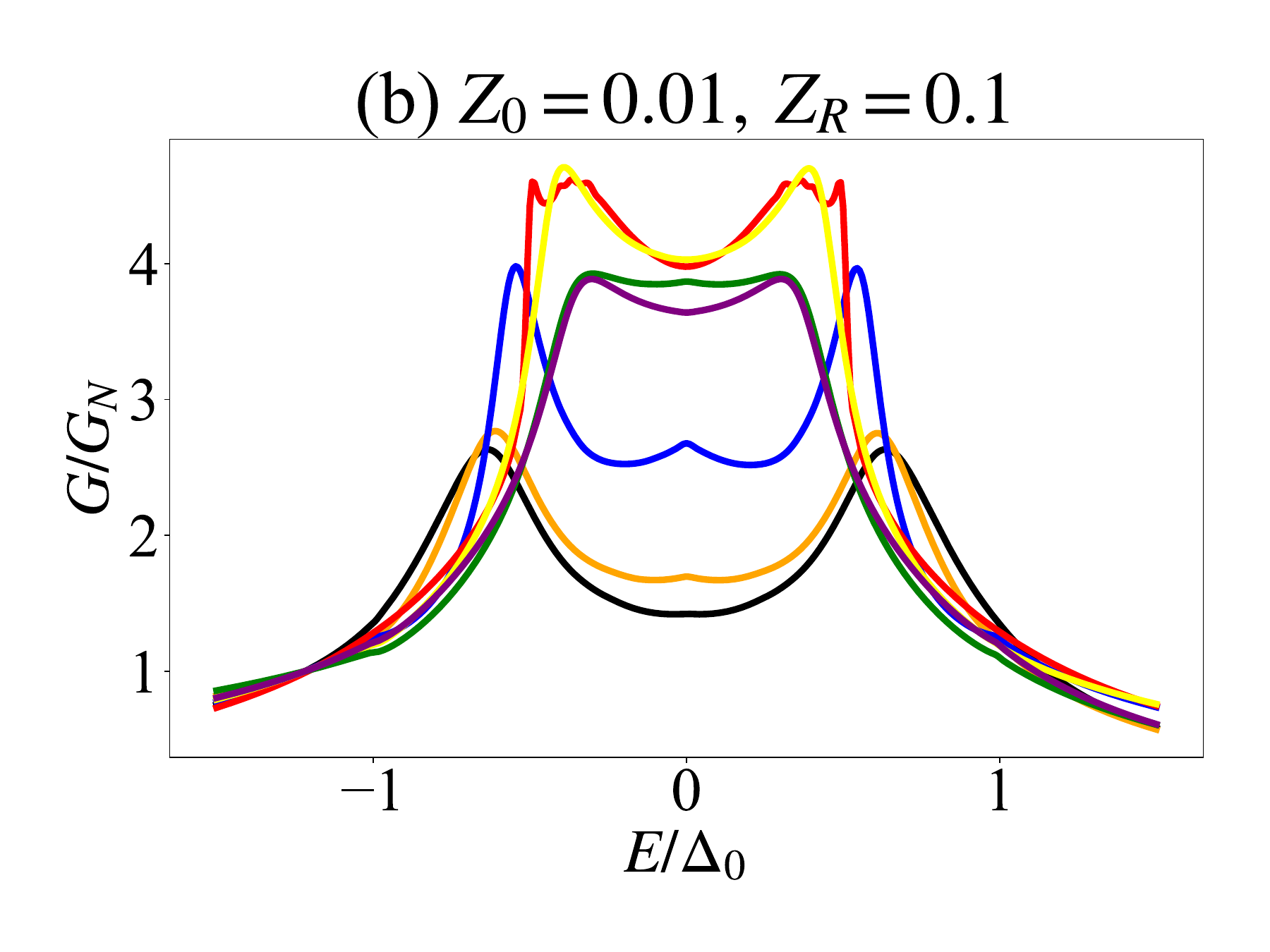}
\hspace{-7mm}
\includegraphics[scale=0.21]{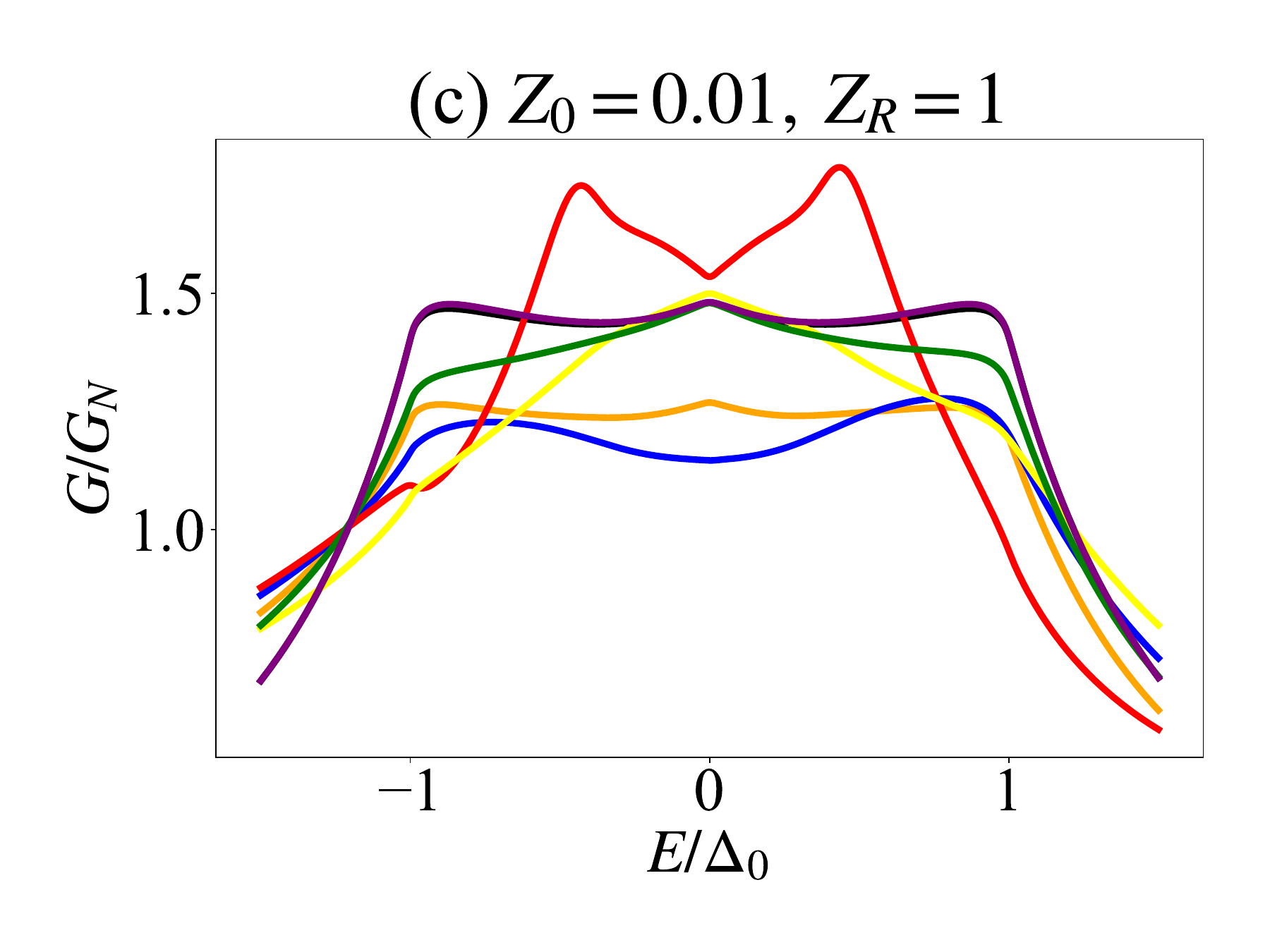}}
\hspace{-6mm}
\centerline{
\hspace{-6mm}
\includegraphics[scale=0.212]{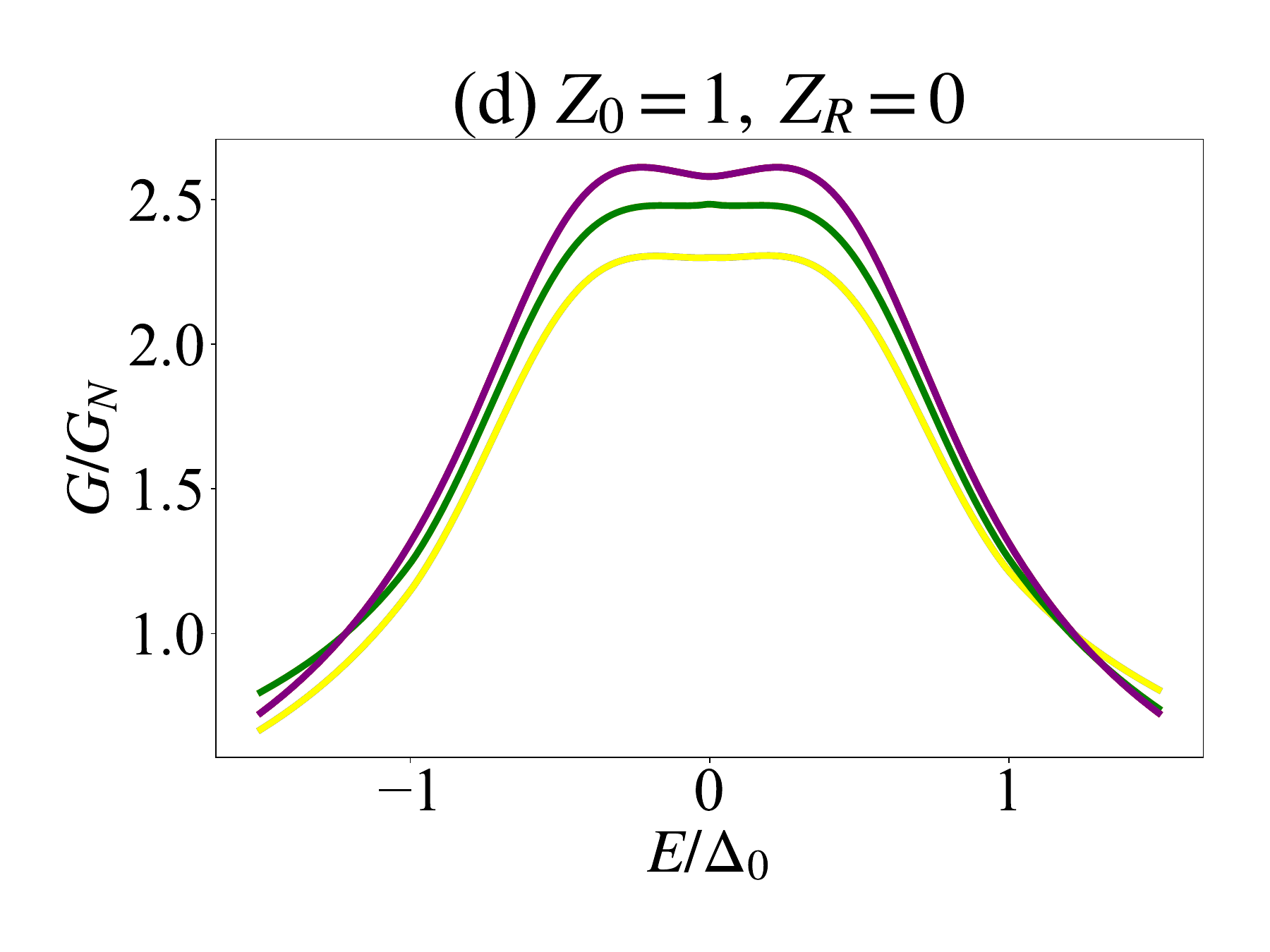}
\hspace{-6mm}
\includegraphics[scale=0.212]{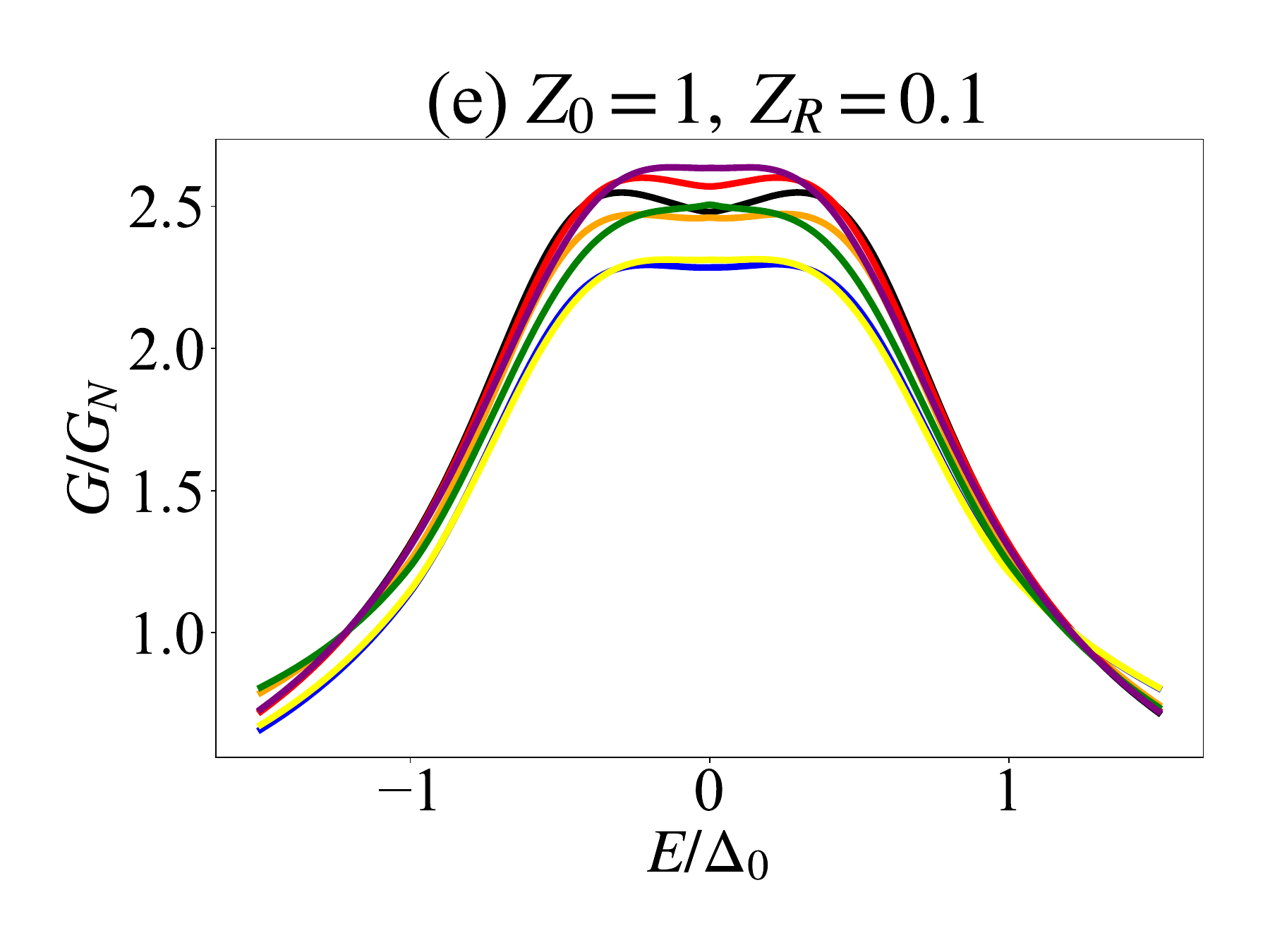}
\hspace{-6mm}
\includegraphics[scale=0.212]{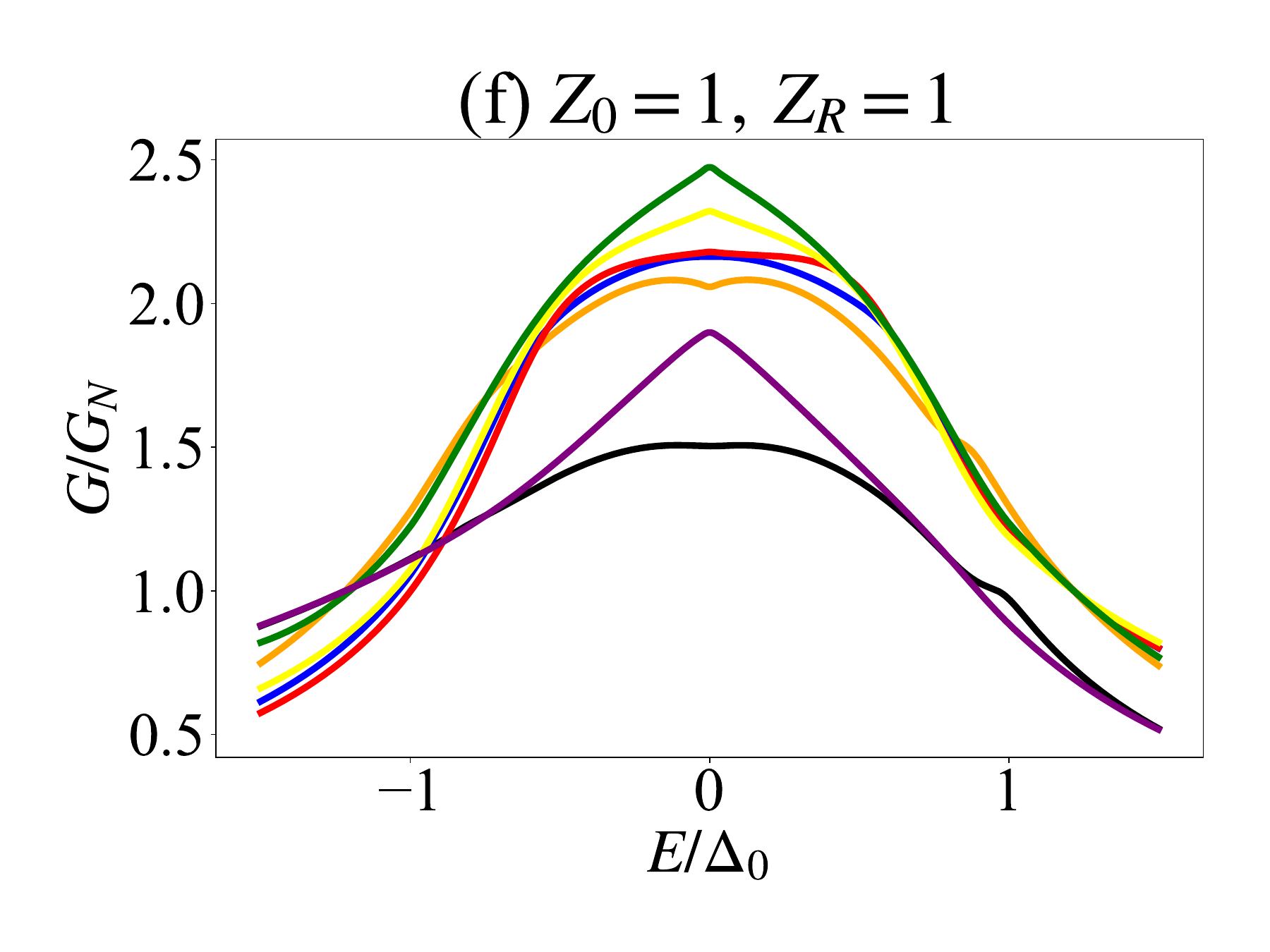}}
\caption{
Differential conductance $G/G_N$ as a function of  $E/\Delta_0$ for AM /$p_x$-wave SC /AM system considering different AM orientations $\theta_m$.
Panels (a)–(c) correspond to a nearly transparent ($Z_0 = 0.01$) while panels (d)–(f) are for opaque barrier ($Z_0 = 1$)  considering different values of  $Z_R$. 
}
\label{fig2}
\end{figure*}
The backward-propagating spinors are related to the forward-propagating ones by particle-hole symmetry, 
\begin{equation} 
\Lambda_{e\sigma}^{-}=\mathcal{C}\Lambda_{h\sigma}^{+},\qquad \Lambda_{h\sigma}^{-}=\mathcal{C}\Lambda_{e\sigma}^{+}, \label{eq21} 
\end{equation} where, $\mathcal{C}=\tau_x\mathcal{K}$ is the BdG charge-conjugation operator. The quasiparticle amplitudes $u$ and $v$ are define as \begin{equation} 
u = \frac{1}{\sqrt{2}}\sqrt{1+\frac{\Omega}{E}}, \qquad v = \frac{1}{\sqrt{2}}\sqrt{1-\frac{\Omega}{E}} 
\label{eq22} 
\end{equation} 
with $\Omega = \sqrt{E^2 - \Delta^2}$. The scattering amplitudes are obtained by imposing the boundary conditions derived from the BdG equation, $\check{\mathcal H}_{\mathrm{BdG}}\Psi = E\Psi$. These conditions follow from integrating the BdG equation across each AM/TSC interface over an infinitesimal region \([-\epsilon,\epsilon]\) and taking the limit \(\epsilon \to 0\). The hermiticity of the Hamiltonian operator is preserved by enforcing the antisymmetrization of the AM contribution. Accordingly, the BdG wave function is required to be continuous at the interfaces. For an incident AM channel $\sigma$, this yields 
\begin{align} 
\Psi^{\mathrm{L}}_{\mathrm{AM},\sigma}(0^-) &= \Psi_{\mathrm{TSC},\sigma}(0^+), \label{eq23}
\\ \Psi_{\mathrm{TSC},\sigma}(L^-) &= \Psi^{\mathrm{R}}_{\mathrm{AM},\sigma}(L^+). 
\label{eq24} 
\end{align} 

In addition, current conservation across the interfaces leads to velocity-jump conditions of the form
\begin{align} \hat v_x^{\mathrm{S}}\Psi_{\mathrm{TSC},\sigma}(0^+) - \hat v_x^{\mathrm{AM},L}\Psi^{\mathrm{L}}_{\mathrm{AM},\sigma}(0^-) &= \hat{\mathcal Z}_L \,\Psi^{\mathrm{L}}_{\mathrm{AM},\sigma}(0^-), \label{eq25}\\ \hat v_x^{\mathrm{AM},R}\Psi^{\mathrm{R}}_{\mathrm{AM},\sigma}(L^+) - \hat v_x^{\mathrm{S}}\Psi_{\mathrm{TSC},\sigma}(L^-) &= \hat{\mathcal Z}_R \,\Psi_{\mathrm{TSC},\sigma}(L^-). \label{eq26} 
\end{align} 
Here, the velocity operators are defined as~\cite{Zutic1999}
\begin{align} 
\hat v_x^{\mathrm{AM},j} &= \frac{\partial \hat{\mathcal H}_0^{(j)}}{\partial k_x} = \frac{\hbar^2 k_x}{m}\,\hat{I} + h_0\big(\alpha_1 k_y+\alpha_2 k_x\big)\, (\hat{\mathbf n}_j\!\cdot\!\boldsymbol{\sigma}),\\ \hat v_x^{\mathrm{S}} &= \frac{\partial}{\partial q_x}\big(\xi_{\mathbf k}\tau_z\big) = \frac{\hbar^2 q_x}{m}\,\tau_z, 
\end{align} 
where $\tau_z$ acts in particle-hole space. Here, \(\hat{\mathcal Z}_{L(R)}\) denote the interface scattering matrices defined as 
\begin{equation} 
\hat{\mathcal Z}_j \;=\; \, Z_0 \hat{I} - Z_R\, k_y\, (\hat{\mathbf n}_j\!\cdot\!\boldsymbol{\sigma}) 
\label{eq29} 
\end{equation} 
where, $Z_0 = \frac{2mU_0}{\hbar^2}$ and $Z_R = \frac{2mU_R}{\hbar^2}$ respectively represent the barrier strength and the strength of RSOC measured in terms of Fermi energy. 

The scattering amplitudes are
obtained by solving the boundary conditions
Eqs.~(\ref{eq23})–(\ref{eq26}). The spin-resolved differential
conductance is evaluated within a velocity-conserving scattering
formalism consistent with the AM band structure. Owing to the
momentum and spin-dependent dispersion of AMs, current
conservation requires explicit inclusion of the corresponding group
velocities.

\section{spin valve effect in AM/TSC/AM hybrid}  
 The AM/TSC/AM hybrid realizes a spin valve effect that
is fundamentally distinct from conventional magnetization-
based devices. Here the spin valve effect arises in the absence of net magnetization and
is instead governed by the momentum-dependent spin splitting intrinsic
to AMs, combined with equal-spin triplet pairing and interfacial RSOC. In this system the
control parameter is the relative orientation of the N\'eel vectors in
the two AM regions, while the spin selectivity originates
from momentum-dependent spin splitting rather than a uniform exchange
field. 

Unlike FM-based devices that rely on magnetic field switching, the AM response can be tuned continuously by gating RSOC and rotating the relative N\'eel-vector orientation, while the absence of stray fields is advantageous for coherent superconducting devices and nanoscale integration. In this section we quantify the resulting spin-dependent
transport in terms of spin-resolved conductance, spin polarization, and
TMR.
\begin{figure*}[t]
\centering
\hspace{-6mm}
\vspace{-3mm}
\centerline{
\includegraphics[scale=0.21]{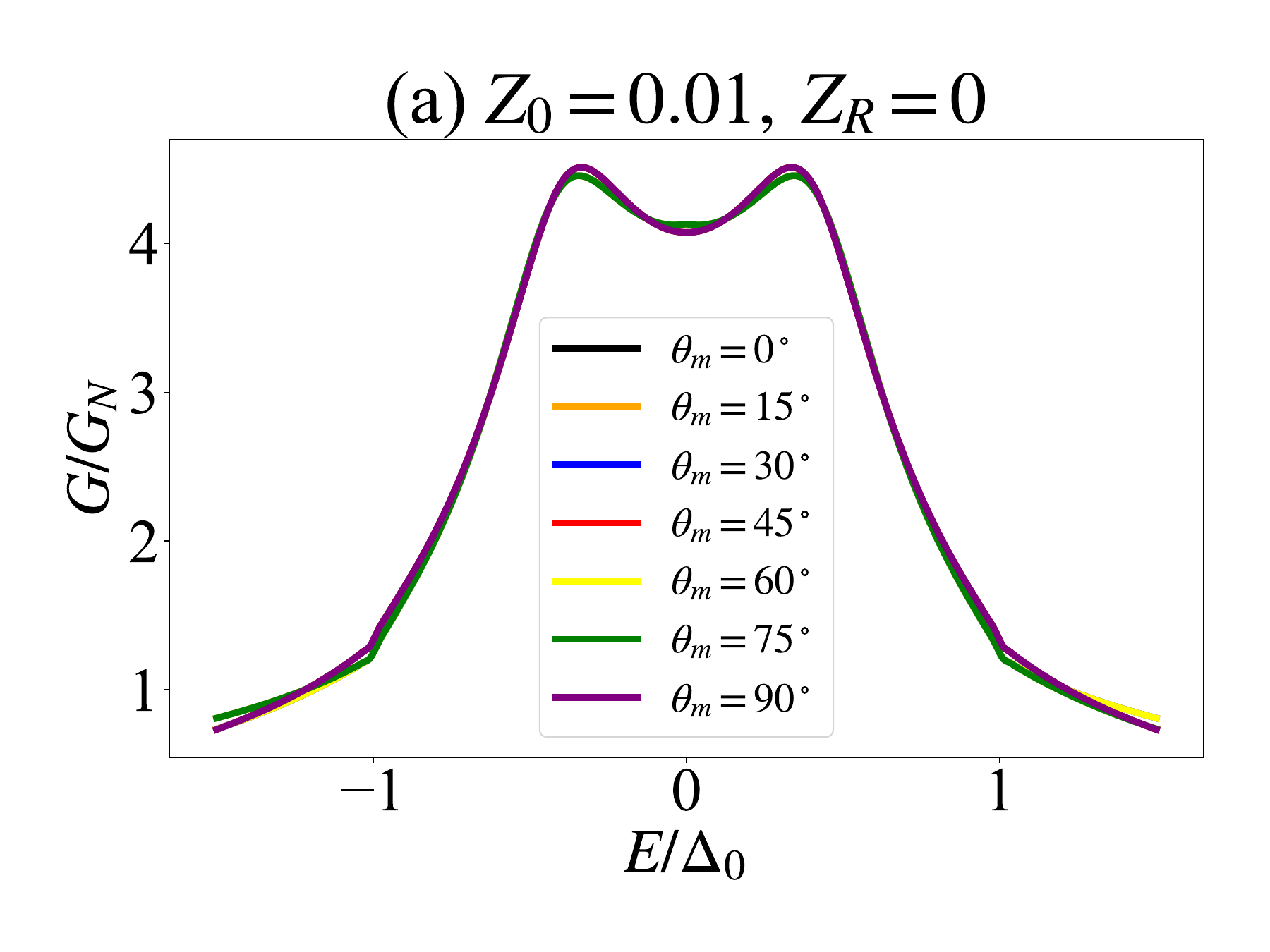}
\hspace{-6mm}
\vspace{-3mm}
\includegraphics[scale=0.21]{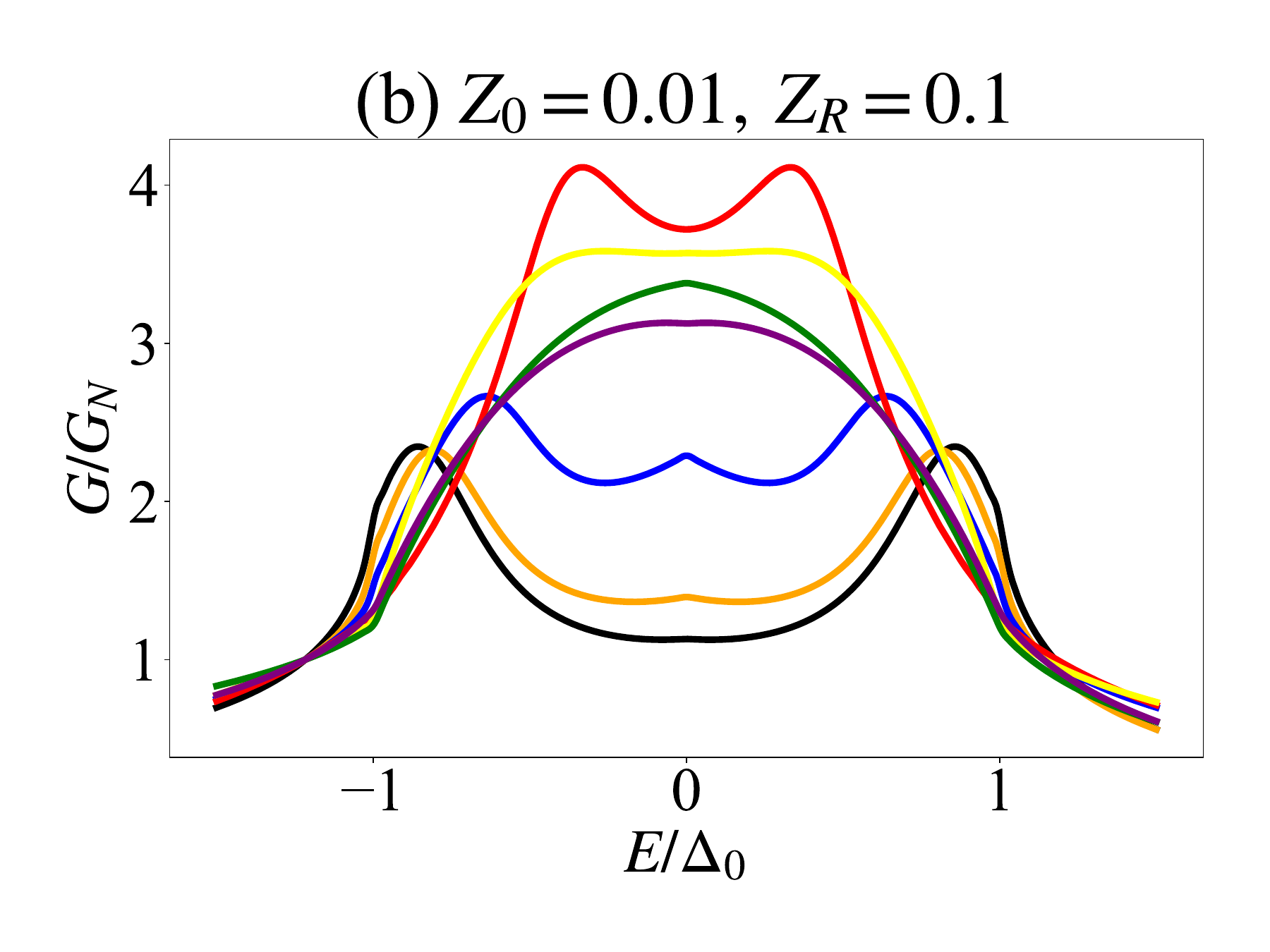}
\hspace{-7mm}
\includegraphics[scale=0.21]{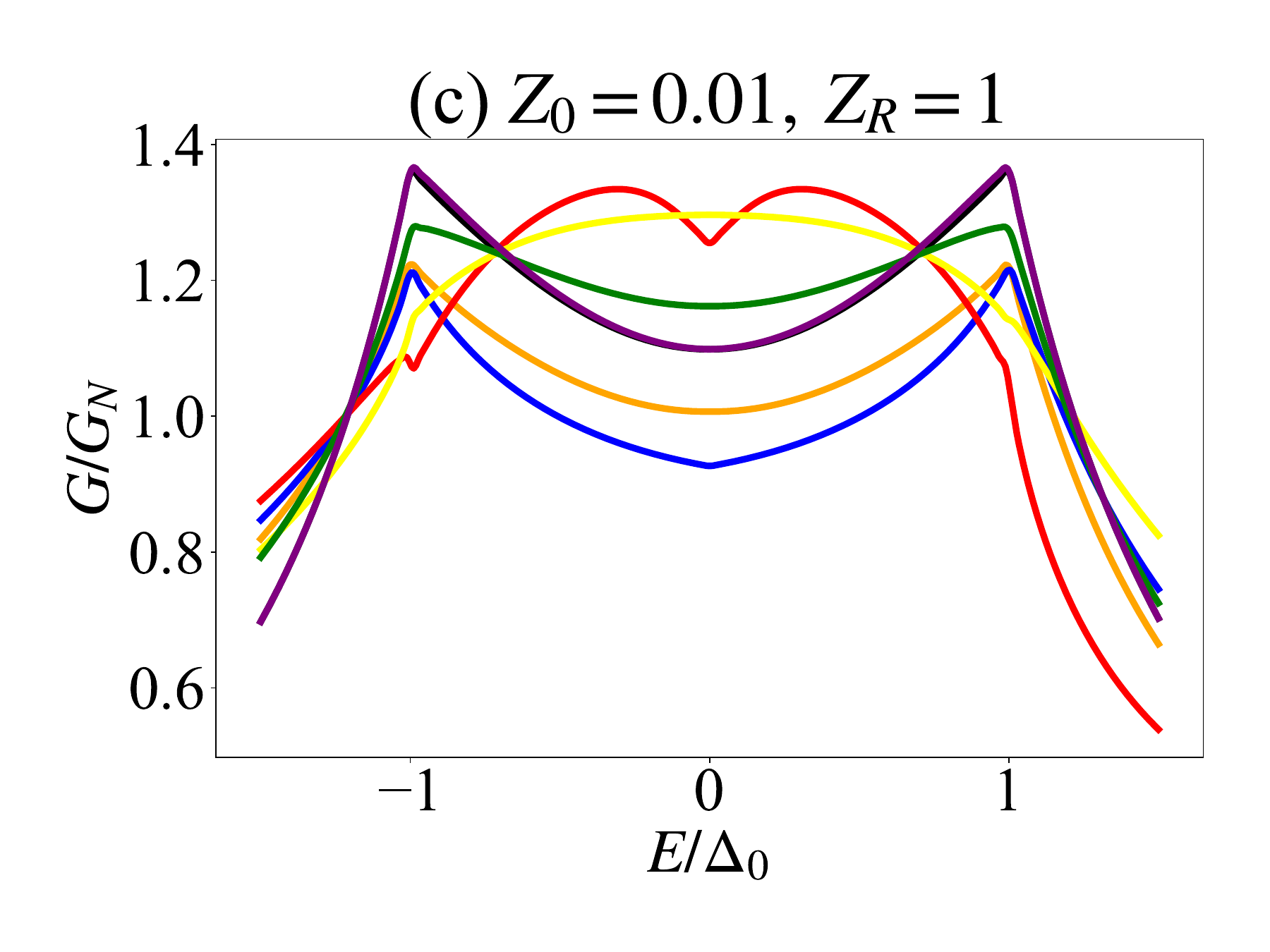}}
\hspace{-6mm}
\centerline{
\hspace{-6mm}
\includegraphics[scale=0.212]{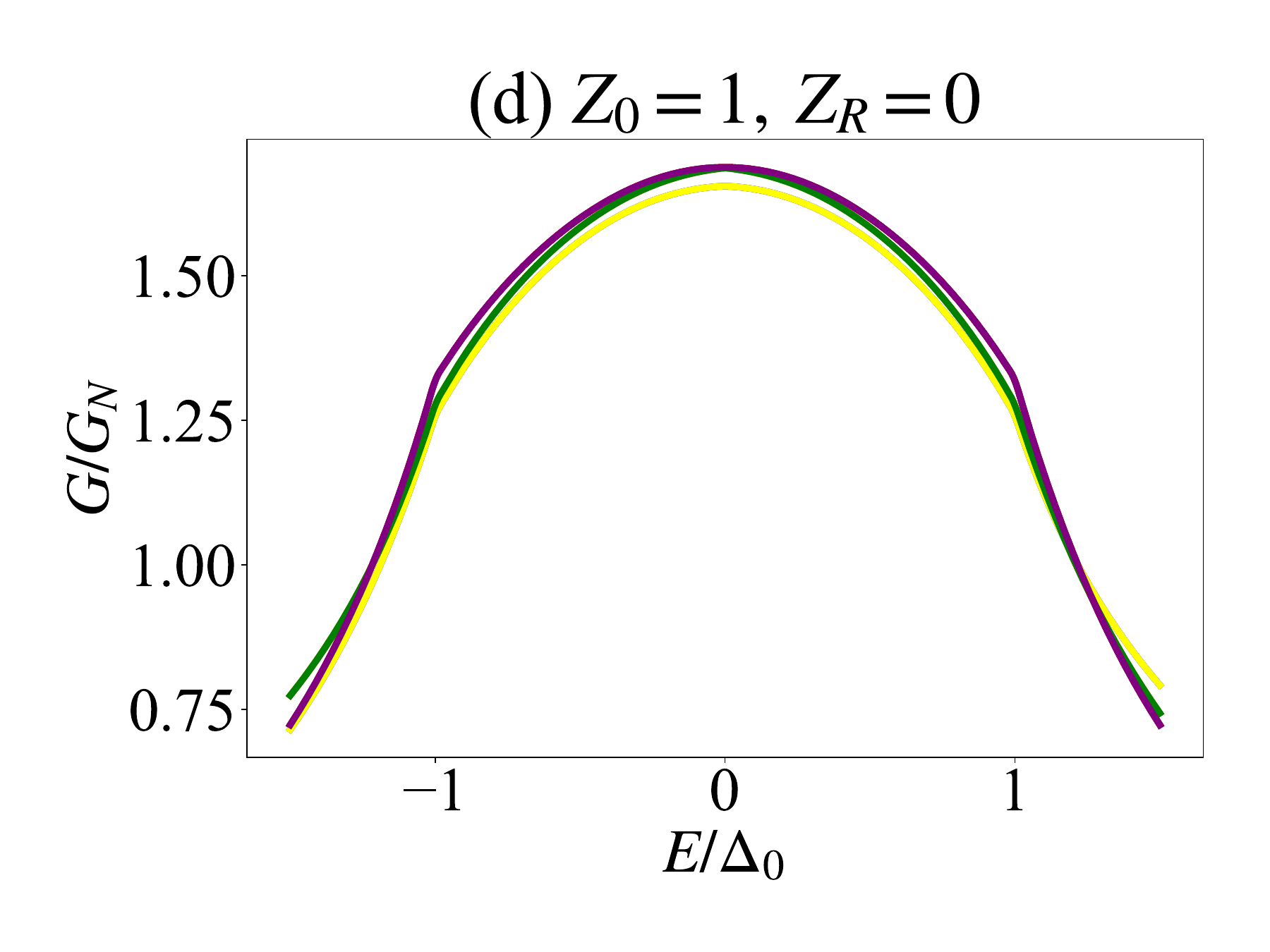}
\hspace{-6mm}
\includegraphics[scale=0.212]{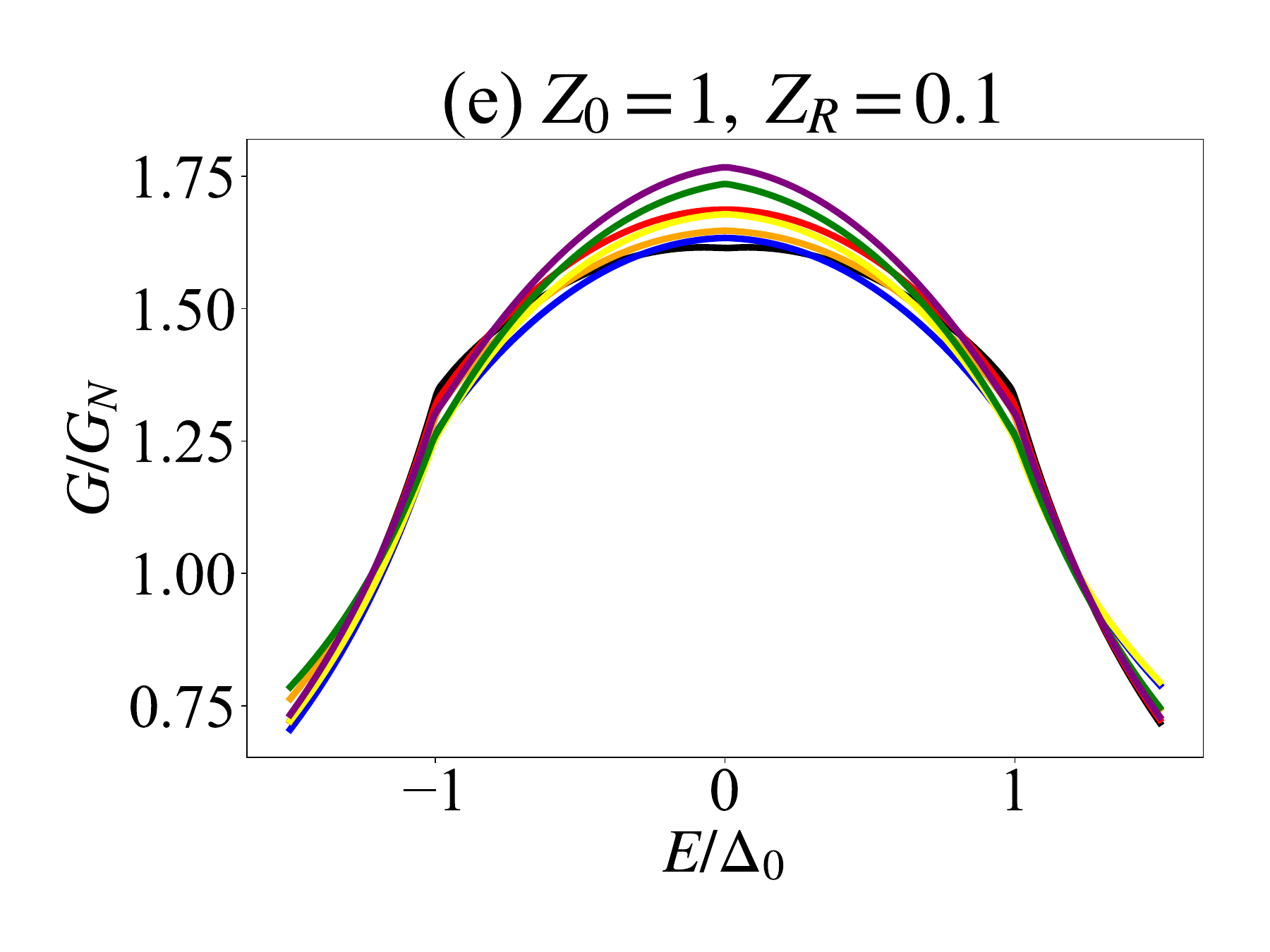}
\hspace{-6mm}
\includegraphics[scale=0.212]{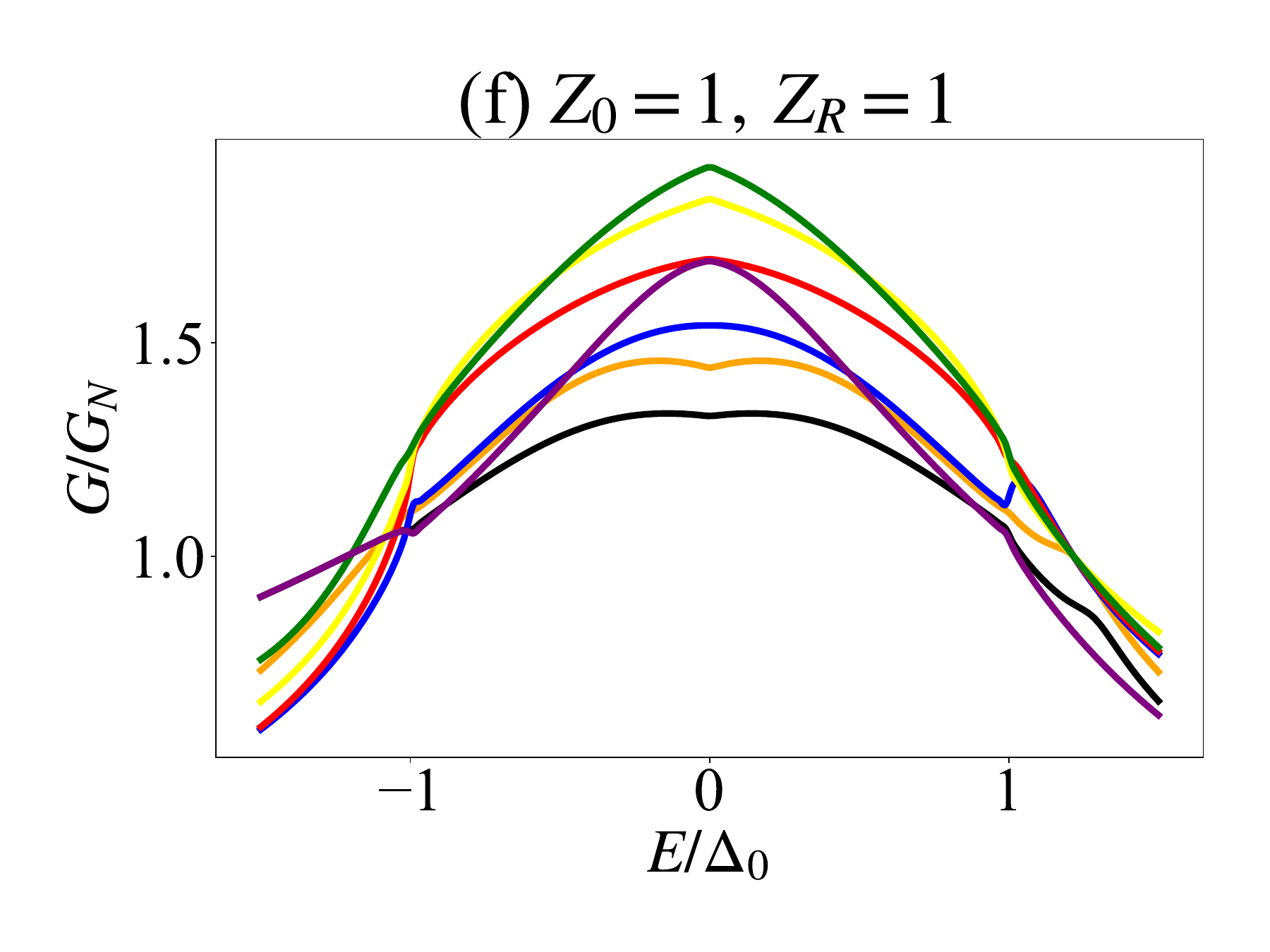}}
\caption{
Differential conductance $G/G_N$ as a function of  $E/\Delta_0$ for AM /$p_x+ip_y$-wave SC /AM system considering different AM orientations $\theta_m$.
Panels (a)–(c) correspond to a nearly transparent ($Z_0 = 0.01$) while panels (d)–(f) are for opaque barrier ($Z_0 = 1$)  considering different values of  $Z_R$. 
}
\label{fig3}
\end{figure*}
\subsection{Spin-resolved differential conductance}
For an electron incident from the left AM lead in eigenchannel
$\sigma=\pm$ at excitation energy $E$ and transverse momentum $k_y$, the
angle-resolved conductance takes the form~\cite{Blonder1982,Zutic1999} 
\begin{align}
\mathcal{G}_{\sigma}(E,k_y)
&=
1
-
\sum_{\sigma'}
\left| b_{\sigma\sigma'}(E,k_y)\right|^2
\frac{v^{\mathrm{AM}}_{e,\sigma'}(E,k_y)}
     {v^{\mathrm{AM}}_{e,\sigma}(E,k_y)}
\nonumber\\
&\quad
+
\sum_{\sigma'}
\left| a_{\sigma\sigma'}(E,k_y)\right|^2
\frac{v^{\mathrm{AM}}_{h,\sigma'}(E,k_y)}
     {v^{\mathrm{AM}}_{e,\sigma}(E,k_y)} ,
\label{eq30}
\end{align}
where $a_{\sigma\sigma'}$ and $b_{\sigma\sigma'}$ denote Andreev and
normal reflection amplitudes, respectively. The velocity ratios ensure
charge-current conservation across the junction.

The normalized spin-resolved differential conductance is obtained by
integrating over all propagating transverse modes,
\begin{equation}
\frac{G_\sigma(E)}{G_N}
=
\frac{
\displaystyle
\int_{-\pi/2}^{\pi/2}
d\theta\,\cos\theta\;
\mathcal{G}_{\sigma}\big(E,k_y(\theta)\big)
}{
\displaystyle
\sum_{\sigma'=\pm}
\int_{-\pi/2}^{\pi/2}
d\theta\,\cos\theta\;
\mathcal{G}^{(N)}_{\sigma'}\big(k_y(\theta)\big)
},
\label{eq31}
\end{equation}
where $G_N$ is the normal-state conductance of the same junction with
superconductivity suppressed. 

The total normalized differential conductance is finally given by
\begin{equation}
\frac{G(E)}{G_N}
=
\sum_{\sigma=\pm}
\frac{G_\sigma(E)}{G_N}.
\label{eq32}
\end{equation}

Eqs.~(\ref{eq30}) - (\ref{eq32}) constitute a
natural generalization of the Blonder–Tinkham–Klapwijk (BTK) formalism to
the AM regions by resolving transport in the AM eigenbasis rather than fixed spin projections, with velocity prefactors encoding the momentum-dependent spin splitting of the AM bands.

Fig.~\ref{fig2} shows the variation of differential conductance $(G/G_N)$ with  bias energy $(E/\Delta_0)$ of an
AM/$p_x$-wave SC/AM junction for different N\'eel vector misalignment angles ($\theta_m$). For a nearly transparent barrier with barrier strength $Z_0=0.01$, it is observed that all $\theta_m$ traces nearly collapse onto a single curve with
a pronounced double-peak structure near $E/\Delta_0 \rightarrow 1$ in absence of interfacial RSOC
($Z_R=0$) as seen from Fig.~\ref{fig2}(a). This behavior reflects dominant Andreev reflection and weak sensitivity
to $\theta_m$ in absence of interfacial RSOC, as equal-spin
triplet pairs couple efficiently to AM eigenchannels. The double-peak structure reflects the nodal nature of nonchiral 
$p_x$ pairing. Furthermore, the sign change of the gap under specular reflection generates zero-energy Andreev bound states (ABS), which strongly enhance subgap conductance.  In presence of RSOC, a pronounced $\theta_m$ dependence is observed from Figs.~\ref{fig2}(b) and \ref{fig2}(c). Interfacial RSOC mixes the AM eigenchannels and generates spin-dependent phase shifts between normal and Andreev reflection processes, thereby converting the relative N\'eel-vector orientation into a tunable control parameter for quasiparticle interference and spin filtering. It is observed that moderate RSOC with
$Z_R = 0.1$ generates additional subgap resonances and suppressions depending on the misalignment, while strong RSOC i.e., $Z_R = 1$ leads to broadened peaks and display a nonmonotonic angular behavior. In case of a opaque barrier with barrier strength $Z_0 = 1$, the
conductance is significantly reduced and found to be angle independent for
$Z_R = 0$ as seen from Fig.~\ref{fig2}(d). However, finite RSOC reintroduces angular sensitivity via spin mixing as observed from Figs.~\ref{fig2}(e) and \ref{fig2}(f). Notably,  the conductance becomes asymmetric for sufficiently large $Z_R$
under $E\rightarrow -E$, indicating that the combined effect of RSOC and the
momentum-dependent AM exchange field that breaks the effective particle-hole
symmetry of the transport spectra, even though the underlying BdG
Hamiltonian remains particle-hole symmetric.
\begin{figure*}[hbt]
\centerline
\centerline{ 
\includegraphics[scale=0.26]{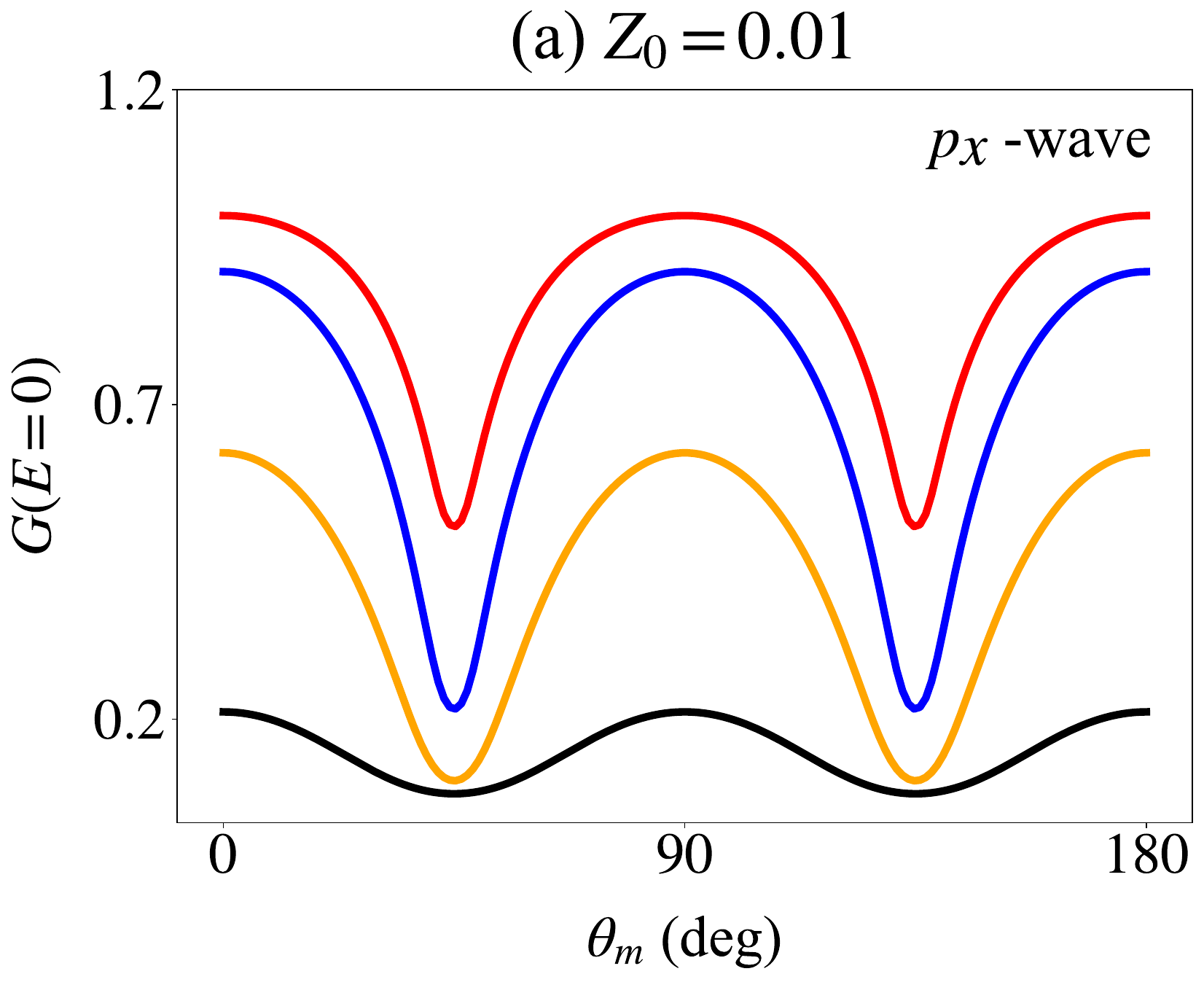}
\hspace{3mm}
\includegraphics[scale=0.26]
{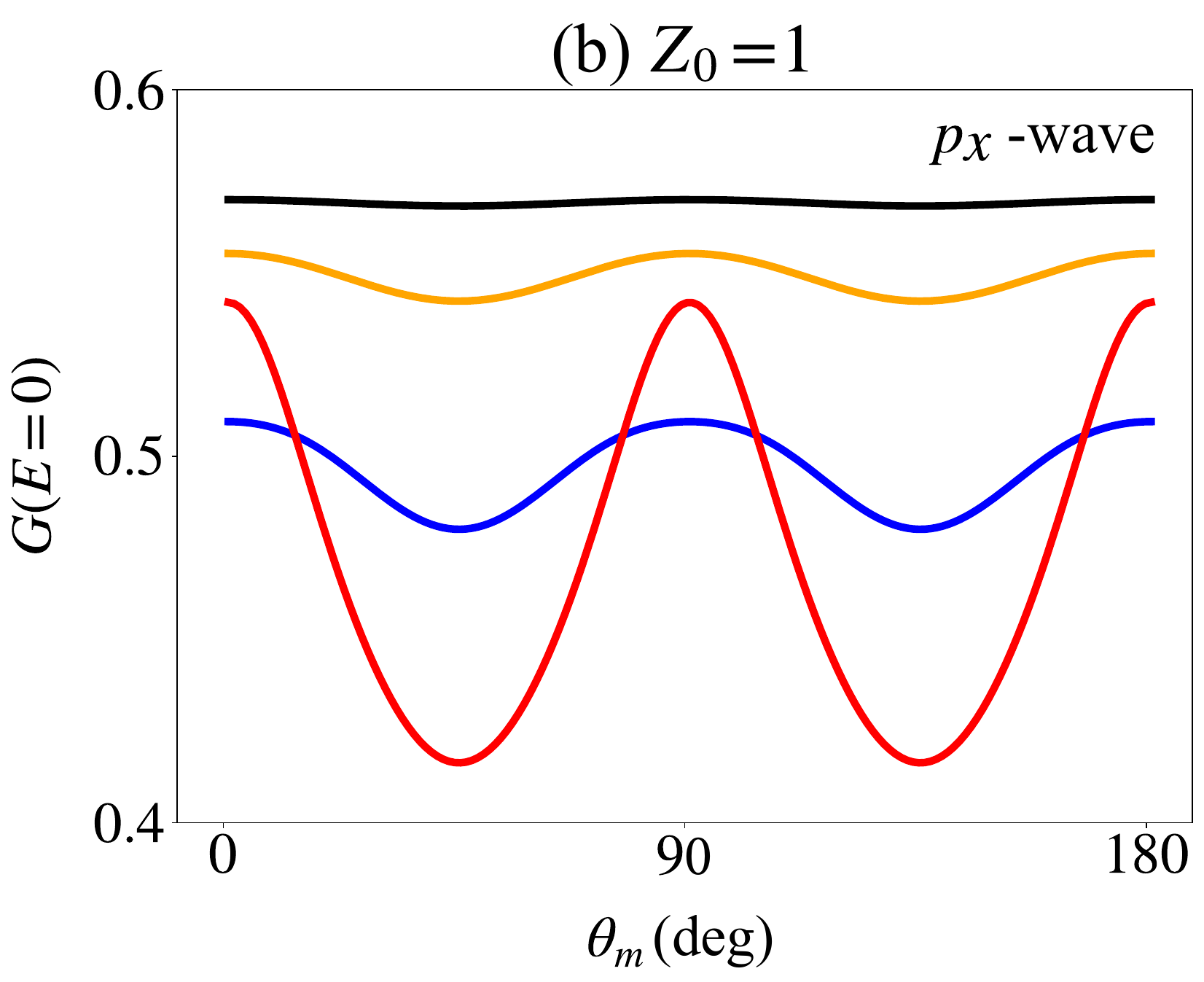}
\centerline{ 
\includegraphics[scale=0.26]
{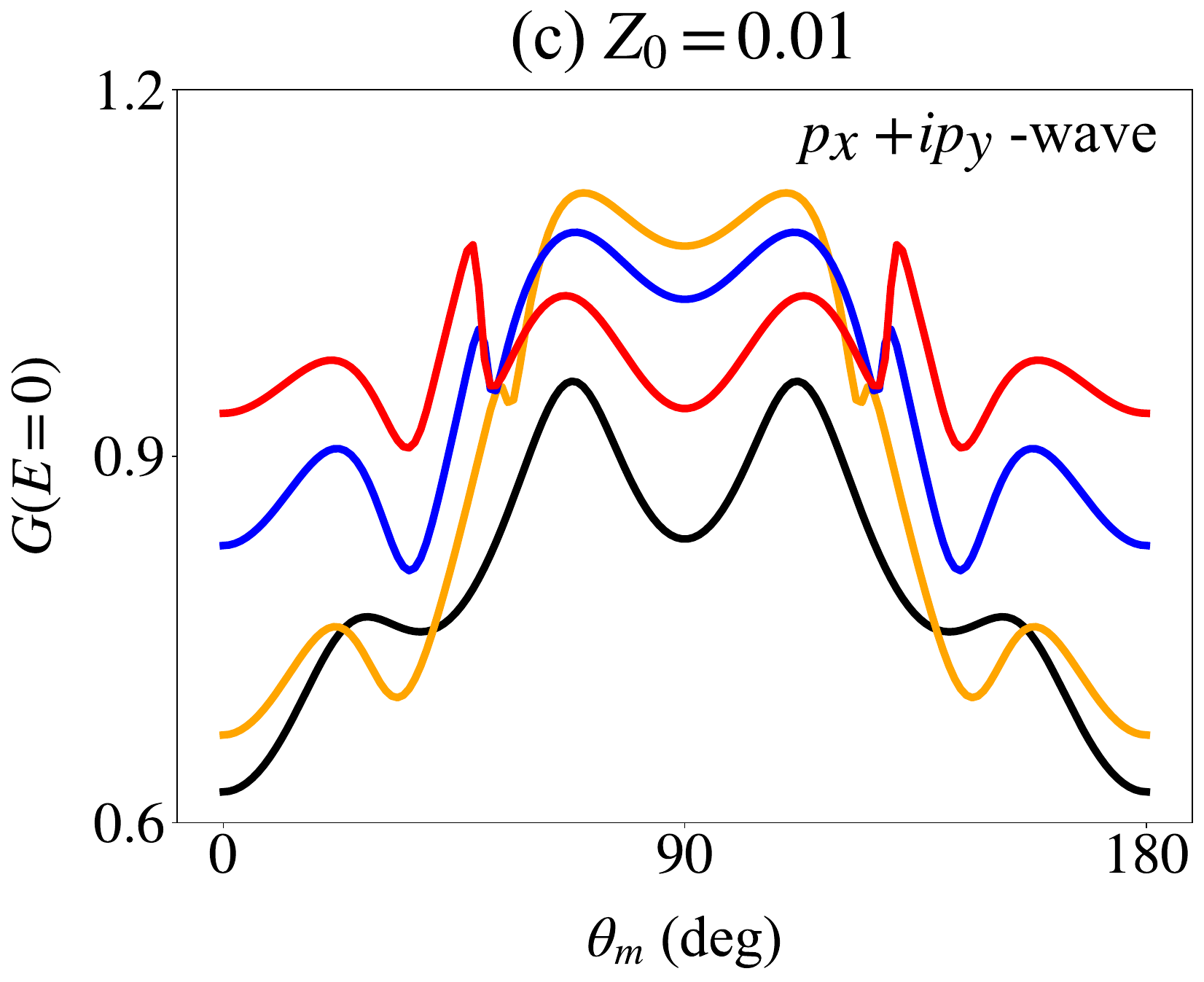}
\hspace{3mm}
\includegraphics[scale=0.26]
{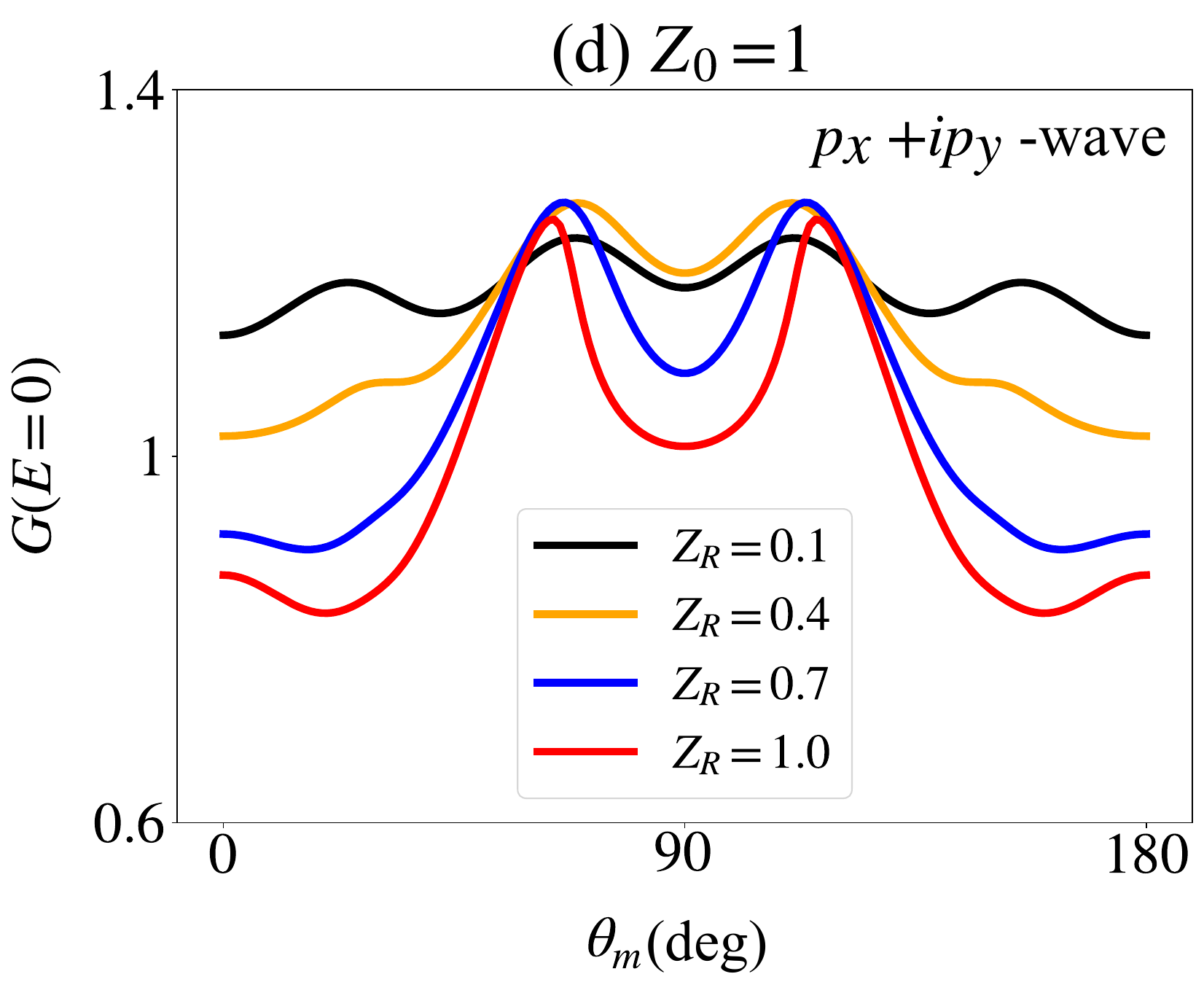}
}
}
\caption{
Zero-bias conductance as a function of the AM orientation angle $\theta_m$ for different $Z_R$ values.
Panels (a) and (b) correspond to the non-chiral $p_x$-wave superconductor, while panels (c) and (d) show the chiral $p_x + i p_y$-wave case.
The upper (lower) panels are for weak (strong) barrier strength $Z_0=0.01$ ($Z_0=1$).
}
\label{fig4}
\end{figure*}

Fig.~\ref{fig3} present the corresponding conductance for a chiral
$p_x+ip_y$-wave superconductor, highlighting significant differences
relative to the nonchiral case in Fig.~\ref{fig2}. In the transparent regime i.e., $Z_0 = 0.01$, the conductance exhibits a smooth central maxima with reduced edge peaks and weak $\theta_m$ dependence for $Z_R = 0$ as seen from Fig.~\ref{fig3}(a). It 
indicates the fully gapped nature of the chiral state and the absence
of sign-changing reflections for individual trajectories. However, a finite RSOC
introduces angular dependence primarily through
amplitude modulation rather than sharp resonances as seen from Figs.~\ref{fig3}(b) and \ref{fig3}(c). However, even for strong  
RSOC the subgap conductance remains comparatively smooth
and predominantly single-peaked. This behavior contrasts with the ABS-dominated response of the nodal $p_x$
case. In the opaque regime, i.e., $Z_0=1$, the
chiral junction exhibits a broad tunneling peak centered near zero bias,
with only moderate $\theta_m$ dependence even at large $Z_R$. While
particle-hole asymmetry $G(E)\neq G(-E)$ again develops at finite RSOC. Moreover, the magnitude of the conducatnace is weaker than in the nodal $p_x$ case, consistent with
the greater robustness of chiral pairing against spin-dependent
interference. Overall, the comparison between Figs.~\ref{fig2} and ~\ref{fig3} demonstrates
that the spin valve response of AM/TSC/AM junctions is strongly
pairing symmetry-dependent: nodal $p_x$ pairing amplifies RSOC-induced
spin mixing into sharp, angle-sensitive subgap features, whereas chiral
$p_x+ip_y$ pairing yields a smoother, more resilient conductance
spectrum.

\begin{figure*}[hbt]
\centering
\centerline{ 
\includegraphics[scale=0.28]{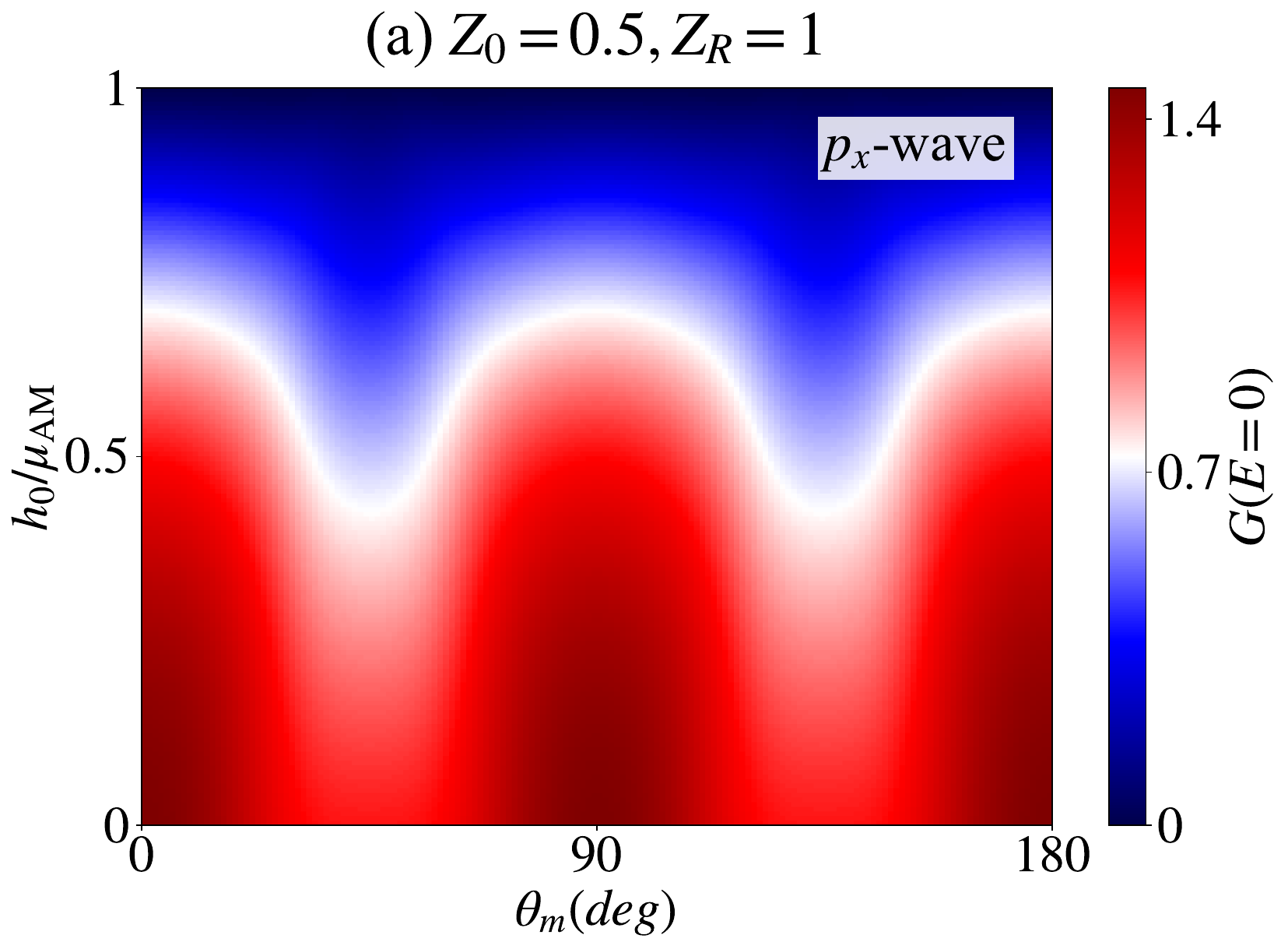}
\hspace{1mm}
\includegraphics[scale=0.28]
{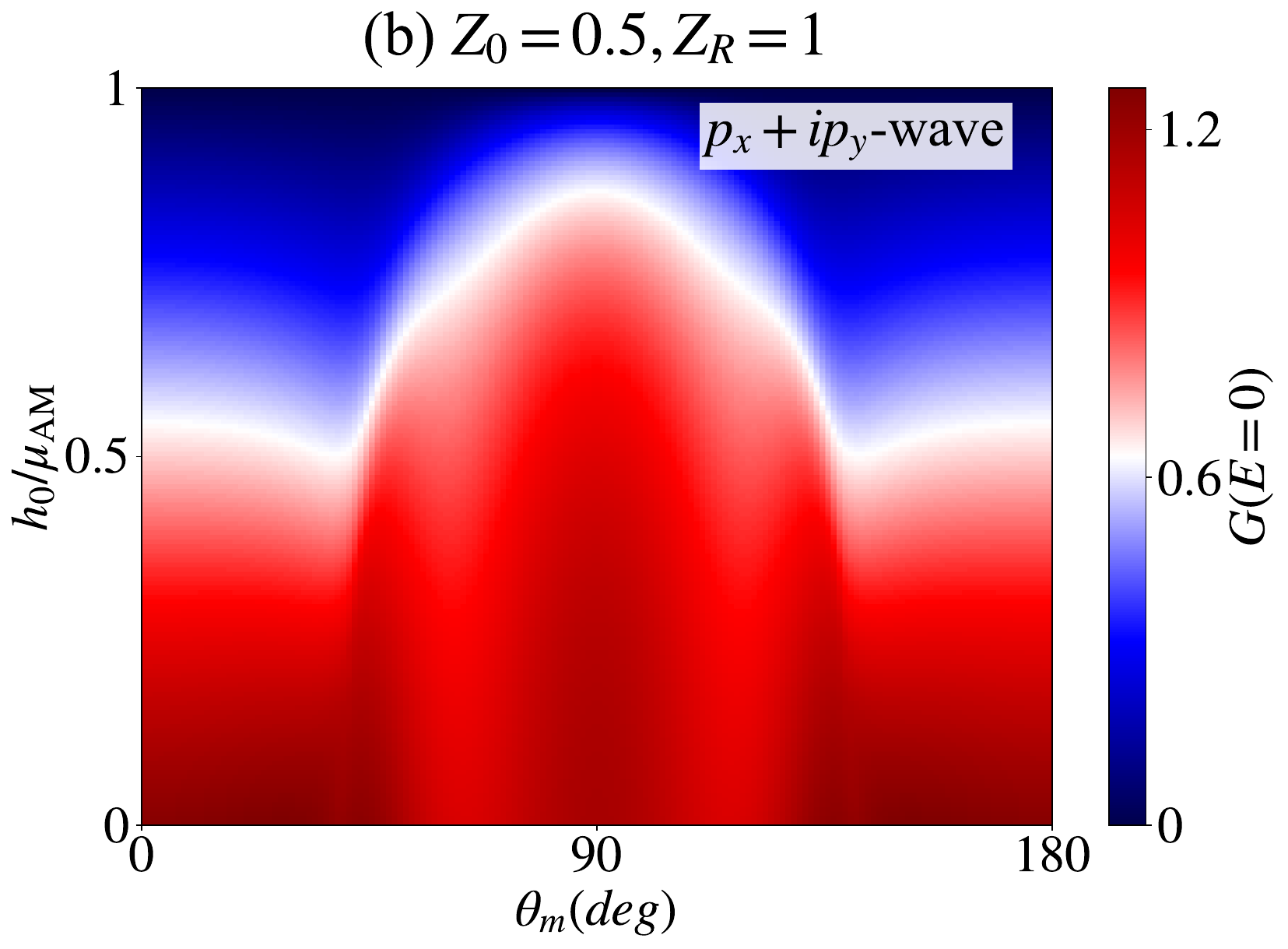}
}
\centerline{ 
\includegraphics[scale=0.28]{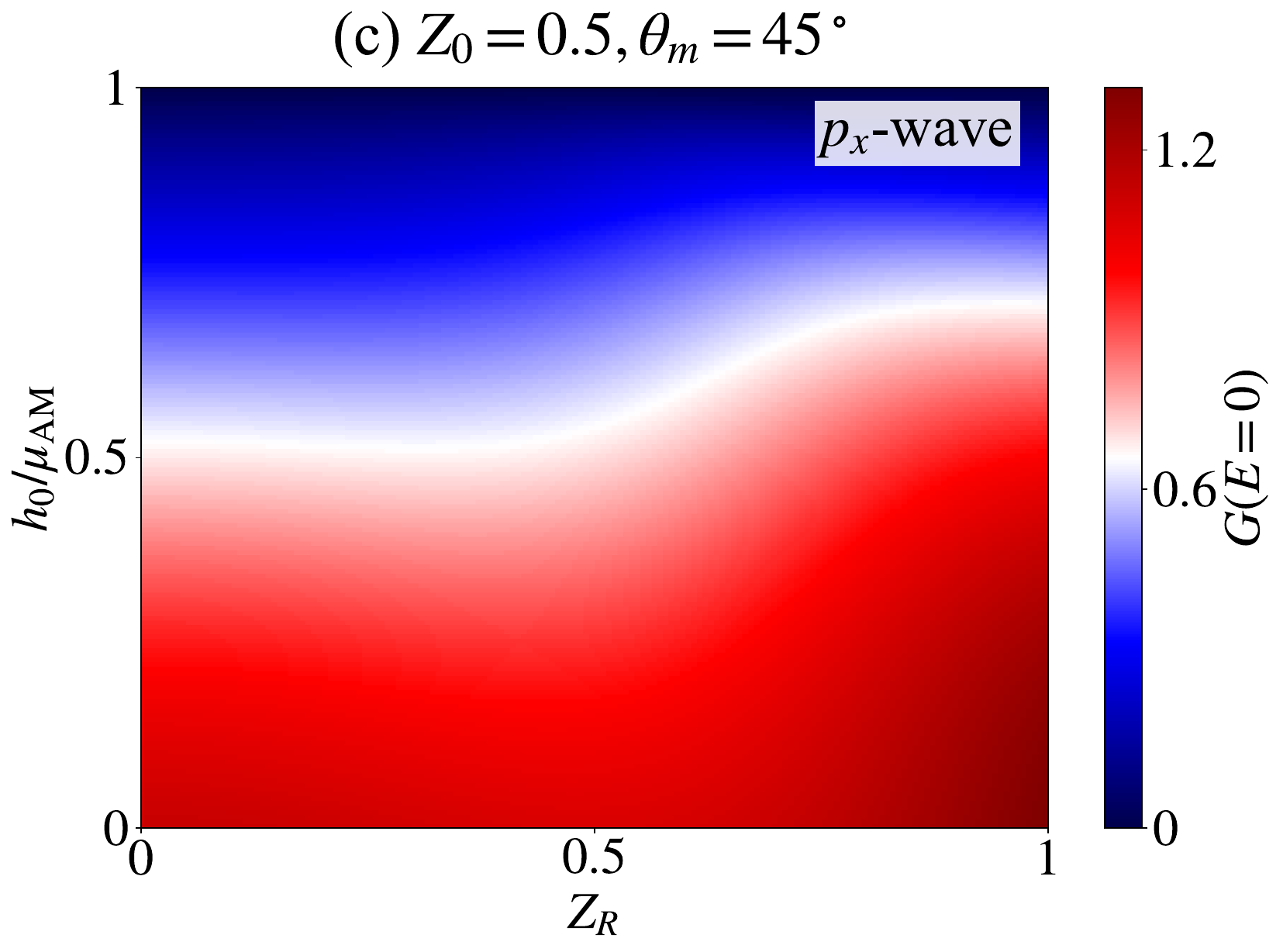}
\hspace{1mm}
\includegraphics[scale=0.28]
{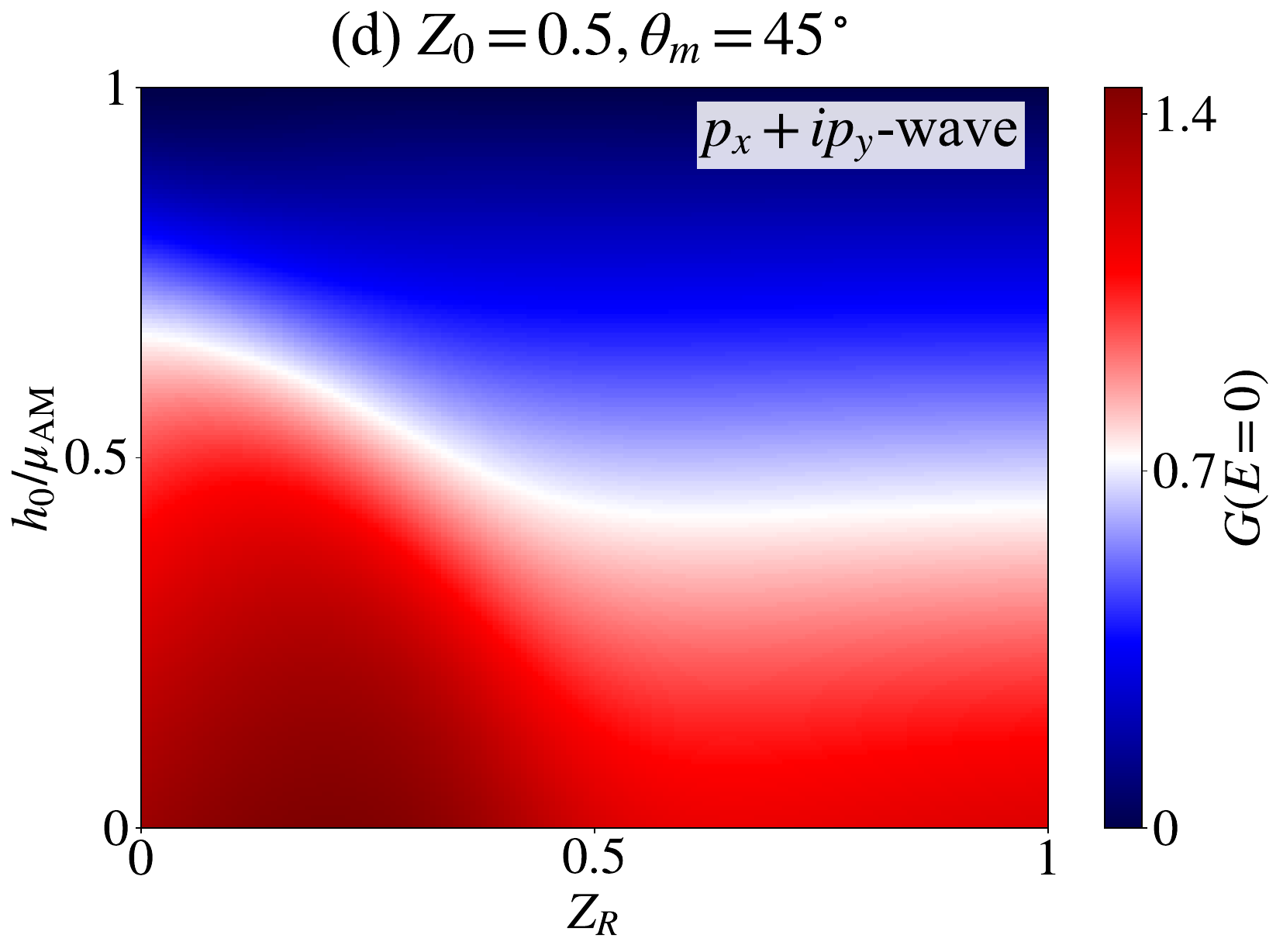}
}
\caption{
(Top panel) (a,b) Zero-bias conductance  as a function of $\theta_m$ and $h_{\rm AM}/\mu$ for $Z_R=0.01$ and $Z_0=5$.
(Bottom panel) (c,d) Zero-bias conductance  as a function of $Z_R$ and $h_{\rm AM}/\mu$ for $Z_0=0.5$  and $\theta_m=45^\circ$. Left panels correspond to a nodal $p_x$-wave superconductor, and right panels to a chiral $p_x+ip_y$ state. }
\label{fig5}
\end{figure*}

The bias-dependent conductance spectra in Figs.~\ref{fig2} and \ref{fig3} provide the comprehensive insight into the quasiparticle transport in AM/TSC junction. But the symmetry-dependent signatures can be better visualized in the zero-energy limit. The zero-bias conductance directly probes the low-energy Andreev reflection and its sensitivity to interfacial spin mixing, which governs the spin valve behavior of the system. Also, by reducing the energy-resolved information to a single experimentally accessible quantity, the zero-bias limit enables a clearer and more direct comparison between nodal $p_x$ and chiral $p_x+ip_y$ pairing states, while emphasizing the role of RSOC based mixing in AM/TSC/AM transport. Fig.~\ref{fig4} presents the zero-bias conductance $G(0)$ as a function of the
N\'eel-vector misalignment angle $\theta_m$ for different values of $Z_R$. Plots in panel (a) and (b) correspond to the nonchiral $p_x$-wave
while panel (c) and (d) are for chiral $p_x+ip_y$-wave. For $Z_0=0.01$,
$G(0)$ exhibits a pronounced and highly non-monotonic angular dependence,
with deep minima at near $\theta_m\simeq 45^\circ$ and
$135^\circ$ as seen from  Fig.~\ref{fig4}(a). This strong modulation reflects the combined effects of
nodal $p_x$ pairing and momentum-dependent spin splitting in the AM
leads. Moreover, at zero bias Andreev reflection is governed by interference
between trajectories experiencing opposite signs of the $p_x$ gap,
while RSOC-induced spin rotation makes the effective overlap between
left and right AM eigenchannels strongly $\theta_m$ dependent. This characteristics further enhances with the increase in $Z_R$, indicating that spin mixing at the
interfaces amplifies destructive interference for certain
misalignment. For $Z_0=1$, the overall
amplitude of $G(0)$ is reduced and the angular dependence becomes smoother,
but a clear $\theta_m$ modulation still persist, especially at larger $Z_R$ as depicted from Fig.~\ref{fig4}(b).
This demonstrates that even when direct Andreev reflection processes are suppressed,
RSOC-mediated spin-selective tunneling retains a finite spin valve
response in the nodal $p_x$ case.

Plots in panel \ref{fig4}(c) and \ref{fig4}(d) display the corresponding zero-bias conductance for the
chiral $p_x+ip_y$ superconductor, revealing qualitatively different
behavior. For $Z_0=0.01$, $G(0)$ remains
finite and relatively large for all $\theta_m$, with oscillatory but
smoother angular dependence compared to the $p_x$ case as seen from Fig.~\ref{fig4}(c). The absence of
nodes in the chiral gap suppresses the strong destructive interference
present for $p_x$ pairing, while interface-localized chiral states
contribute robust low-energy spectral weight. This behavior follows from RSOC-induced spin mixing discussed above. For $Z_0=1$, the angular dependence becomes more
pronounced and develops maxima near $\theta_m\simeq 90^\circ$ for larger
$Z_R$ as observed from Fig.~\ref{fig4}(d). This behavior reflects the interplay between spin-dependent
tunneling and chiral pairing, where RSOC-enhanced spin mixing selectively
favors certain N\'eel-vector configurations without producing the sharp
suppression seen in the nodal case. Overall, Fig.~\ref{fig4} highlights that the
zero-bias spin valve response of AM/TSC/AM junctions is strongly sensitive
to the superconducting pairing symmetry.  Nodal $p_x$ pairing gives deep,
RSOC-controlled angular minima, whereas chiral $p_x+ip_y$ - wave pairing produces
a more robust and smoother $\theta_m$ dependence.

\begin{figure*}[t]
\centering
\centerline{ 
\includegraphics[scale=0.2]{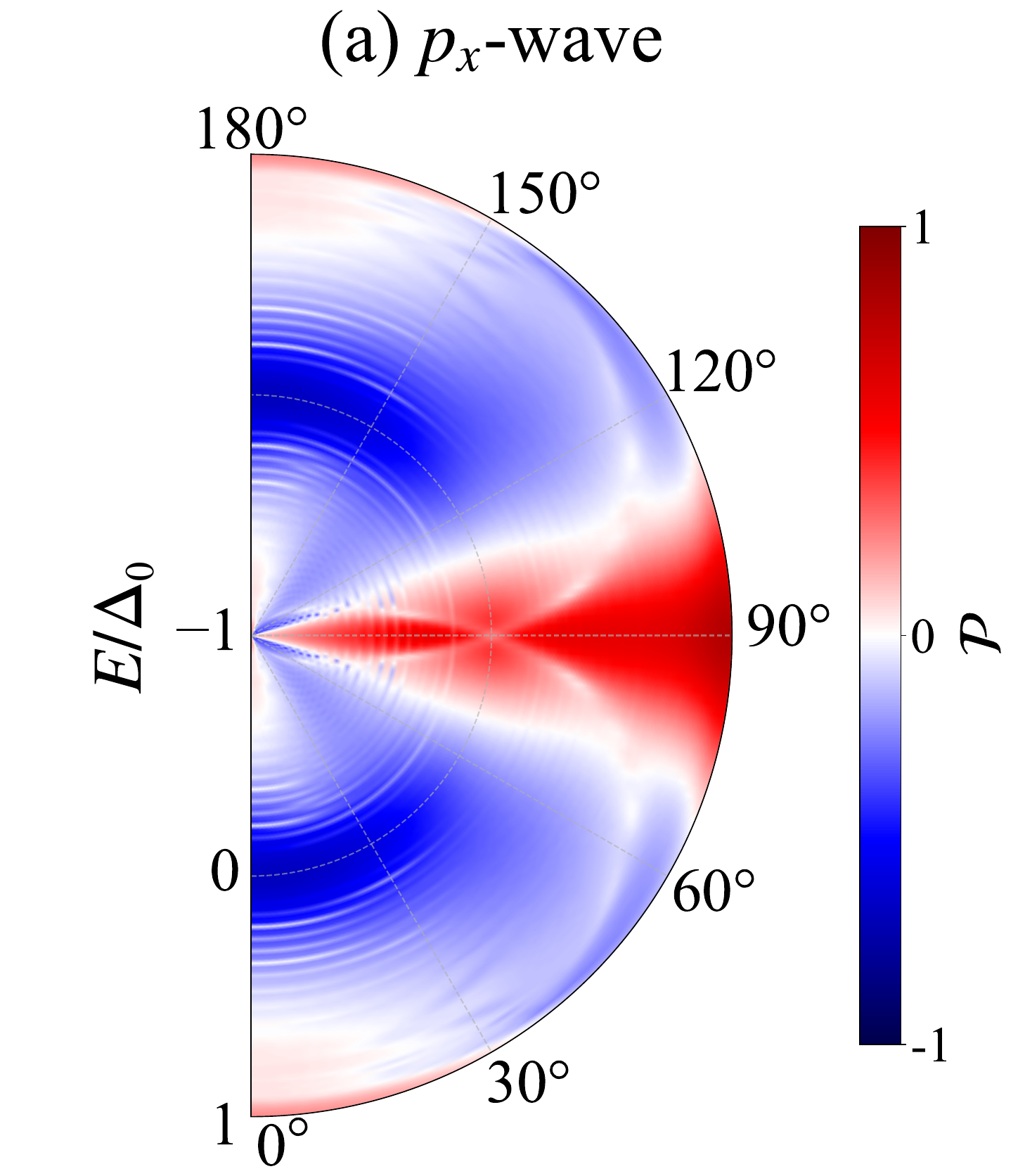}
\hspace{-5mm}
\includegraphics[scale=0.2]{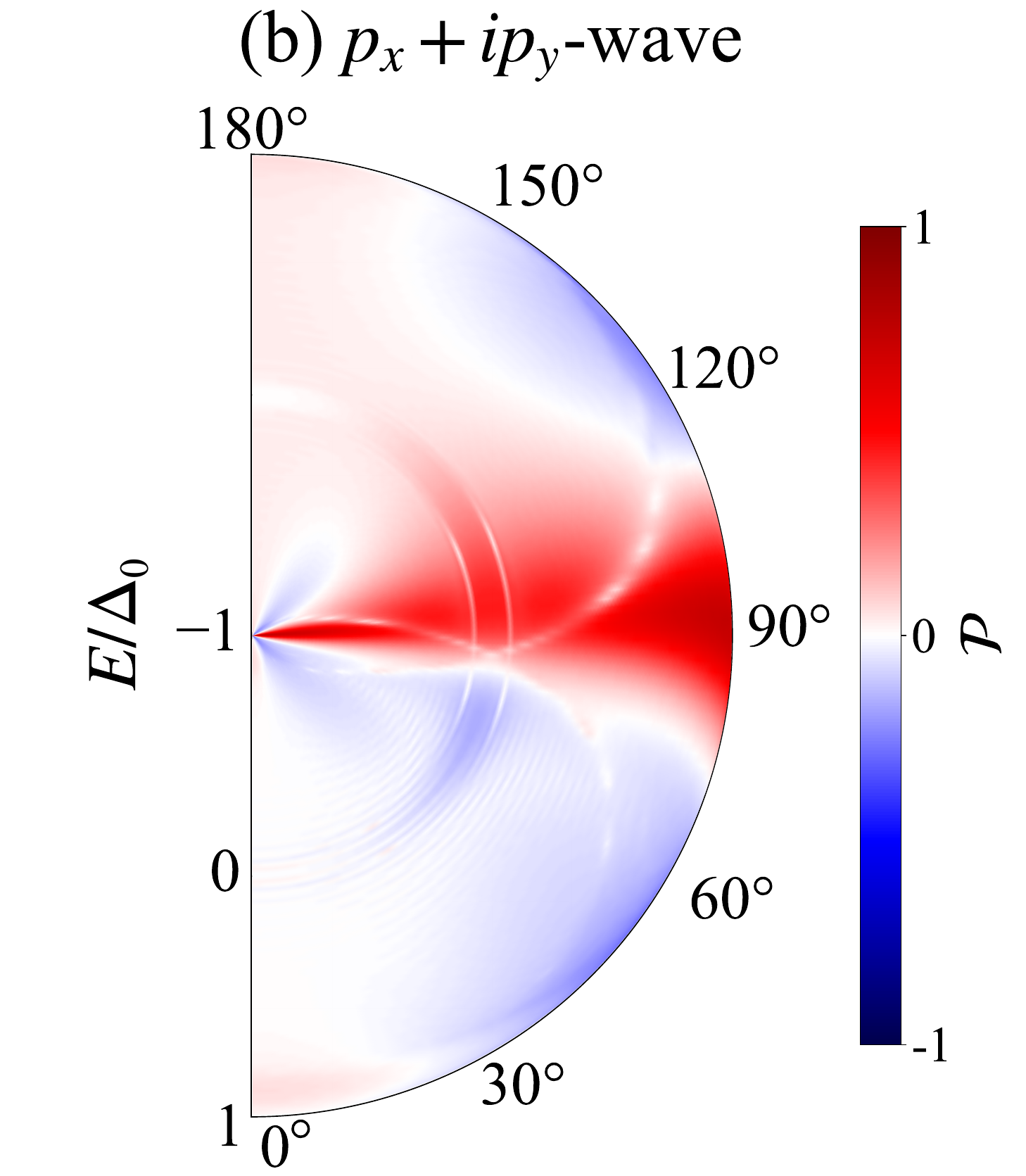}
\hspace{-5mm}
\includegraphics[scale=0.2]{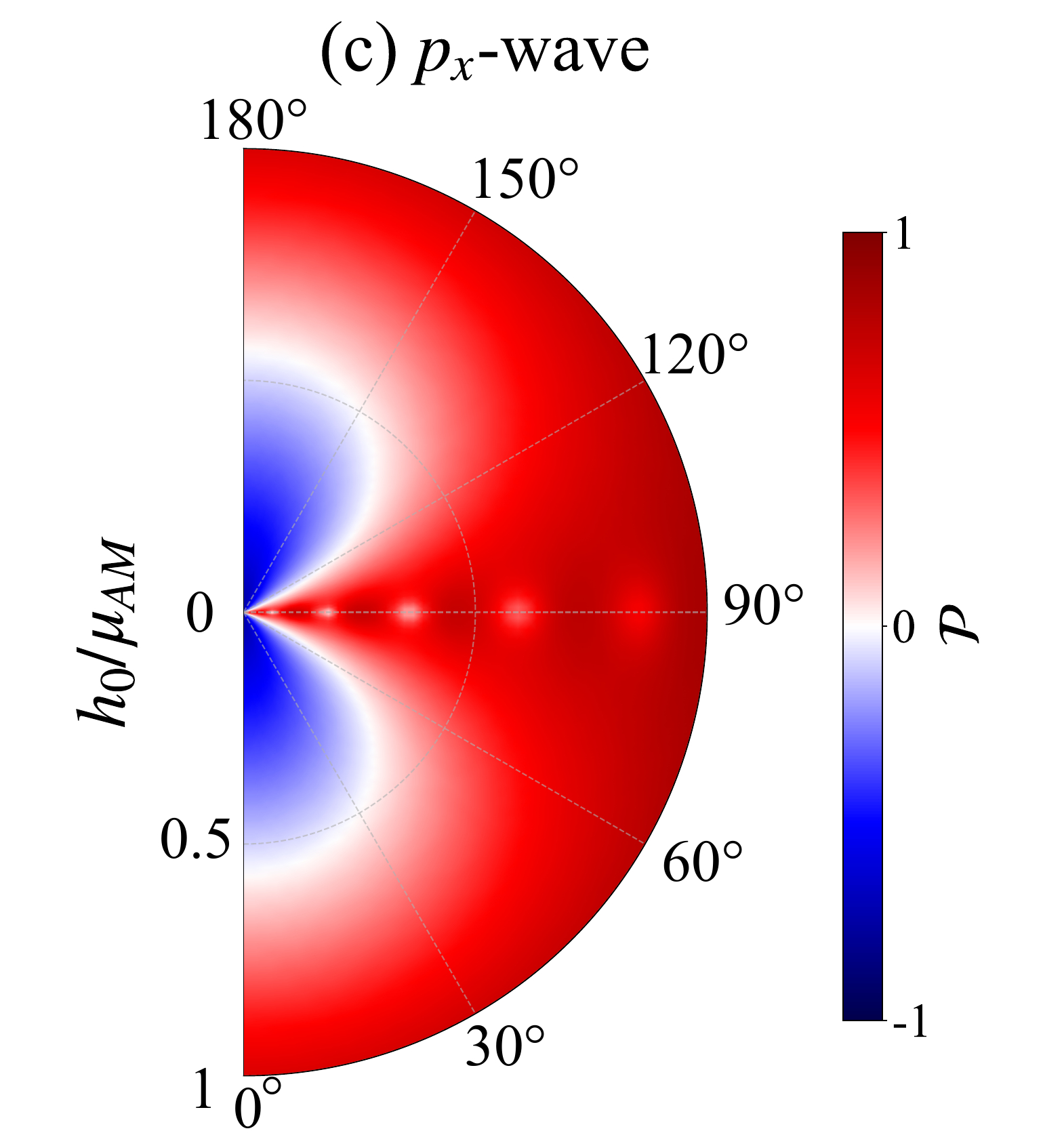}
\hspace{-8mm}
\includegraphics[scale=0.2]{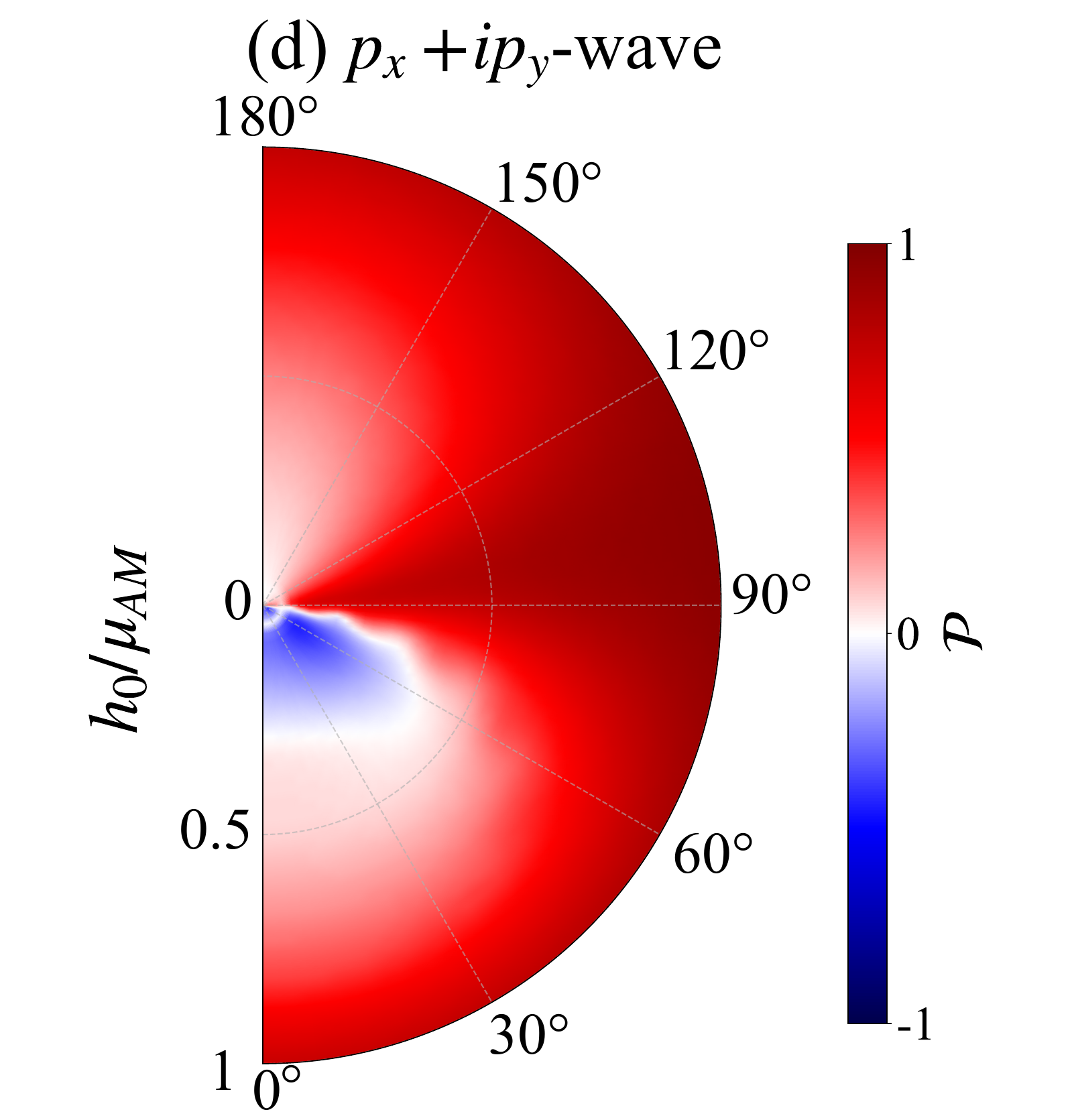}
}
\caption{
Panels (a,b) show Polar maps of the angle-resolved spin polarization ($\mathcal{P}$) as a function of $E/\Delta$ and $\theta_{m}$ for for $p_x$ and chiral $p_x+i p_y$ - wave 
superconductors respectively considering $Z_0 = 0.1$, $Z_R = 0.5$ and $h_0/\mu_\text{AM} = 0.1$. 
Panels (c,d) display the zero-bias polarization
as a function of $h_0/\mu_{\mathrm{AM}}$ and $\theta_m$. or for $p_x$ and chiral $p_x+i p_y$ - wave 
superconductors respectively considering $Z_0 = 0.1$, $Z_R = 0.5$. Red (blue) indicates
positive (negative) polarization.
}
\label{fig6}
\end{figure*}

Although Fig.~\ref{fig2} - Fig.~\ref{fig4}, reveal pronounced conductance asymmetry and spin valve effect in our proposed geometry, it is necessary to understand the combined effect of AM exchange strength, relative orientation of N\'eel vector mismatch and RSOC. This is most transparently achieved in the zero-energy limit, where transport is governed by low-energy Andreev processes that are highly sensitive to
the superconducting phase structure and spin-selective scattering at the interfaces. 
Fig.~\ref{fig5} present the variation zero bias conductance with the $h_{\rm AM}/\mu$, $\theta_m$ and $Z_R$ considering $Z_0 = 0.5$.   In panels (a) and (b), the conductance is shown with $\theta_m$ and $h_{\rm AM}/\mu$, the angular modulation originates from the relative orientation between the momentum-dependent AM spin texture and the spin structure of the triplet Cooper pairs. For the nodal $p_x$ state, from Fig.~\ref{fig5}(a) the zero-bias response is dominated by the ABS discussed above. These states are highly sensitive to spin-dependent phase shifts induced by altermagnetism, so increasing $h_{\rm AM}$ progressively spin-splits the ABS and reduces their spectral overlap at zero energy. This results in a gradual suppression of $G(0)$, as indicated by the extended low-conductance (blue) regions. The oscillatory dependence on $\theta_m$ reflects constructive or destructive alignment between the injected spin polarization and the triplet $d$-vector, consistent with the spin valve behavior and RSOC as already observed in Fig.~\ref{fig2} and Fig.~\ref{fig3} . In contrast, Fig.~\ref{fig5}(b) shows the chiral $p_x+ip_y$ case, where the bulk gap is fully open and low-energy transport is dominated by topological edge modes. These modes are more resilient to pair-breaking but still sensitive to spin filtering, $G(0)$ exhibits a broader high-conductance region and a weaker suppression with $h_{\rm AM}$. This behavior indicate that coupling to chiral edge channels partially compensates the spin splitting induced by altermagnetism. 

Fig.~\ref{fig5}(c) and \ref{fig5}(d) display the dependence of $G(0)$ on $Z_R$ considering $\theta_m=45^\circ$. Here the evolution reflects the competition between RSOC and AM strength. For the $p_x$ state, with the increase in $Z_R$ enhances confinement at the interface, sharpening the Andreev reflection and increasing their zero-energy spectral weight, so $G(0)$ grows with $Z_R$ despite reduced transmission, indicating midgap surface states. The residual variation with $h_{\rm AM}$ shows that strong AM exchange strength still spin-polarizes these states and limits their contribution. In the chiral case, the conductance pattern is smoother because the dominant carriers are edge modes extending along the interface as seen from Fig.~\ref{fig5}(d). Moreover, increasing $Z_R$ leads to moderate localization of these modes and produces a broad conductance maximum at intermediate RSOC strengths. Overall, Fig.~\ref{fig5} demonstrates that zero-bias transport in the AM/TSC/AM structure, governed by the interplay of AM exchange strength, orientation, pairing-symmetry–dependent surface states, and interface scattering. It provide a direct spectroscopic signature of both the unconventional triplet order and the spin-textured nature of altermagnetism, in agreement with the spin valve characteristics discussed in Fig.~\ref{fig2} - Fig.~\ref{fig4}.

\begin{figure*}[hbt]
\centerline
\centerline{ 
\includegraphics[scale=0.25]{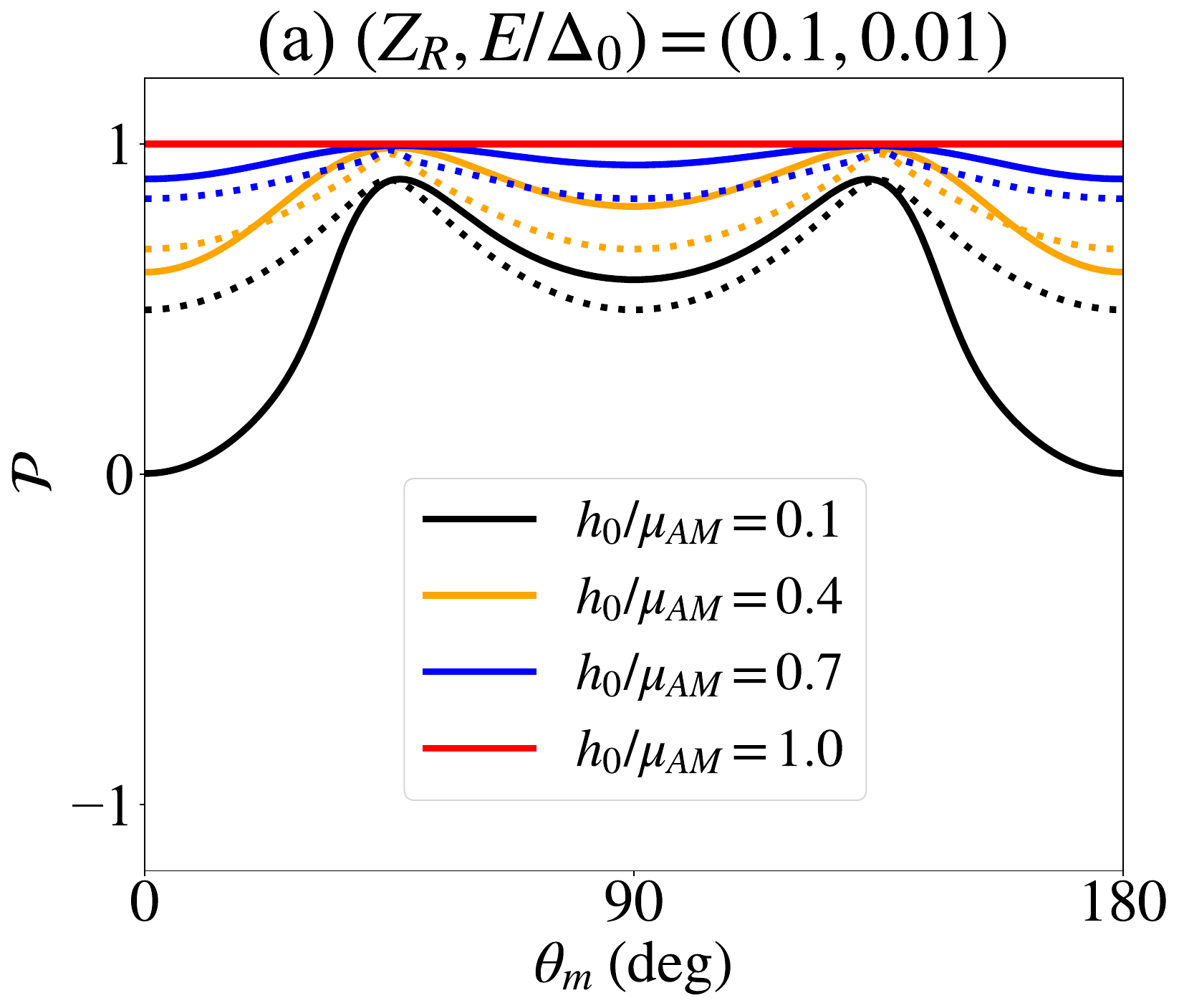}
\hspace{3mm}
\includegraphics[scale=0.25]
{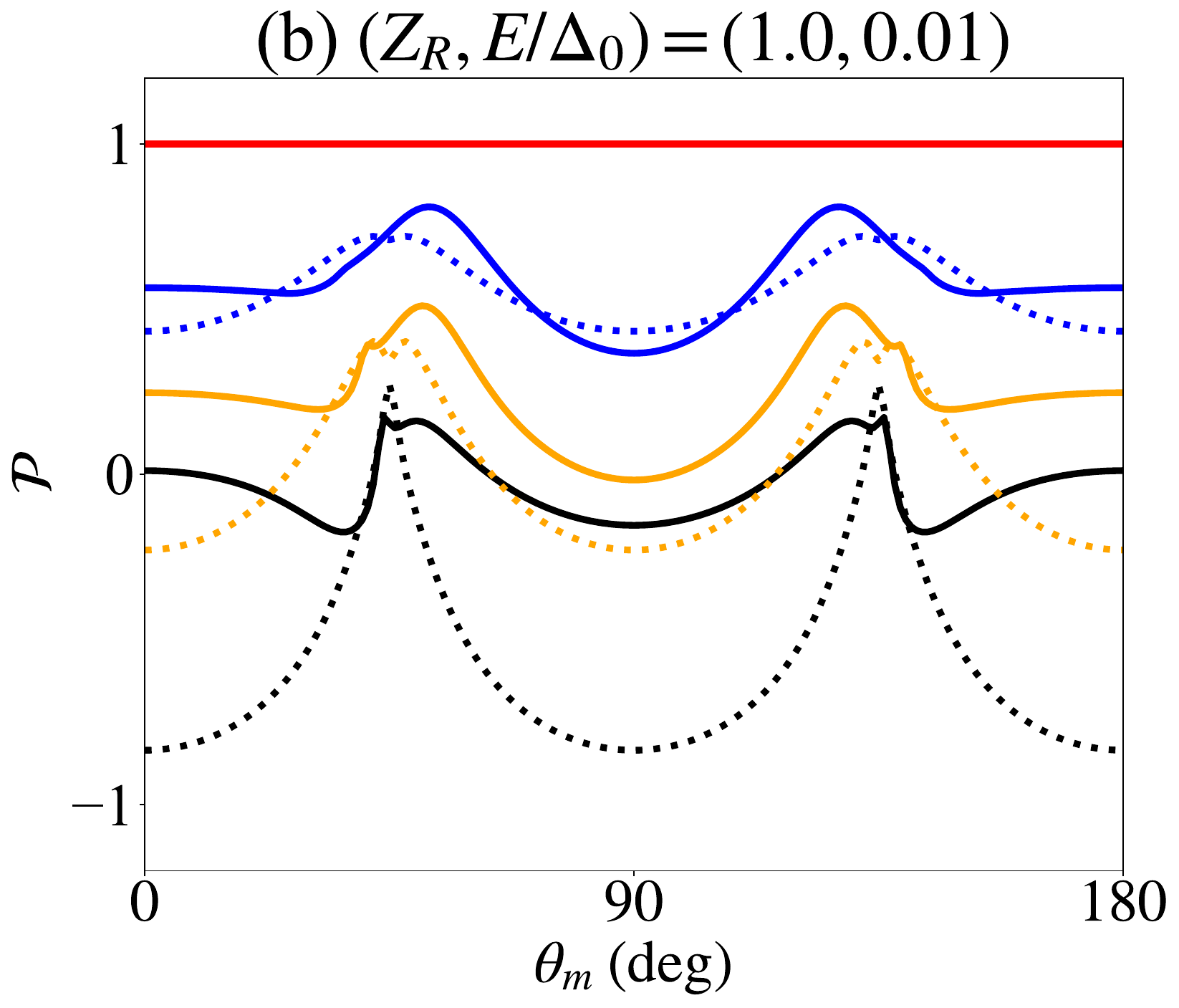}}
\centerline
\centerline{ 
\includegraphics[scale=0.25]{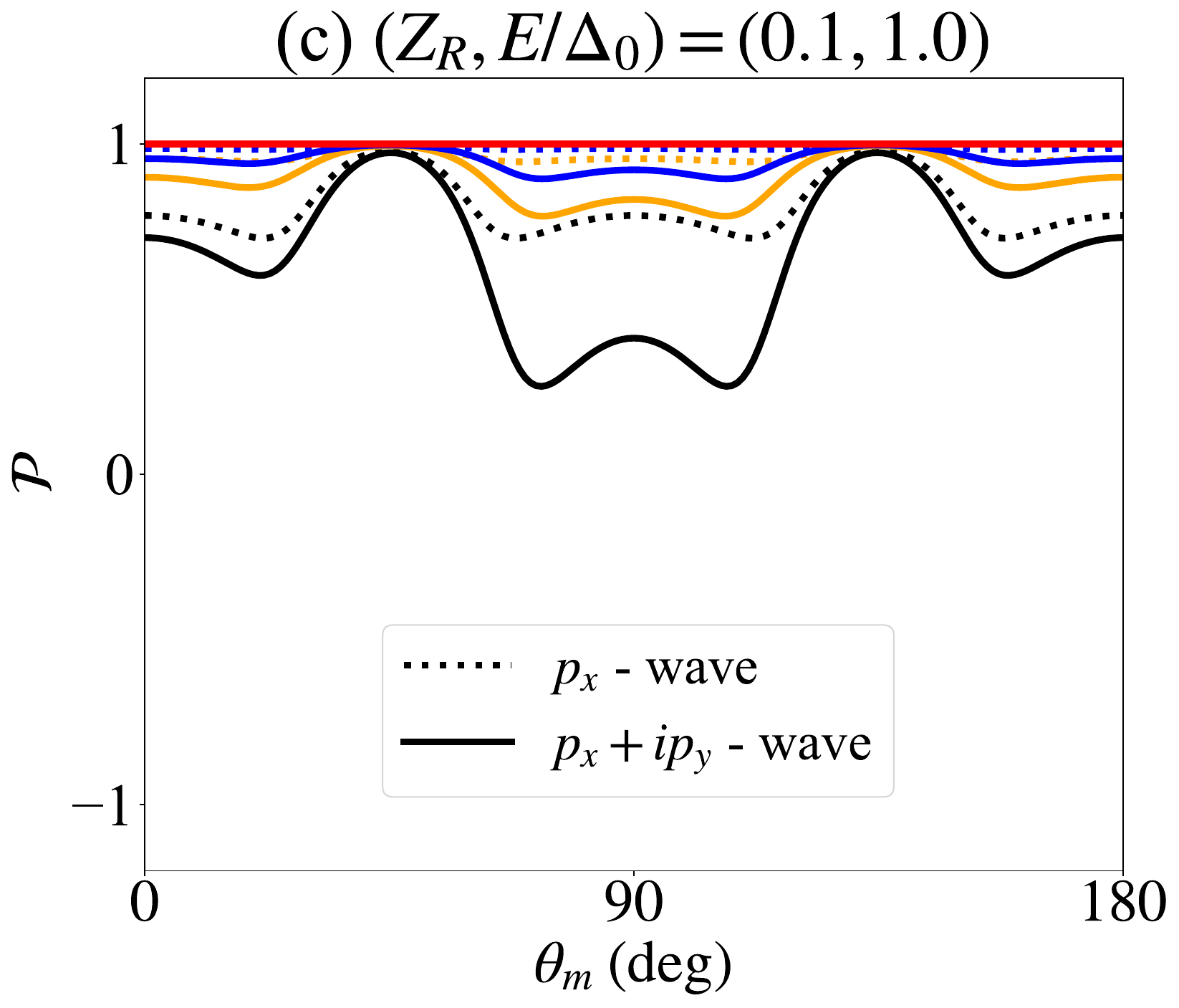}
\hspace{3mm}
\includegraphics[scale=0.25]
{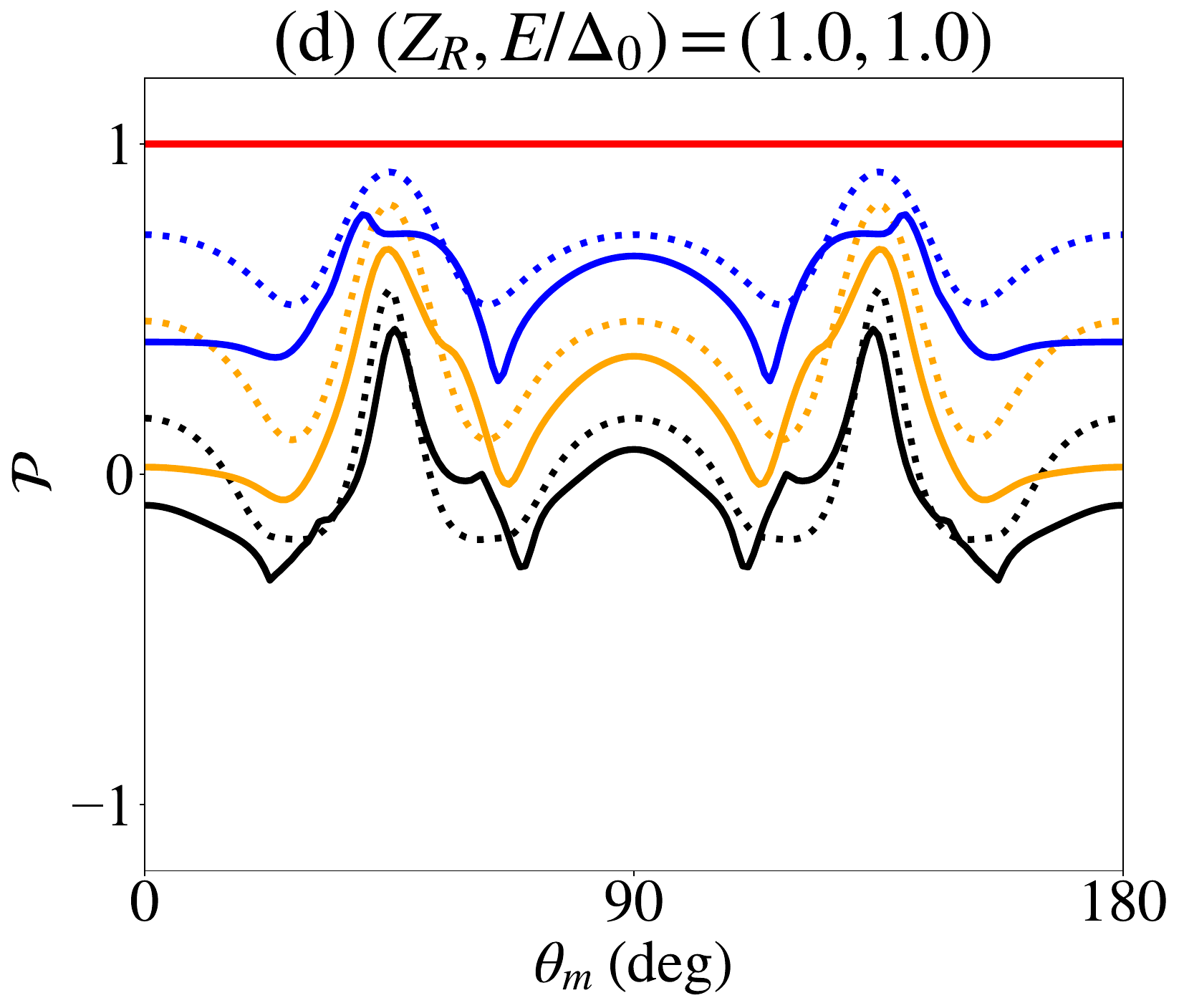}}
\caption{Variation of $\mathcal{P}$ with $\theta_m$ for different values of $h_0/\mu_{\mathrm{AM}}$. Panels (a)-(d)
are computed for the parameter pairs $(Z_R,E/\Delta_0)=(0.1,0.01)$, $(1.0,0.01)$,
$(0.1,1.0)$ and $(1.0,1.0)$, respectively (top/bottom rows correspond to weak/strong
barrier strength $Z_0=0.01$/$Z_0=1$). 
Dotted (solid) lines correspond to the nodal $p_x$ (chiral $p_x+i p_y$) - wave SC. }
\label{fig7}
\end{figure*}

\subsection{Spin polarization}
While the conductance spectra discussed above capture the role of pairing symmetry, RSOC, and the AM exchange jointly to control charge transport but it is necessary to understand whether the transmitted current acquires a finite spin polarization. In altermagnets, spin polarization arises from momentum-dependent spin splitting of the electronic bands rather than from uniform magnetization. The combined action of superconducting proximity and interfacial spin mixing due to RSOC can therefore convert the spin-textured AM band structure into a net spin-polarized current, making the spin polarization a direct and experimentally relevant measure of merit for the spin valve functionality of the AM/TSC/AM junction. The spin polarization of the charge current at excitation energy $E$  within linear response is defined as~\cite{Zutic1999,Eschrig2011},
\begin{equation}
\mathcal{P}(E)
\equiv
\frac{G_{+}(E)-G_{-}(E)}
     {G_{+}(E)+G_{-}(E)} ,
\label{eq33}
\end{equation}
where $G_\sigma$ are the spin resolved conductances can be obtained from Eq. (\ref{eq31}), with $\sigma=\pm$ label the two AM eigenchannels.
Using the angle-resolved conductance 
Eq.~\eqref{eq33} can be written explicitly
as
\begin{equation}
\mathcal{P}(E)
=
\frac{
\displaystyle
\int_{-\pi/2}^{\pi/2}
d\theta\,\cos\theta\,
\Big[
\mathcal{G}_{+}\big(E,k_y(\theta)\big)
-
\mathcal{G}_{-}\big(E,k_y(\theta)\big)
\Big]
}{
\displaystyle
\int_{-\pi/2}^{\pi/2}
d\theta\,\cos\theta\,
\Big[
\mathcal{G}_{+}\big(E,k_y(\theta)\big)
+
\mathcal{G}_{-}\big(E,k_y(\theta)\big)
\Big]
}.
\label{eq34}
\end{equation}
\begin{figure*}[hbt]
\centering
\centerline{
\hspace{15mm}
\includegraphics[scale=0.25]{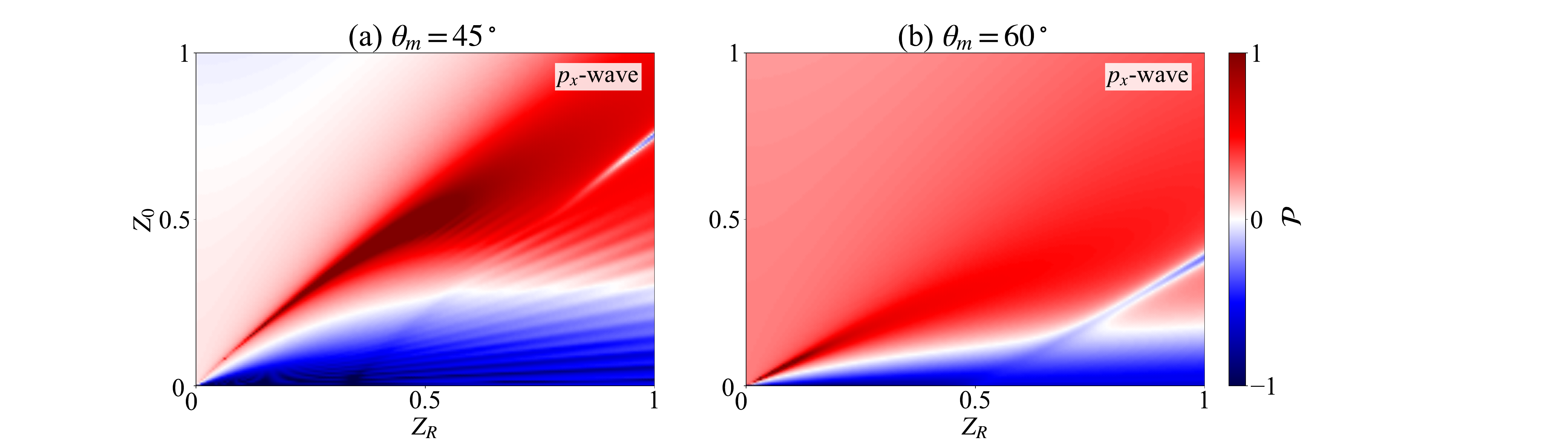}}
\centering
\centerline{
\hspace{15mm}
\includegraphics[scale=0.25]
{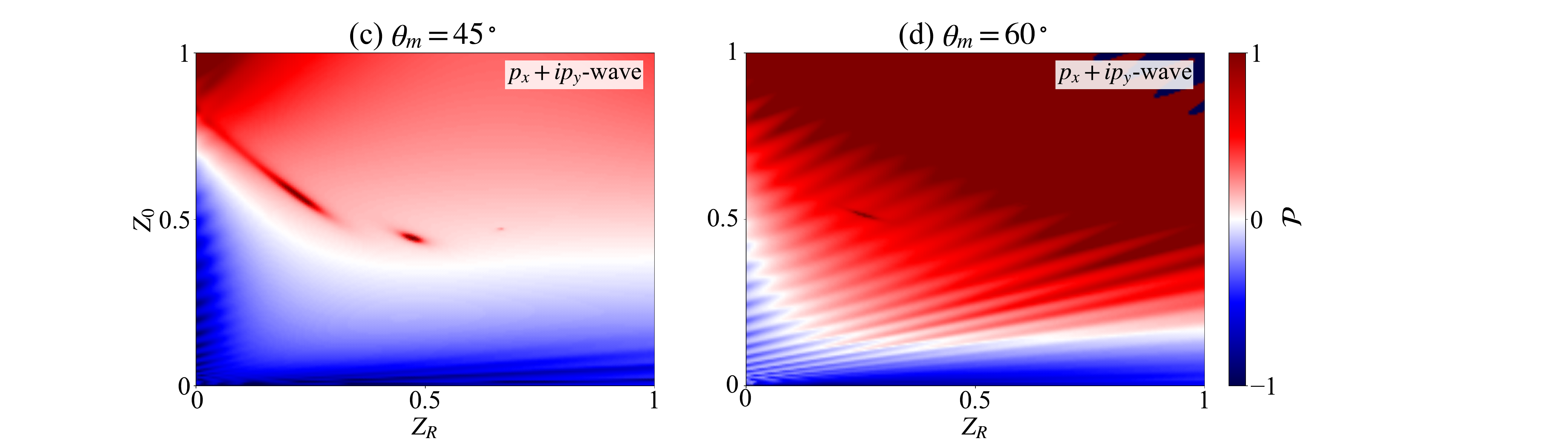}}
\caption{
Zero-bias spin polarization $\mathcal{P}$ in an AM/TSC/AM junction as a
function of $Z_R$ and $Z_0$ for fixed
$\theta_m=45^\circ$ and $60^\circ$. Panels (a,b)
correspond to the nodal $p_x$-wave superconductor, while panels (c,d) show the
chiral $p_x+i p_y$ case.
}
\label{fig8}
\end{figure*}

Fig.~\ref{fig6} presents the angle and energy-resolved spin polarization in the AM/TSC/AM junction, illustrating how the orbital symmetry of triplet pairing is imprinted onto spin-selective transport. Panels \ref{fig6}(a) and \ref{fig6}(b) show polar maps of $\mathcal{P}(E/\Delta,\theta_m)$ for $Z_0=0.1$, $Z_R=0.5$, and $h_0/\mu_{\rm AM}=0.1$ for nodal $p_x$ and chiral $p_x+ip_y$ superconductors, respectively. For the nodal $p_x$ state, $\mathcal{P}(E/\Delta,\theta_m)$ exhibits pronounced sign changes and strong angular modulation as seen from Fig.~\ref{fig6}(a). This behavior originates from zero-energy ABS formed due to the sign change of the order parameter under specular reflection. These midgap states are highly sensitive to spin-dependent phase shifts induced by the momentum-dependent AM exchange field and interfacial RSOC. Consequently, small variations of $\theta_m$ strongly modify the interference between spin-polarized electron and hole amplitudes, leading to alternating regions of positive and negative polarization in the $(E,\theta_m)$ plane. The sharp contrast near $E\approx0$ indicates that low-energy spin polarization is dominated by ABS whose spin texture is efficiently filtered by the relative orientation of the N\'eel vector. In contrast, the chiral $p_x+ip_y$ state exhibits a smoother polarization pattern with weaker angular modulation and reduced sign reversals as observed from Fig.~\ref{fig6}(b). Moreover, it is found to be highly asymmetric in this scenario. Since the bulk is fully gapped, low-energy transport is governed by chiral edge modes rather than flat midgap ABS. These topological modes are robust against moderate spin-dependent perturbations and therefore display a less oscillatory polarization response under rotations of $\theta_m$. The reduced role of trajectory-dependent interference is consistent with the smoother conductance spectra discussed earlier.

Polar plots in panels \ref{fig6}(c) and \ref{fig6}(d) display the corresponding zero-bias spin polarization as a function of $h_0/\mu_\text{AM}$ and $\theta_m$. For the $p_x$ state, $\mathcal{P}(E=0)$ changes sign over a wide region of the $(\theta_m, h_{\rm AM}/\mu_{\rm AM})$ plane as seen from Fig.~\ref{fig6}(c), indicating that the spin character of the zero-energy Andreev levels can be inverted by tuning either the N\'eel-vector misalignment or the magnitude of the AM exchange splitting. This strong sensitivity originates from the spin-dependent phase shifts accumulated by quasiparticles undergoing repeated Andreev reflections in the presence of momentum-dependent AM spin splitting. Increasing $h_{\rm AM}$ progressively lifts the degeneracy of the midgap states and redistributes their spectral weight between the two AM eigenchannels, which directly translates into sign reversals of the spin-polarized current. By contrast, the chiral case  exhibits a predominantly positive polarization over most of the parameter space, with a comparatively weak dependence on $h_{\rm AM}$ as observed from Fig.~\ref{fig6}(d). This behavior reflects the robustness of chiral edge modes against spin-selective pair breaking. Although interfacial RSOC and AM exchange still induce spin filtering, the topological nature of the low-energy modes suppresses strong polarization reversals. Consequently, the polarization remains smoother and more uniform compared to the nodal case. Overall, Fig.~\ref{fig6} shows that spin polarization in AM/TSC/AM junctions is a sensitive probe of pairing symmetry: nodal $p_x$ pairing yields strongly angle- and energy-dependent polarization with frequent sign reversals, whereas chiral $p_x+ip_y$ pairing produces a smoother, more robust response dominated by edge-state transport.

Fig.~\ref{fig7} present the angular dependence of the zero-bias spin polarization
$\mathcal{P}(0,\theta_m)$ for different $Z_R$, $E/\Delta_0$, $Z_0$, and
$h_0/\mu_{\mathrm{AM}}$. Panels (a) and (b) correspond to low bias
($E/\Delta_0=0.01$) with weak and strong RSOC, respectively, while panels (c)
and (d) show the high-bias case ($E/\Delta_0=1$). The upper (lower) row
corresponds to a nearly transparent (opaque) interface represented by $Z_0=0.01$ ($Z_0=1$).
Dotted (solid) curves denote nodal $p_x$ (chiral $p_x+ip_y$) pairing. For the nodal $p_x$ state, the gap changes sign under specular reflection, and satisfy Tanaka–Kashiwaya sign-change condition,
$\Delta(k_x,k_y)\to -\Delta(-k_x,k_y)$, and produce zero-energy surface ABS~\cite{Tanaka2000}. At low bias, these ABS dominate transport and exhibit strong angular
modulation, frequent sign reversals, and large polarization amplitudes that
increase with the increase in $h_0/\mu_{\mathrm{AM}}$ as seen from Figs. \ref{fig7}(a) and \ref{fig7}(b).
The odd-momentum AM exchange field together with interfacial RSOC induces spin-dependent phase accumulation along time-reversed trajectories, rendering the ABS spectrally spin-imbalanced and enabling efficient spin filtering. In contrast, the chiral $p_x+ip_y$ state exhibits smoother angular profiles with fewer sign reversals. This is consistent with transport governed by chiral edge modes rather than trajectory-dependent ABS interference. With the increase in $Z_R$ interfacial spin mixing enhances. Thus it  reshapes the angular structure in both pairing states and broadening ABS-related features in the nodal case. At higher energies for larger $Z_0$ the ABS contribution is strongly suppressed and quasi-particle tunneling dominates,leading to reduced angular behavior and a more smoother polarization profile as observed from Figs.~\ref{fig7}(c) and \ref{fig7}(d). Overall, Fig.~\ref{fig7} highlights two symmetry-controlled regimes:
an ABS-dominated regime for nodal pairing with strong, angle-dependent
spin polarization, and a topological-mode-dominated regime for
chiral pairing with a more robust polarization. Moreover, the
interfacial RSOC, junction transparency, and AM exchange strength control the
crossover between these regimes and the magnitude of the measurable
$\mathcal{P}(\theta_m)$.

\begin{figure*}[t]
\centerline
\centerline{ 
\includegraphics[scale=0.2]{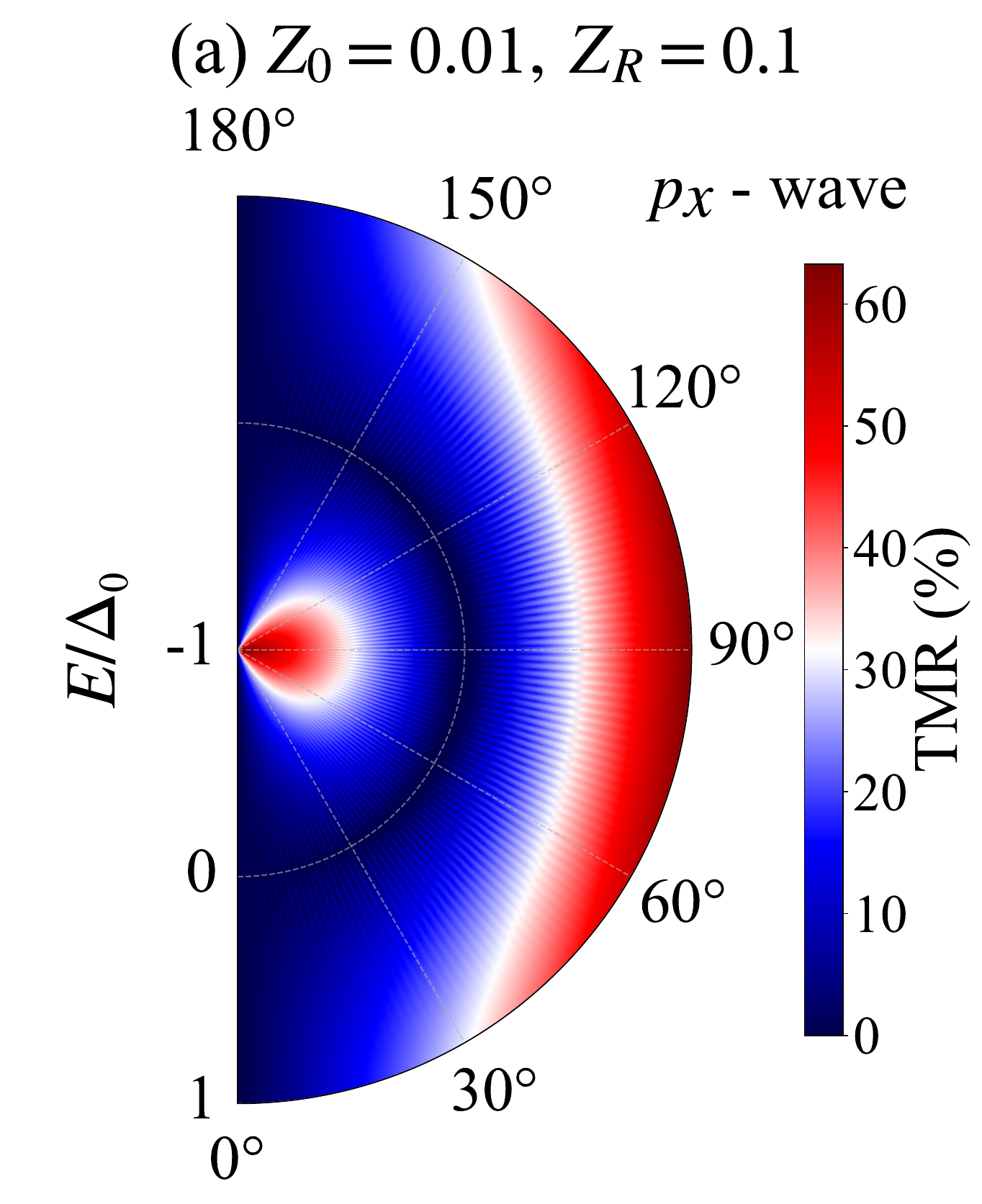}
\hspace{-5mm}
\includegraphics[scale=0.2]{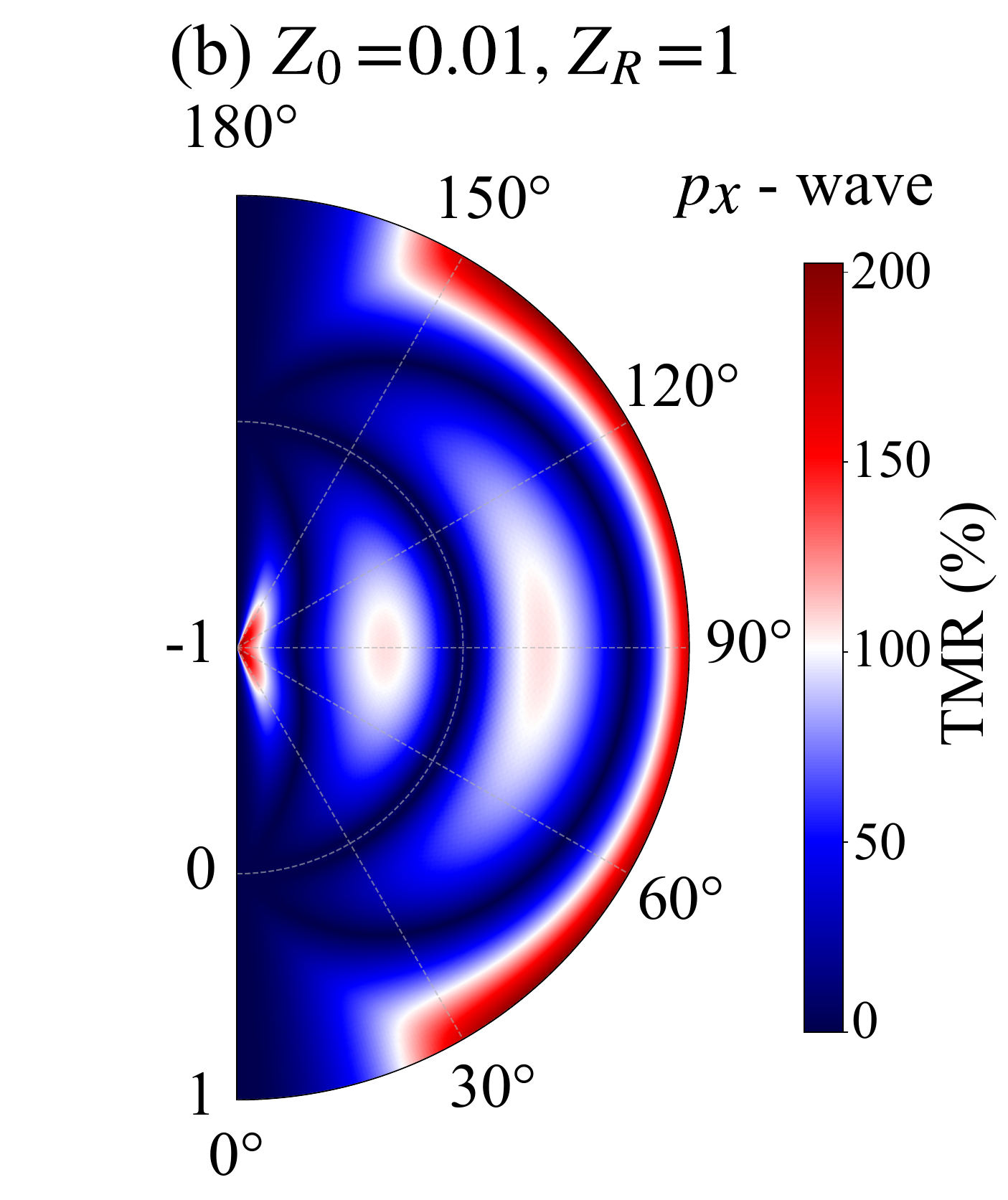}
\hspace{-5mm}
\includegraphics[scale=0.2]{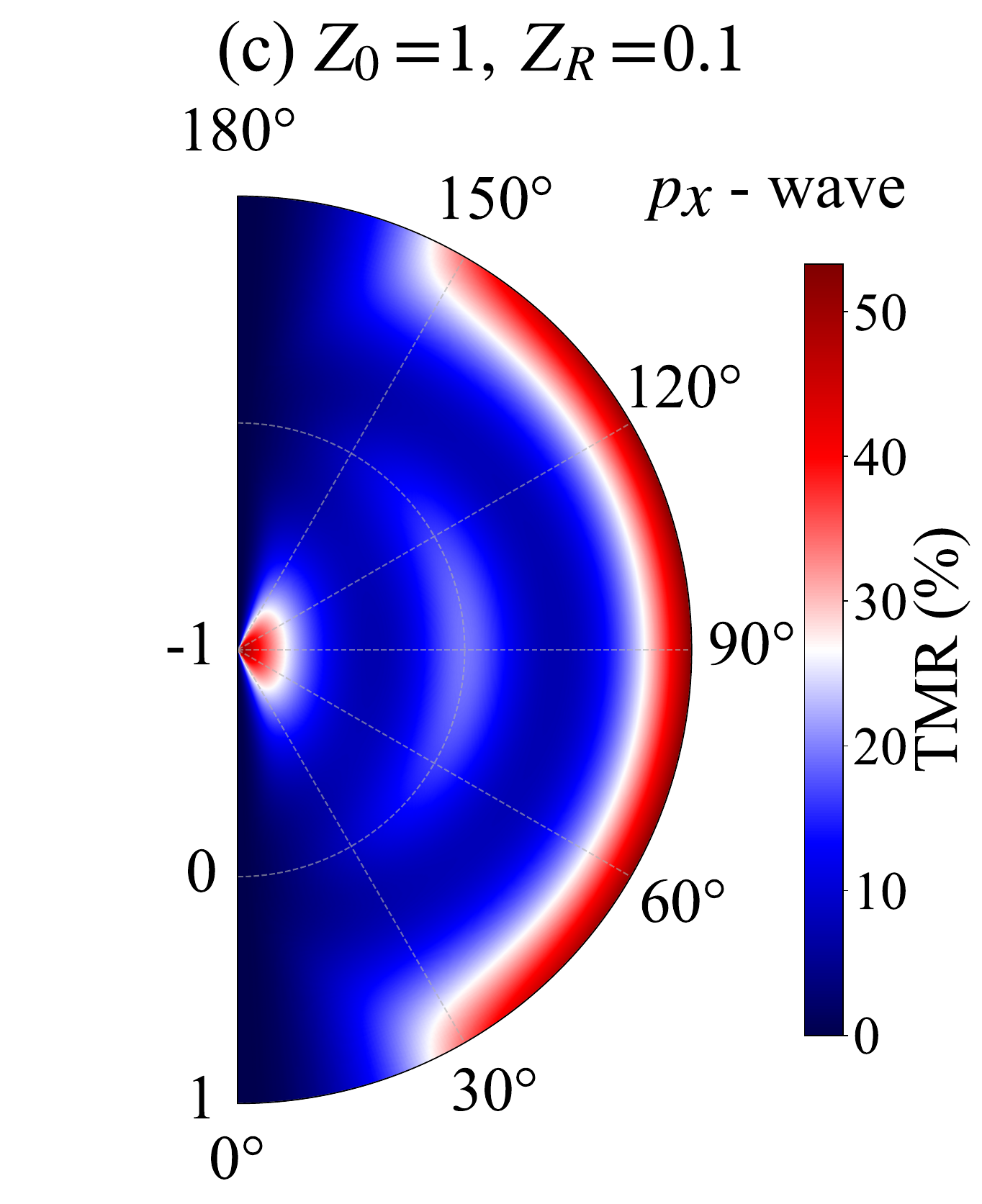}
\hspace{-5mm}
\includegraphics[scale=0.2]{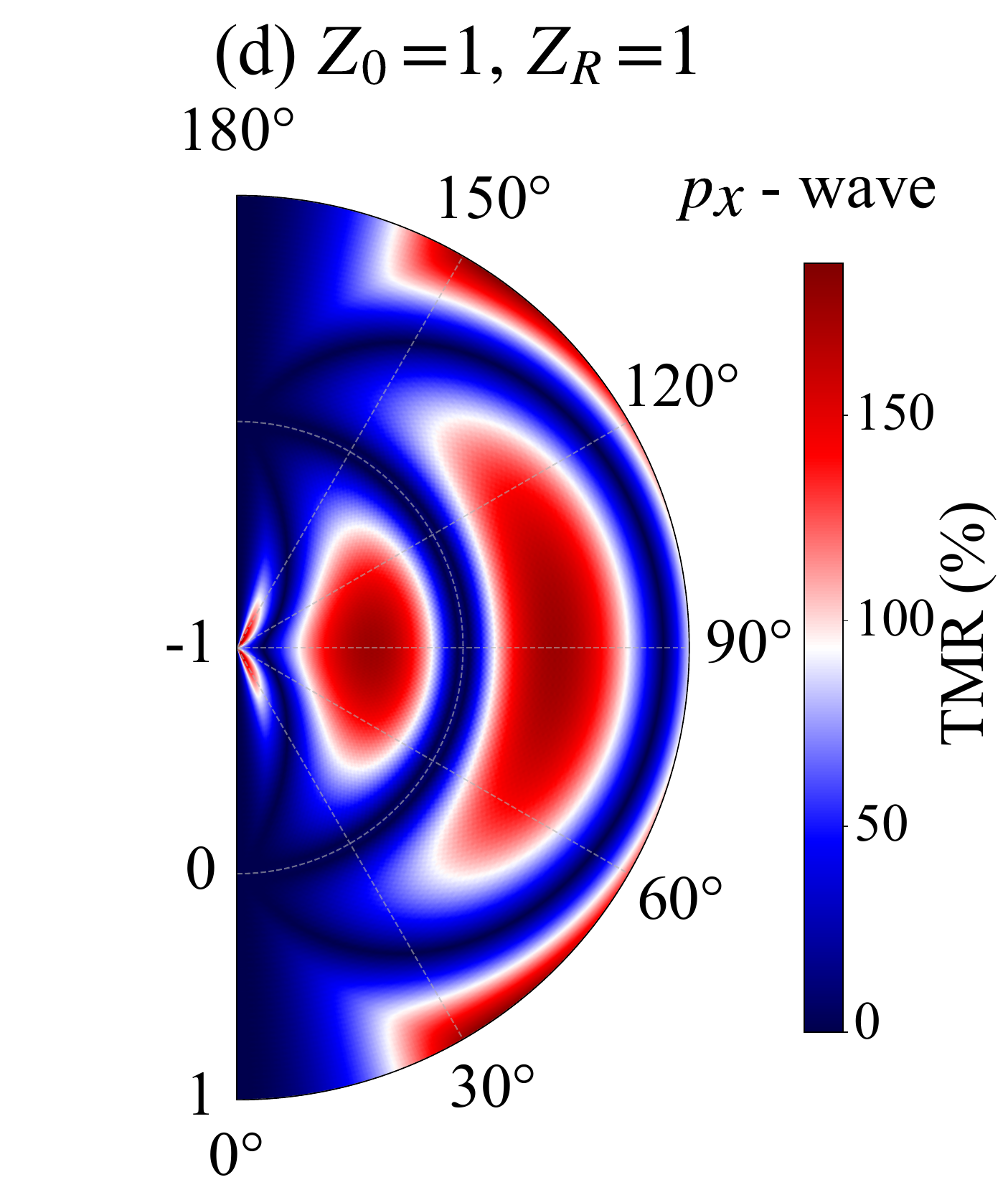}
}
\centerline{ 
\includegraphics[scale=0.2]{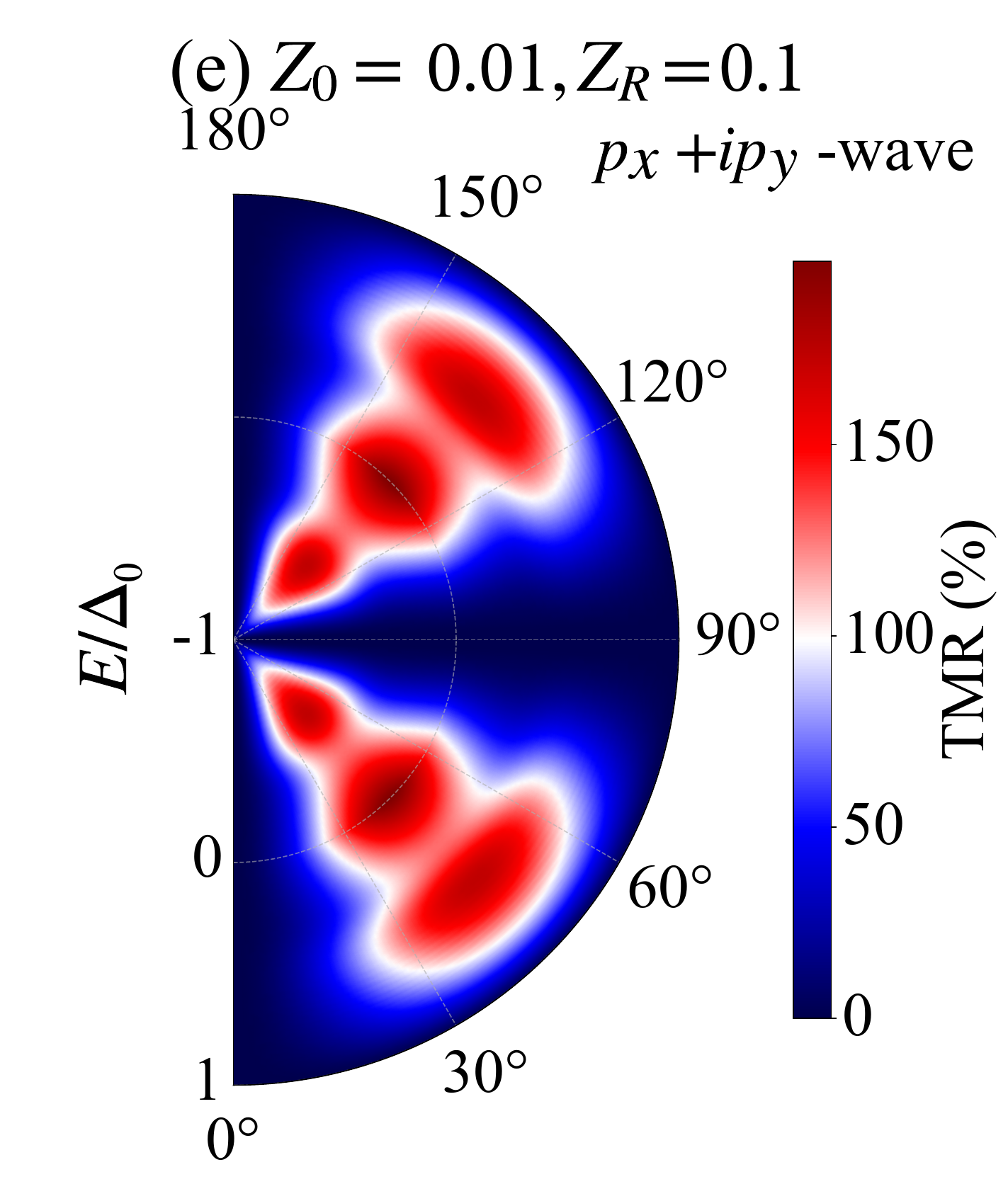}
\hspace{-8mm}
\includegraphics[scale=0.2]{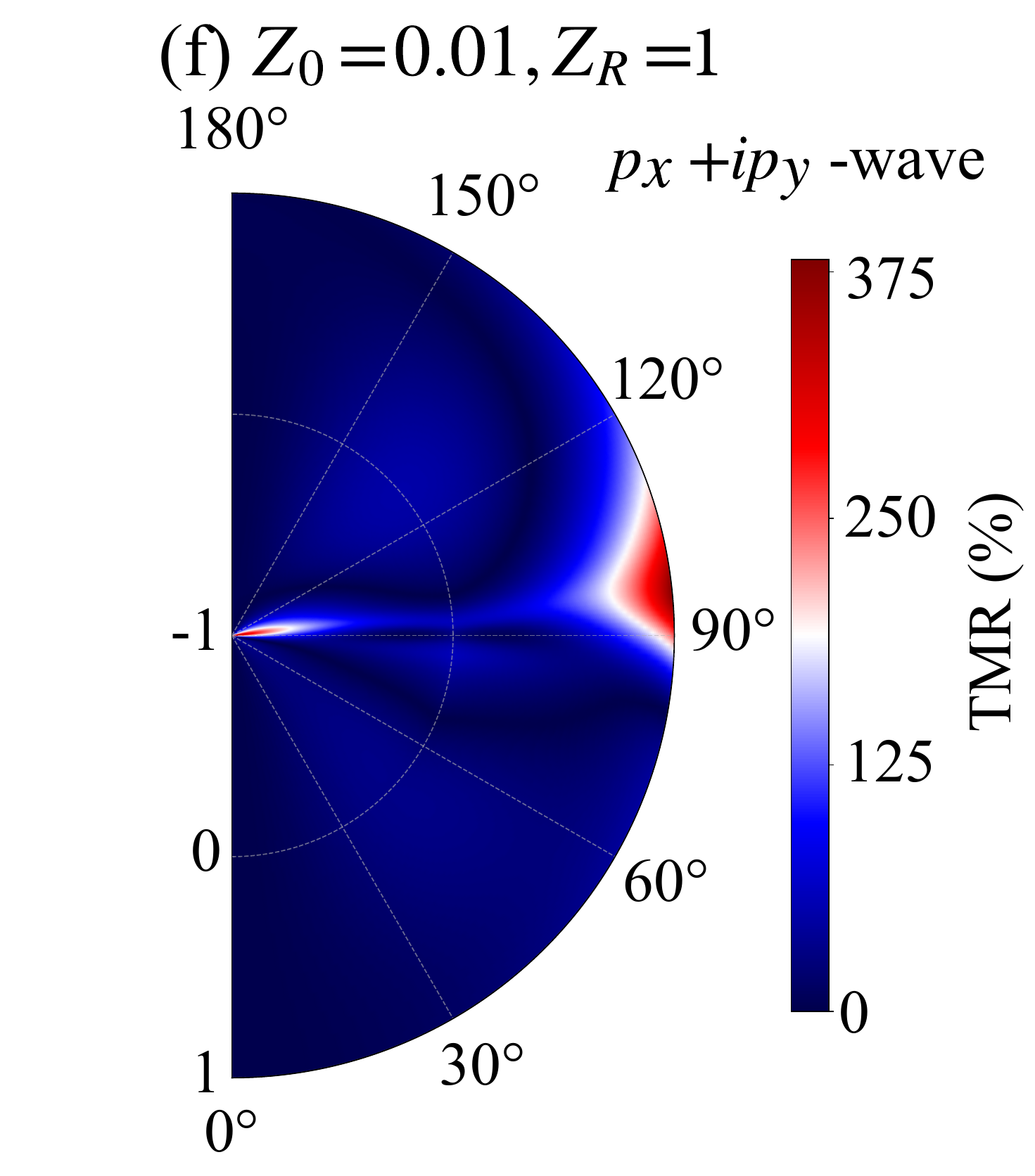}
\hspace{-8mm}
\includegraphics[scale=0.2]{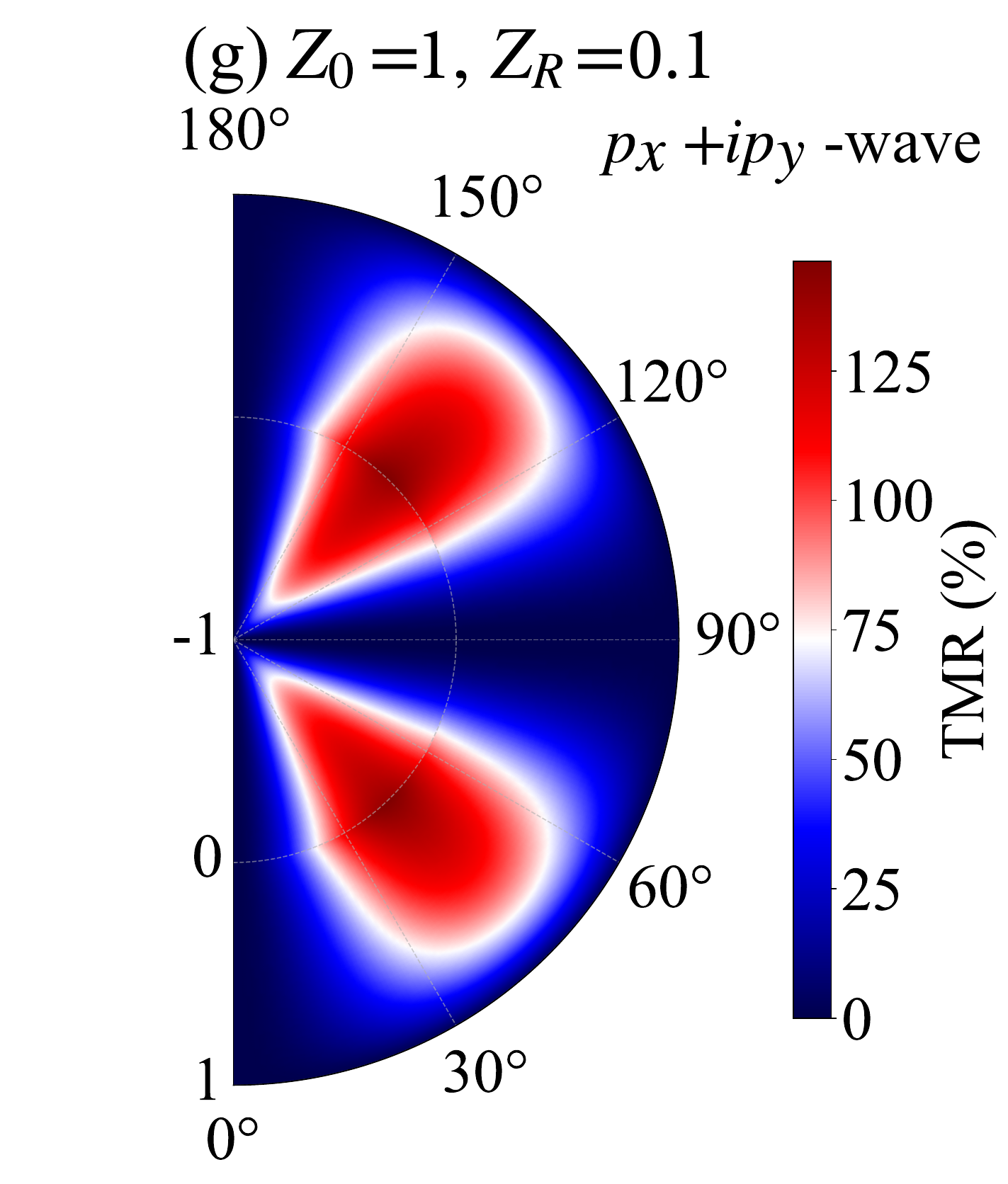}
\hspace{-8mm}
\includegraphics[scale=0.2]{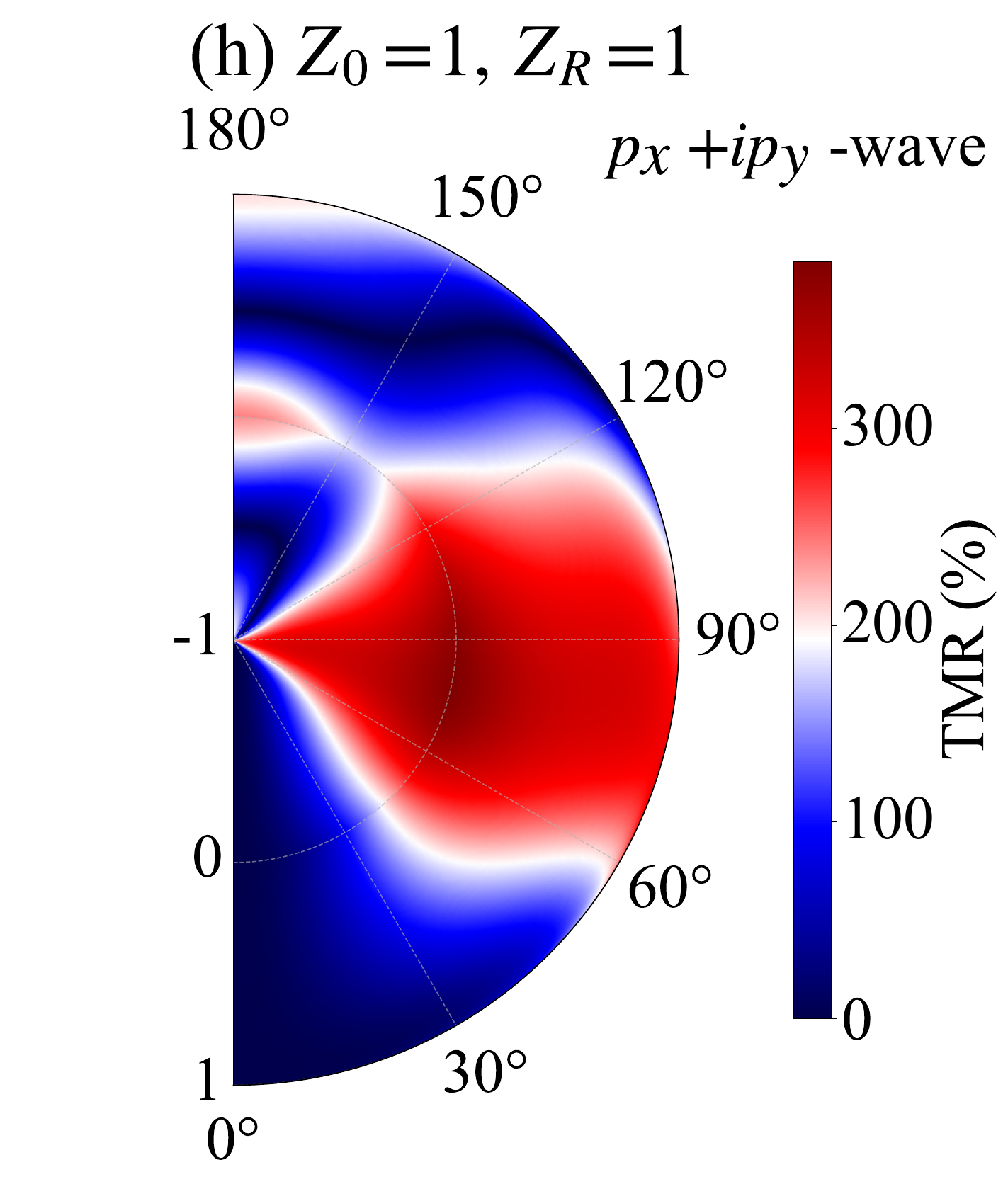}
}
\caption{
Polar maps of the TMR as a function of quasiparticle energy \(E/\Delta\) and relative AM orientation angle \(\theta_{m}\) for non-chiral \(p\)-wave SC (top panel) and chiral \(p_x+ip_y\)-wave SC (bottom panel). Plots in panel (a) and (b) correspond to $Z_R = 0.1$ and $1$ respectively with $Z_0 = 0.01$  while panel (c) and (d) respectively is for $Z_R = 0$ and $1$ with $Z_0 = 1$.}
\label{fig9}
\end{figure*}

Fig.~\ref{fig8} present the dependence of zero-bias spin polarization on $Z_R$ and $Z_0$ for $\theta_m=45^\circ$ and
$60^\circ$, complementing the angle and energy resolved results in
Figs.~\ref{fig6} and \ref{fig7}. Panels (a) and (b) correspond to the nodal $p_x$-wave case, while panels (c) and (d) show the chiral $p_x+ip_y$-wave case. For the nodal $p_x$-wave case, $\mathcal{P}(0)$ displays
strongly anisotropic regions with frequent sign reversals across the $(Z_R, Z_0)$
plane as seen from Figs.~\ref{fig8}(a) and ~\ref{fig8}(b). This reflects the crossover from ABS-dominated transport at low $Z_0$ to quasiparticle tunneling at large $Z_0$, together with RSOC-enhanced spin mixing that reshapes the interference between spin-conserving and spin-flip Andreev channels. The resulting phase-sensitive interference produces sharp boundaries between regions of opposite polarization. By contrast, the chiral $p_x+ip_y$ state exhibits broader and more uniform polarization patterns with comparatively few sign reversals as observed from Figs.~\ref{fig8}(c) and \ref{fig8}(d). Since
specular reflection does not induce a sign change of the complex gap, conventional sign-change ABS are absent, and the zero-bias response is governed predominantly by
chiral edge and propagating subgap modes~\cite{Tanaka2000}. These modes are polarized smoothly by the
combined action of the AM exchange field and interfacial RSOC, leading
to a more robust polarization landscape that evolves gradually with $Z_R$ and
$Z_0$. This highlights the qualitative distinction between localized ABS in nodal pairing and extended edge modes in the chiral state~\cite{Kallin2012}. In summary, Fig.~\ref{fig8} shows that zero-bias spin polarization in AM/TSC/AM junctions is jointly tunable by interfacial RSOC and barrier strength, with a strong dependence on pairing symmetry. Nodal $p_x$ pairing yields ABS-dominated, strongly anisotropic polarization with frequent sign reversals, whereas chiral $p_x+ip_y$ pairing exhibits a smoother, edge-mode–dominated response governed by spin mixing and the AM spin texture.

\subsection{Tunneling magnetoresistance (TMR)}
The spin valve response of the AM/TSC/AM junction is quantified by the tunneling
magnetoresistance (TMR), which measures the change in conductance upon
rotating the relative orientation of the N\'eel vectors in the two AM
leads. The total differential conductance at energy $E$ and misalignment angle $\theta_m$ is defined as 
\begin{equation}
G(E,\theta_m)
=
\sum_{\sigma=\pm}
\frac{e^2}{h}
\int_{-\pi/2}^{\pi/2}
d\theta\,\cos\theta\;
\mathcal{G}_{\sigma}\big(E,k_y(\theta);\theta_m\big),
\label{eq35}
\end{equation}
where the dependence on $\theta_m$ enters through the scattering
amplitudes via the rotation between the AM eigenbases of the two regions. Using the collinear configuration $\theta_m=0$ as a reference, the TMR can be
defined as~\cite{Cheng2013}
\begin{equation}
\mathrm{TMR}(E,\theta_m)
=
\frac{G(E,\theta_m)-G(E,0)}{G(E,0)} .
\label{eq36}
\end{equation}

The angular dependence of $G(E,\theta_m)$ can be understood by expressing
the overlap between left and right AM eigenspinors as
\begin{equation}
\mathcal{S}_{\sigma'\sigma}(\theta_m)
=
\langle
\chi^{R}_{\sigma'}
|
\chi^{L}_{\sigma}
\rangle ,
\label{eq37}
\end{equation}
The Eq. (\ref{eq37}) quantifies the probability amplitude for an incoming quasiparticle in AM eigenchannel $\sigma$ on the left to couple to eigenchannel $\sigma'$ on the right. Because the two AM spin frames are related by a relative rotation of angle $\theta_m$, the overlap matrix $\mathcal{S}_{\sigma'\sigma}$ forms a unitary rotation $U(\theta_m)$ in spin space.

Thus the scattering matrix in the misaligned configuration is obtained from the reference configuration, by a basis transformation
\begin{equation}
    \mathcal{S}
    \label{eq38}
    (\theta_m)=U^\dagger(\theta_m)\mathcal{S}_0U(\theta_m)
\end{equation} 
where, $\mathcal{S}_0$ is the scattering matrix in the common AM basis. Thus varying $\theta_m$ rotates the spin basis in which normal and Andreev scattering processes are resolved. Consequently, the conductance acquires an angular dependent characteristics through interference between spin-conserving and spin-mixing channels, controlled by the relative orientation of the AM N\'eel vectors.

Building on the ABS and edge mode-dominated transport regimes discussed above, we analyze the TMR as a function of $E/\Delta_0$ and $\theta_m$ in Fig.~\ref{fig9}. Plots ~\ref{fig9}(a)-\ref{fig9}(d) in the top panel corresponds to the nodal $p_x$ state while plots ~\ref{fig9}(e)–\ref{fig9}(h) in the bottom panel are for chiral $p_x+i p_y$ state. The four columns sweep interface characteristics between high transparency and weak RSOC with $(Z_0,Z_R)=(0.01,0.1)$, strong RSOC at high transparency with $(Z_0,Z_R)=(0.01,1)$, the opaque barrier with weak RSOC with $(Z_0,Z_R)=(1,0.1)$, and opaque barrier and strong RSOC with $(Z_0,Z_R)=(1,1)$. The color scale denotes the relative conductance change between the two magnetic configurations defining the TMR. For the nodal $p_x$ pairing, the gap parameter changes sign under specular reflection~\cite{Tanaka2000}, resulting zero-energy surface ABS as already discussed in Figs.~\ref{fig6}-~\ref{fig8}. These states dominate subgap transport and are highly sensitive to spin-dependent phase shifts: the odd momentum AM exchange produces different phase accumulation in the two AM channels, while interfacial RSOC mixes spins and redistributes weight between spin-conserving and spin-flip Andreev processes. Their interplay produces large ring like angle and energy-dependent TMR near the zero bias for $Z_0 \ll 1$ as seen from Fig.~\ref{fig9}(a) and \ref{fig9}(b).  This is predominantly due to coherent multiple Andreev reflections. For $Z_0 = 1$, coherent Andreev cycles are suppressed, the ABS become spectrally sharper and more localized near zero energy, and the TMR reaches larger peak amplitudes due to strong spin selectivity of tunneling as illustrated from Fig.~\ref{fig9}(c) and \ref{fig9}(d). By contrast, the chiral $p_x+ip_y$ state does not support sign-change–induced ABS for specular reflection. Subgap transport is instead governed by chiral edge and propagating modes, leading to broader, lobe-like TMR patterns with smoother energy dependence as seen from Figs.~\ref{fig9}(e)–\ref{fig9}(h). Increasing $Z_R$ enhances interfacial spin mixing and polarizes the chiral modes, shifting the angular position of the TMR lobes without generating midgap resonances. Increasing $Z_0$ sharpens the energy selectivity of the response by driving the junction toward the tunneling regime, while preserving the smoother structure characteristic of the chiral state.  Overall, Fig.~\ref{fig9} shows that TMR provides symmetry-resolved behavior nodal $p_x$ pairing yields ABS-dominated, strongly structured near-zero-bias response, whereas chiral $p_x+ip_y$ pairing produces smoother, edge-mode–dominated features. RSOC and $Z_0$ act as electrical and structural knobs that control spin mixing and coherence, enabling large, symmetry-selective TMR in AM/TSC spin valves without ferromagnetic electrodes.

\begin{figure*}[t]
\centerline{ 
\includegraphics[scale=0.27]{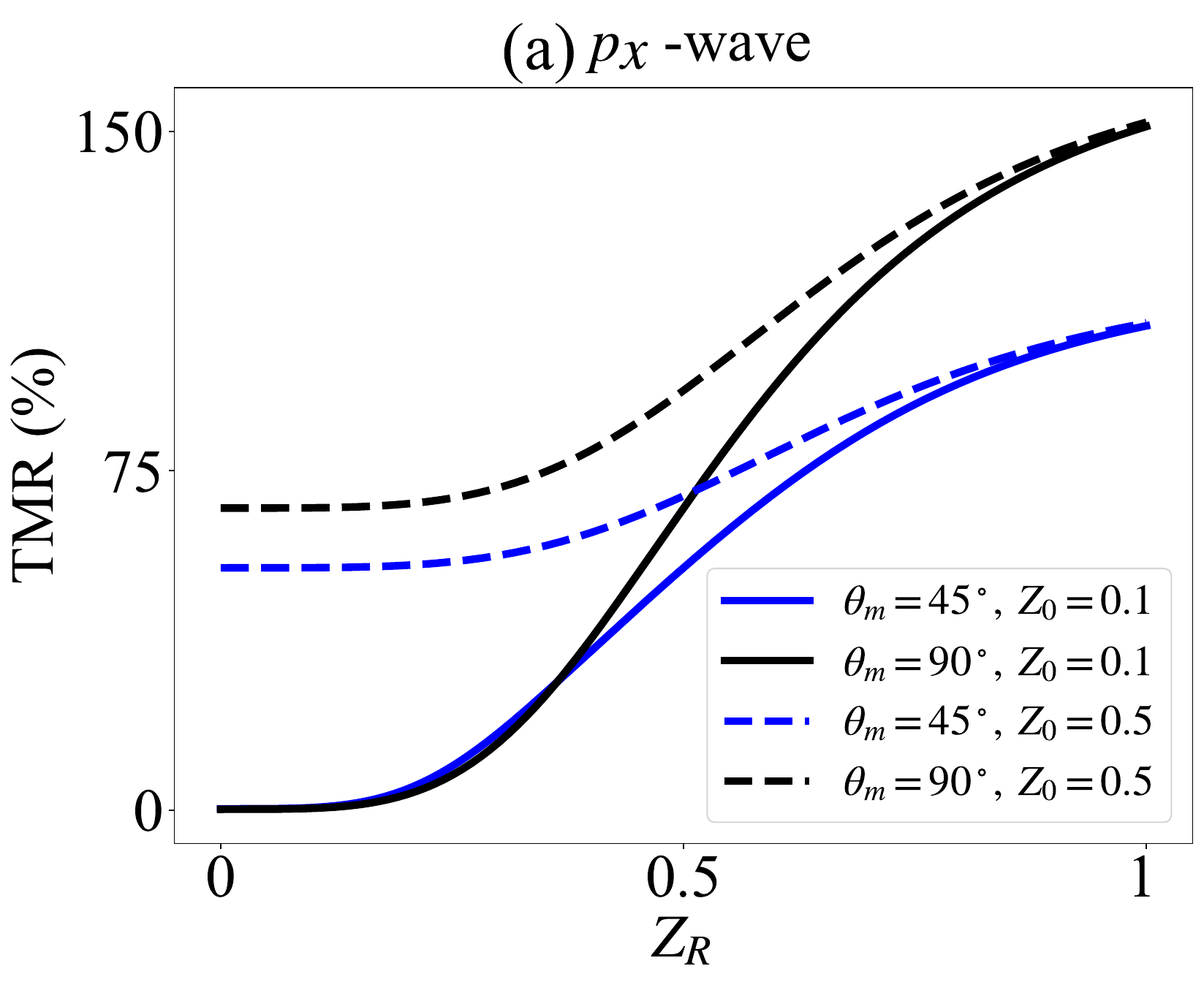}
\hspace{5mm}
\includegraphics[scale=0.27]{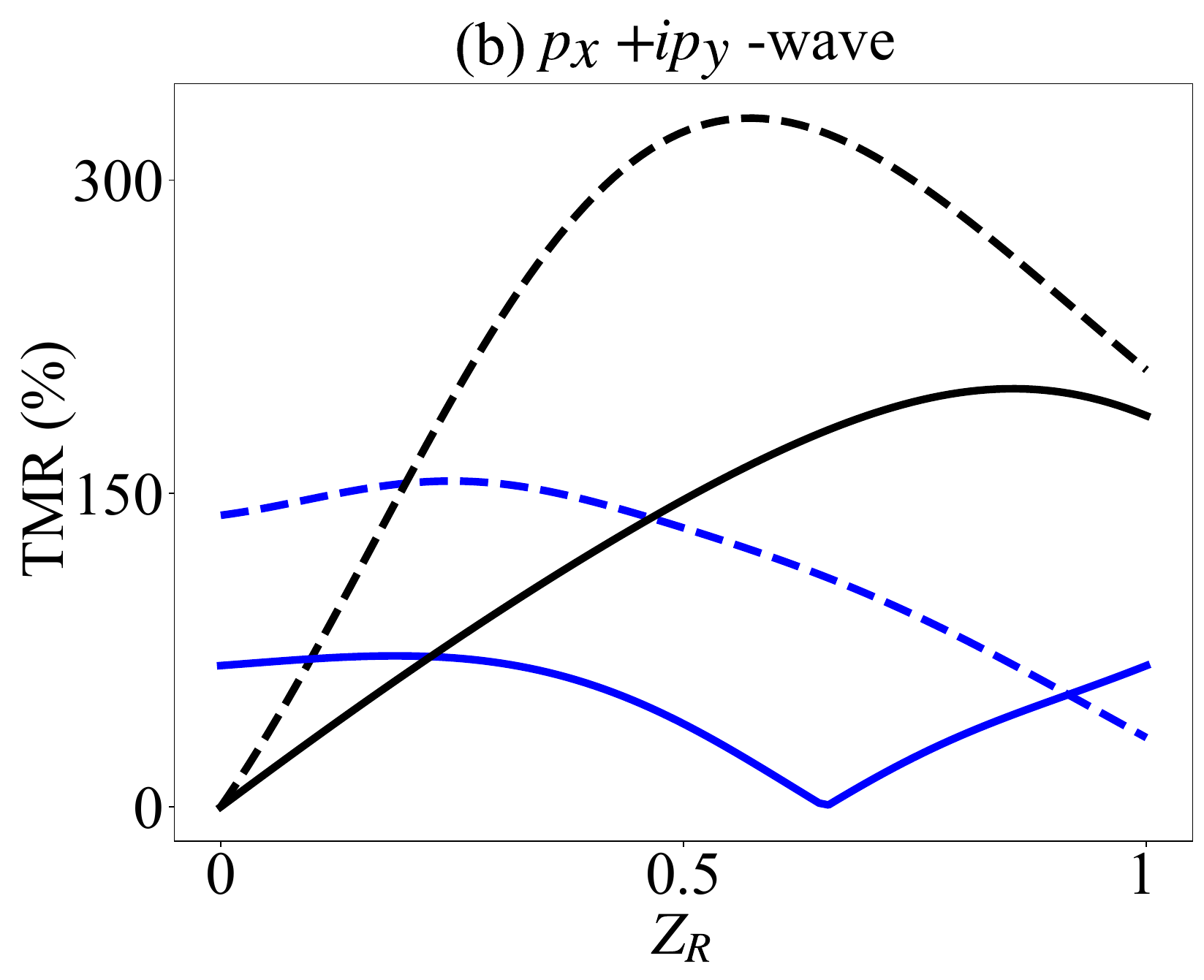}
}
\caption{TMR as a function of interfacial $Z_R$ for (a) $p_x$-wave and (b) chiral
$p_x+i p_y$ -  wave SC in an AM/TSC/AM junction. Solid and dashed curves corresponds to $Z_0 = 0.1$ and $Z_0 = 0.5$ respectively for different $\theta_m$ values.}
\label{fig10}
\end{figure*}
We note three key features: (i) the appearance of ring or arc-like interference patterns indicate coherent angle resolved resonances. For a given $\theta_m$ and $E$, the phase accumulated
on a full Andreev cycle can meet a constructive-interference condition, producing a
conductance maximum in one magnetic configuration but not the other, thereby produce large TMR. (ii) The AM provides momentum-dependent spin splitting, enabling strong spin-selective coupling to interface states without net magnetization. Thus we observe large TMR without any conventional ferromagnetic leads. (iii) Finally, the interplay of $Z_R$ and $Z_0$ controls whether the interface acts predominantly as a spin-preserving or spin-mixing scatterer: at low $Z_0$, RSOC mainly redistributes spectral weight among Andreev channels and generates complex interference patterns; at high $Z_0$, RSOC controls the spin selectivity of tunneling resonances and sets the sign and magnitude of TMR.

Fig.~\ref{fig10} shows the dependence of the TMR on $Z_R$ for different $\theta_m$ and $Z_0$ values. Panel (a) corresponds to $p_x$-wave while panel (b) is for chiral $p_x+i p_y$-wave SC. For the nodal $p_x$-wave SC, a monotonic increase of TMR with $Z_R$ is observed which reflects the growing spin selectivity of interface transport
mediated by sign change induced ABS. The order
parameter reverses sign under
specular reflection, enforcing the formation of zero-energy ABS that dominate
subgap conductance as already discussed. These ABS acquire spin-dependent spectral weights in the presence of AM exchange field as seen from Fig.~\ref{fig10}(a). With the increase in $Z_R$, the coupling between spin channels enhances and amplifies the imbalance between transmission probabilities in the two magnetic configurations. Thus it result in steady enhancement of TMR. Moreover, it is observed that larger $Z_0$ sharpen the energy selectivity of ABS-mediated transport, resulting a higher TMR, while the dependence on $\theta_m$ reflects
the angular structure of the AM exchange field.
In contrast, the chiral $p_x+i p_y$ -wave SC exhibits a qualitatively different and non-monotonic dependence of TMR on $Z_R$ as illustrated from Fig.~\ref{fig10}(b). In this case the subgap transport involves chiral edge modes and
spin-polarized propagating quasiparticles whose interference is strongly
affected by RSOC as already seen from Figs.~\ref{fig6}-\ref{fig9}. For $Z_0 = 0.1$, increasing  $Z_R$ enhances TMR by enabling additional
spin-dependent scattering channels. However, at larger $Z_R$, excessive
spin mixing reduces the contrast between the two magnetic configurations,
leading to suppression of TMR for certain $\theta_m$ and
$Z_0$. This competition produces the nonmonotonic behavior observed in
panel Fig.~\ref{fig10}(b), in sharp contrast to the ABS-dominated response of the nodal
$p_x$ case. Thus it can be concluded that RSOC play a very significant role in  TMR. While RSOC systematically enhances ABS-mediated spin filtering and TMR in the
$p_x$ state, it plays a dual role in the chiral state by both generating and
eventually degrading spin contrast through strong spin mixing. These trends
provide a clear transport-based distinction between nodal and chiral triplet
pairing symmetries in AM/TSC/AM junctions.

Recent experimental progress on AM materials provides realistic platforms for realizing the proposed AM/TSC/AM heterostructures, with candidate systems including $\mathrm{RuO_2}$ and $\mathrm{MnTe}$, which exhibit momentum-dependent spin splitting without net magnetization and are compatible with thin-film growth and heterostructure fabrication. Interfacial RSOC can be engineered via structural inversion asymmetry and electrostatic gating, offering an electrical knob to tune spin mixing and the resulting spin-valve response. Although intrinsic bulk spin-triplet superconductors are rare, equal-spin triplet pairing can be induced by proximity in heterostructures combining conventional superconductors with strong SOC and magnetic order. Together, these advances make the proposed AM/TSC/AM junctions experimentally feasible, with predicted conductance, spin polarization, and TMR providing direct transport probes of AM spin textures and unconventional triplet pairing symmetry.

\section{Conclusion}
In this work, we have presented a comprehensive theoretical study of
spin-resolved transport in altermagnet-triplet-superconductor-altermagnet
(AM/TSC/AM) junctions, with particular emphasis on how orbital symmetry of
triplet pairing, interfacial RSOC, and misalignment of N\'eel vector orientation in the AM regions collectively determine the transport properties. Using a
microscopic Bogoliubov-de Gennes formalism, we systematically analyzed angle-resolved conductance, spin polarization, zero-bias response, and TMR for nodal $p_x$-wave and chiral
$p_x+i p_y$-wave superconducting states. Our results demonstrate that nodal and chiral triplet superconductors exhibit
qualitatively distinct transport behaviours when coupled to AMs with interfacial RSOC. For the nodal $p_x$-state, the sign change of the order parameter under
specular reflection enforces the formation of interface Andreev bound states,
which dominate subgap transport and become strongly spin selective in the
presence of odd momentum AM exchange and interfacial RSOC. This mechanism leads to pronounced angular anisotropy,
giant zero-bias spin polarization, and large, tunable TMR, whose magnitude and
structure can be controlled by the N\'eel vector misalignment angle, barrier
transparency, and RSOC strength. In contrast, chiral 
$p_x+i p_y$ pairing is governed by topological edge modes, yielding smoother conductance spectra and polarization profiles with broader, lobe-like TMR features. As a result, the chiral case
exhibits smoother energy and angular dependence, non-monotonic RSOC sensitivity, and robust but less sharply resonant spin filtering.
Moreover, our results suggest that AMs provide a powerful and conceptually distinct platform for symmetry-controlled spin transport in
superconducting hetero-structures, enabling large spin polarization and TMR
without net magnetization or external magnetic fields. The strong sensitivity
of the transport response to pairing symmetry suggests that AM/TSC/AM junctions
can serve as phase-sensitive probes of triplet superconductivity, while the
electrical tunability afforded by RSOC offers a
route toward magnetization free superconducting spintronic devices.

\end{document}